\documentclass[aps,prd,12pt,nofootinbib,tightenlines,preprintnumbers,superscriptaddress]{revtex4-1}
\usepackage{graphicx}

\usepackage[utf8]{inputenc}

\usepackage{amssymb}
\usepackage{amsmath}
\usepackage{xcolor}
\usepackage{hyperref}
\hypersetup{colorlinks, linkcolor = [rgb]{0,0.0,0.75}, citecolor = [rgb]{0,0.0,0.75}, urlcolor = [rgb]{0,0.0,0.75}}

\makeatletter
\g@addto@macro\bfseries{\boldmath}
\makeatother

\newcommand{\be}{\begin{equation}}
\newcommand{\ee}{\end{equation}}
\newcommand{\ba}{\begin{eqnarray}}
\newcommand{\ea}{\end{eqnarray}}
\newcommand{\la}{\label}

\newcommand{\Tr}{\,\hbox{\rm Tr}}

\newcommand{\<}{\langle}
\renewcommand{\>}{\rangle}

\newcommand{\ycut}{|y|_{\rm cut}}

\newcommand{\amu}{a_\mu}
\newcommand{\ahlbl}{a_\mu^{\rm hlbl}}

\newcommand{\kernel}{\mathcal{\bar L}}
\newcommand{\Max}{\text{max}}

\newcommand{\ahlblpi}{a_\mu^{\mathrm{hlbl};\pi^0} }

\begin{document}
\title{Hadronic light-by-light contribution to $(g-2)_\mu$ \\ from lattice QCD with SU(3) flavor symmetry}

\preprint{MITP/20-036}
\preprint{CERN-TH-2020-109}
      
\author{En-Hung Chao}
\affiliation{PRISMA$^+$ Cluster of Excellence \& Institut f\"ur Kernphysik,
Johannes Gutenberg-Universit\"at Mainz,
D-55099 Mainz, Germany}

\author{Antoine G\'erardin}
\affiliation{Aix Marseille Univ, Universit\'{e} de Toulon, CNRS, CPT, Marseille, France.}

\author{Jeremy R.\ Green}
\affiliation{Theoretical Physics Department, CERN, 1211 Geneva 23, Switzerland}

\author{Renwick J.\ Hudspith} 
\affiliation{PRISMA$^+$ Cluster of Excellence \& Institut f\"ur Kernphysik,
Johannes Gutenberg-Universit\"at Mainz,
D-55099 Mainz, Germany}

\author{Harvey B.\ Meyer} 
\affiliation{PRISMA$^+$ Cluster of Excellence \& Institut f\"ur Kernphysik,
Johannes Gutenberg-Universit\"at Mainz,
D-55099 Mainz, Germany}
\affiliation{Helmholtz~Institut~Mainz,
Johannes Gutenberg-Universit\"at Mainz,
D-55099 Mainz, Germany}

\begin{abstract}
We perform a lattice QCD calculation of the hadronic light-by-light
contribution to $(g-2)_\mu$ at the SU(3) flavor-symmetric point
$m_\pi=m_K\simeq 420\,$MeV.  The representation used is based on
coordinate-space perturbation theory, with all QED elements of the
relevant Feynman diagrams implemented in continuum, infinite Euclidean
space.  As a consequence, the effect of using finite lattices to
evaluate the QCD four-point function of the electromagnetic current is
exponentially suppressed.  Thanks to the SU(3)-flavor symmetry, only
two topologies of diagrams contribute, the fully connected and the
leading disconnected. We show the equivalence in the continuum limit
of two methods of computing the connected contribution, and
introduce a sparse-grid technique for computing the disconnected
contribution. Thanks to our previous calculation of the pion transition form factor,
we are able to correct for the residual finite-size effects and extend the tail of the integrand.
We test our understanding of finite-size effects by using gauge ensembles differing only by their volume.
After a continuum extrapolation based on four lattice spacings, we obtain $\ahlbl = (65.4\pm 4.9 \pm 6.6)\times 10^{-11}$,
where the first error results from the uncertainties on the individual gauge ensembles
and the second is the systematic error of the continuum extrapolation.
Finally, we estimate how this value will change as the light-quark masses are lowered to their physical values.
\end{abstract}

\date{\today}

\maketitle

\section{Introduction}

Electrons and muons carry a magnetic moment aligned with their
spin. The proportionality factor between the two axial vectors is
parameterized by the gyromagnetic ratio $g$. In Dirac's theory, $g=2$,
and for a lepton family $\ell$ one characterizes the deviation of $g$
from this reference value by $a_\ell=(g-2)_\ell/2$. Historically, the
ability of Quantum Electrodynamics (QED) to quantitatively predict
this observable played a crucial role in establishing quantum field
theory as the framework in which particle physics theories are
formulated.

Presently, the achieved experimental precision on the measurement of
the anomalous magnetic moment of the muon~\cite{Bennett:2006fi},
$a_\mu$, is 540\,ppb.  At this level of precision, such a measurement tests not
only QED, but also the effects of the weak and the strong interaction
of the Standard Model (SM) of particle physics. Currently there exists 
a tension of about 3.7 standard deviations between the SM
prediction and the experimental measurement. The status of this test of the SM
is reviewed in~\cite{Jegerlehner:2009ry,Blum:2013xva,Jegerlehner:2017gek,Aoyama:2020ynm}.
At the time of writing, the E989 experiment at Fermilab is performing a new direct
measurement of $a_\mu$~\cite{Grange:2015fou}, and a further experiment
using a different experimental technique is planned at
J-PARC~\cite{Mibe:2011zz}.  The final goal of these experiments is to
reduce the uncertainty on $a_\mu$ by a factor of four.  A commensurate reduction of
the theory error is thus of paramount importance.

The precision of the SM prediction for $a_\mu$ is completely dominated
by hadronic uncertainties. The leading hadronic contribution enters at
second order in the fine-structure constant $\alpha$ via the vacuum
polarization and must be determined at the few-permille level in order
to match the upcoming precision of the direct measurements of
$a_\mu$. The most accurate determination comes from the use of
$e^+e^-\to{\rm hadrons}$ data via a dispersion relation, although
lattice QCD calculations have made significant progress in computing
this quantity from first principles~\cite{Meyer:2018til,Aoyama:2020ynm}.  The
hadronic light-by-light (HLbL) scattering contribution $\ahlbl$, which is of
third order in $\alpha$, currently contributes at a comparable level
to the theory uncertainty budget and is being addressed both by
dispersive and lattice methods;
see~\cite{Blum:2016lnc,Asmussen:2018oip,Colangelo:2017urn} and
references therein.

Our approach for determining $\ahlbl$ is based on coordinate-space perturbation theory where the QED elements of the Feynman diagrams
yielding $\ahlbl$ are precomputed in infinite volume, and only the 
four-point amplitude of the electromagnetic current is actually computed 
on the lattice. Here we compute the lattice contribution at a point in the space of light quark masses
corresponding to QCD with exact SU(3)-flavor (denoted SU(3)$_{\rm f}$ in the rest of this work) symmetry. Furthermore, the sum of the three light quark masses is
approximately the same as in nature. These two conditions leads to a
degenerate mass of pions, kaons and the eta meson of about 420\,MeV.

Our motivation for calculating $\ahlbl$ at the SU(3)$_{\rm f}$-symmetric point
is twofold.  First, the lattice calculation itself is simplified in
that only two out of five classes of Wick contractions contribute, due
to the vanishing trace of the quark electric charge matrix. In addition,
the overall lattice calculation is computationally far cheaper 
than for physical quark masses, so that more tests of systematic
errors can be performed. Second, the interpretation of the results
is simplified: the SU(3)$_{\rm f}$-symmetry reduces the
number of unknown parameters in model estimates based on the exchange
of the lightest mesons. In particular, the transition form factor
(TFF) of the pion, which describes the coupling of the neutral pion to
two virtual photons, has been calculated~\cite{Gerardin:2019vio} on
the lattice ensembles that we use. The TFF of the eta meson coincides
with the TFF of the pion up to a simple overall charge factor. Of the
pseudoscalar mesons, only the TFF of the $\eta'$ remains independent
and is largely unknown at the SU(3)$_{\rm f}$-symmetric point, however
experimental information is available for the two-photon decay width
(which provides the coupling strength to two real photons) and some
experimental results are available for the singly as well as the doubly-virtual form factor~\cite{Gronberg:1997fj,BABAR:2011ad,BaBar:2018zpn}, although only for relatively large virtualities above 1.5~GeV$^2$. 

The simplified connection to model estimates enables our work to provide
a valuable cross-check
for the predictions of hadronic models and dispersive methods; this work is
thereby complementary to lattice calculations directly aiming at
$\ahlbl$ for physical quark masses~\cite{Blum:2019ugy}. At the same
time, this study allows us to learn about the size of various sources
of systematic error, particularly the finite-size effects, and how
well we are able to correct for them semi-analytically.

The rest of this paper is organized as follows: We begin by presenting our methodology in Section \ref{sec:meth}, including the two methods we will investigate for computing the quark-connected contribution to $\ahlbl$. Section \ref{sec:latparms} begins with a description of the lattice ensembles used in this work, as well as an example of the lowest-lying relevant meson spectrum for one of our ensembles. We then discuss the lattice determination of the $\pi^0$ and $\eta$ transition form factors used for the modelling and finite-size correction of our data. Section \ref{sec:intgnd} discusses the various model predictions for the integrand at the SU$(3)_{\rm f}$-symmetric point and confronts these with a selection of our lattice data.
Results for the fully-connected class of Wick
contractions are presented in Section~\ref{sec:res_conn}, and those
for the non-vanishing disconnected class in Section~\ref{sec:disc}.  In both cases, the lattice results are compared to the prediction for the exchange of pseudoscalar mesons at the integrand level. The main result of this paper --- $a_\mu^{\rm hlbl}$ at the SU(3)$_{\rm f}$-symmetric point,
Eq.\ (\ref{eq:final_amu}) --- is obtained in Section~\ref{sec:combined},
which also contains a discussion of how this result will change for physical values of the quark masses.
We summarize our findings and conclude in Section~\ref{sec:concl}, and various technical aspects of the calculation are described in more detail in the appendices (\ref{app:pi0matching}, \ref{app:leploop}, \ref{app:WI}).
 
\section{Methodology}\label{sec:meth}

\begin{figure}[t]
	\includegraphics*[height=0.18\linewidth]{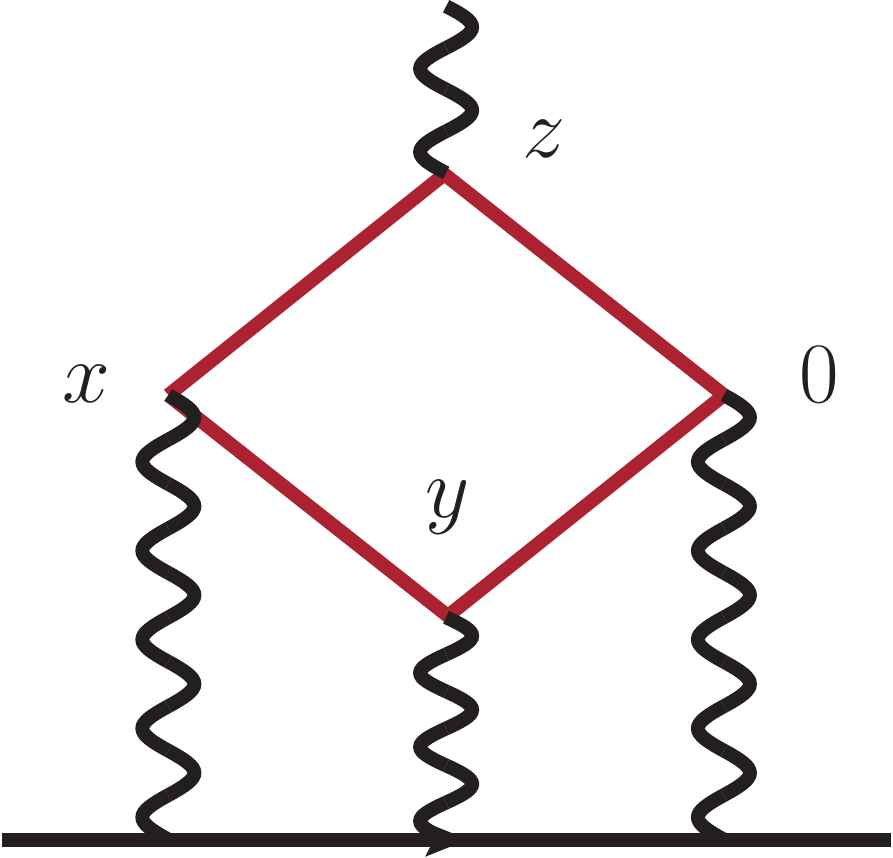} \hspace{2cm}
	\includegraphics*[height=0.18\linewidth]{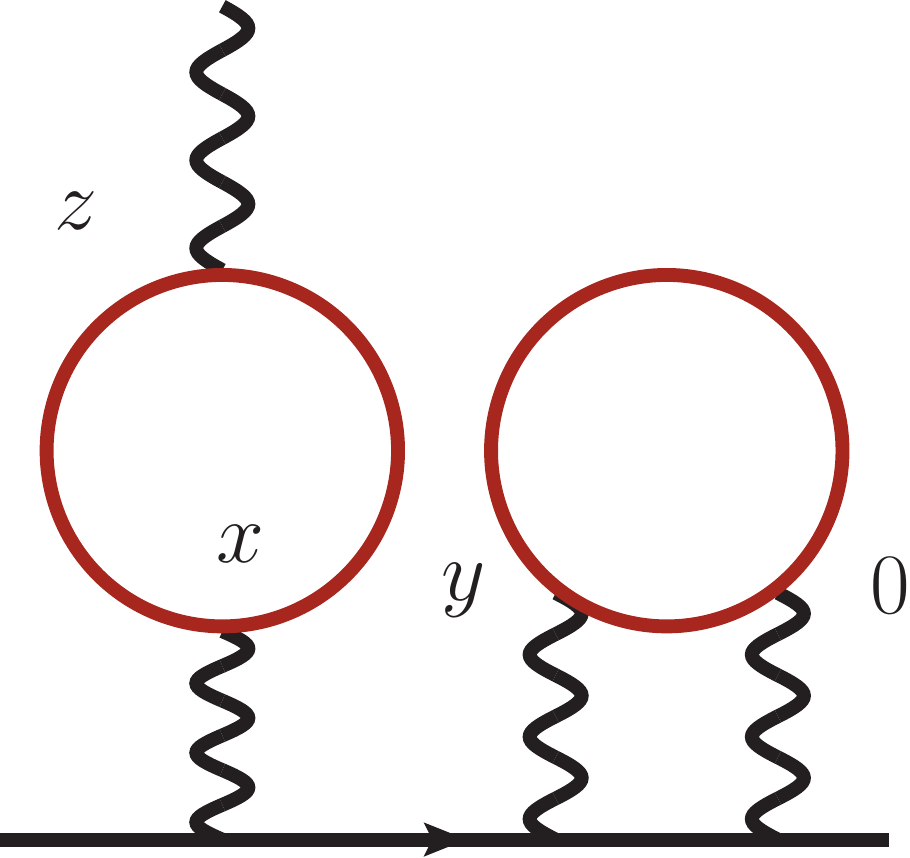} \\
	\includegraphics*[height=0.18\linewidth]{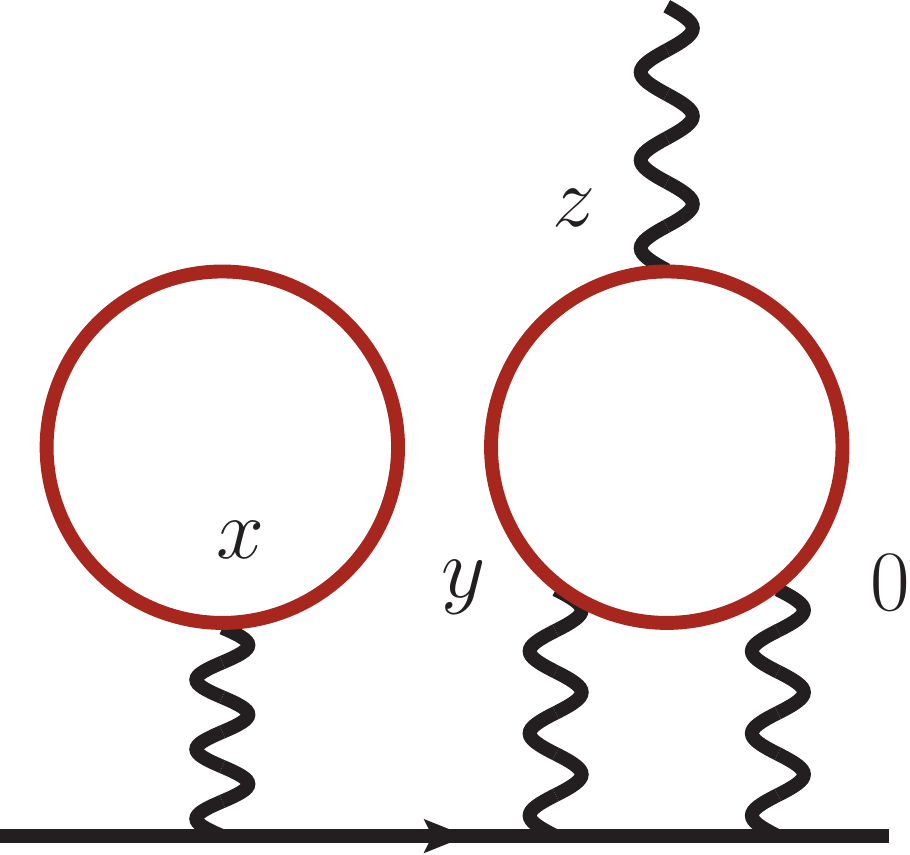} \hspace{1.5cm}
	\includegraphics*[height=0.18\linewidth]{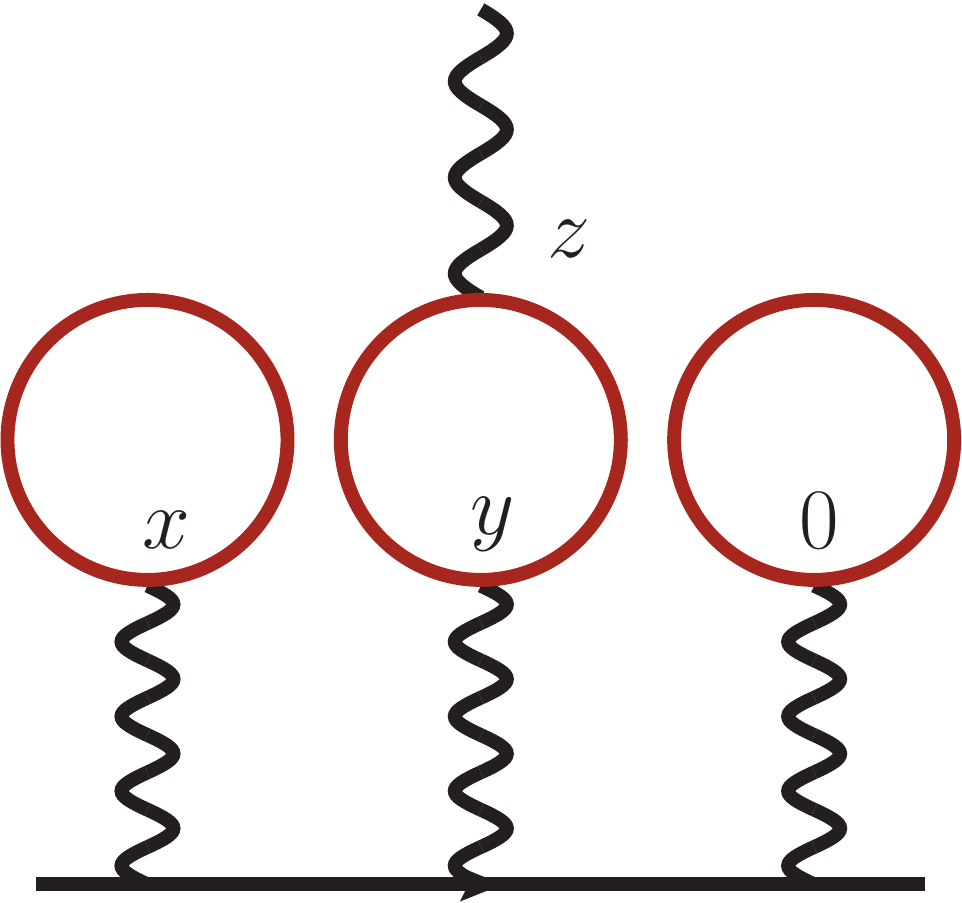} \hspace{1.5cm}
	\includegraphics*[height=0.18\linewidth]{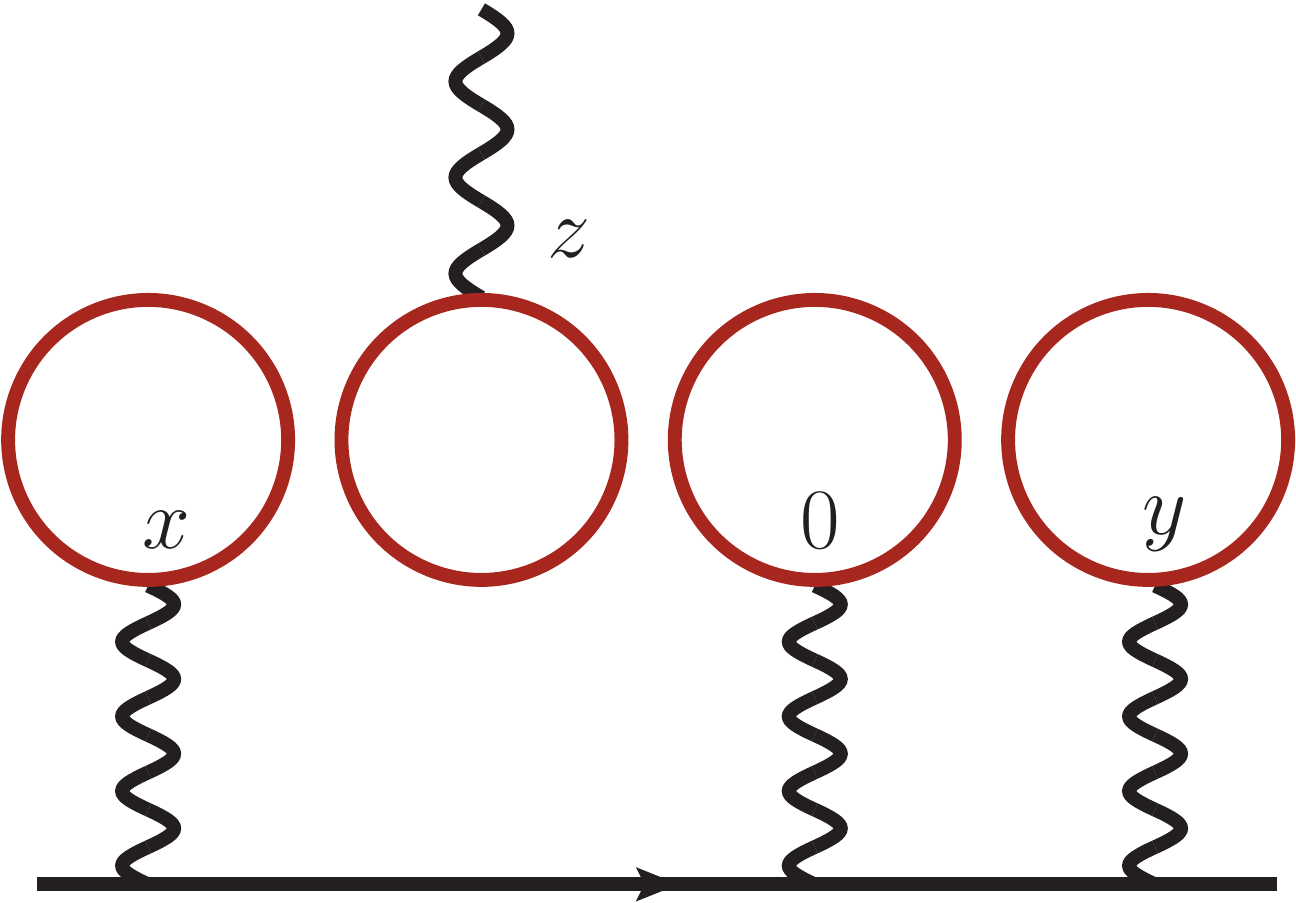} 
			
	\caption{Five Wick-contraction topologies that are necessary for the calculation of $\ahlbl$ as listed in Table~\ref{tab:wick}. For each diagram, nontrivial permutations of the four quark-photon vertices yield the number of contractions listed in that table. The two diagrams in the first line are the dominant ones and the other three vanish in the SU$(3)_{\rm f}$ limit.}
	\label{fig:wicktopo}
\end{figure}

One can compute the light-by-light scattering contribution to the $g-2$ of the muon in position space by performing the integrals
\begin{equation}\label{eq:master}
\ahlbl = \frac{m_\mu e^6}{3}\int d^4y \int d^4x \; \kernel_{[\rho,\sigma];\mu\nu\lambda}(x,y)\;i\widehat\Pi_{\rho;\mu\nu\lambda\sigma}(x,y),
\end{equation}
where $\kernel$ is a QED kernel and $i\widehat\Pi$ is a spatial moment of the connected Euclidean four-point function in QCD,
\begin{eqnarray}
i\widehat \Pi_{\rho;\mu\nu\lambda\sigma}( x, y)  &=& -\int d^4z\,  z_\rho\, \widetilde\Pi_{\mu\nu\sigma\lambda}(x,y,z), \label{eq:pihat}
\\ 
\widetilde\Pi_{\mu\nu\sigma\lambda}(x,y,z)&\equiv&\Big\<\,j_\mu(x)\,j_\nu(y)\,j_\sigma(z)\, j_\lambda(0)\Big\>_{\rm QCD}\,,
\end{eqnarray}
with $j_\mu(x)$ the hadronic component of the electromagnetic current
\begin{equation}
j_\mu(x) = \frac{2}{3} (\overline{u} \gamma_{\mu} u)(x) - \frac{1}{3} (\overline{d} \gamma_{\mu} d)(x) - \frac{1}{3} (\overline{s} \gamma_{\mu} s)(x)  \,.
\end{equation}

The QCD correlation function consists of all the various ways one can contract four vector currents, as shown in Fig.~\ref{fig:wicktopo}, all of which are ``disconnected'' except for the fully-connected contribution. At the flavor-symmetric point, only the upper two topologies contribute. Away from this point it is expected, from large-$N_c$ arguments as well as from numerical evidence by the RBC/UKQCD collaboration~\cite{Blum:2016lnc}, that the remaining topologies are suppressed. The number of contractions for each topology is given in Table~\ref{tab:wick}.

\begin{table}[h!]
\centering
\begin{tabular}{c|c|c|c|c|c}
\toprule
& conn & $2+2$ & $3+1$ & $2+1+1$ & $1+1+1+1$ \\
\hline
$m_l\neq m_s$ & 6 & 3 & 8 & 6 & 1 \\ 
$m_l=m_s$ & 6 & 3 & 0 & 0 & 0 \\ 
\botrule
\end{tabular}
\caption{Number of contractions needed for each type of diagram in Fig.~\ref{fig:wicktopo}.}
\label{tab:wick}
\end{table}

More information on the infinite-volume QED kernel $\kernel(x,y)$ can be found in \cite{mainzHLbL1}. We make use of $O(4)$ symmetry to simplify the integral further,
\begin{equation}
\ahlbl = \int_0^\infty d|y|\, f(|y|),
\label{eq:masterM1}
\end{equation}
where in the starting representation (\ref{eq:master}), 
\begin{equation}\label{eq:integrand}
f(|y|) = \frac{m_\mu e^6}{3} 2\pi^2  |y|^3 \int d^4x \; \kernel_{[\rho,\sigma];\mu\nu\lambda}(x,y)\;i\widehat\Pi_{\rho;\mu\nu\lambda\sigma}(x,y).
\end{equation}
We will often display this integrand and always denote it by $f(|y|)$, even though
in practice we employ modified representations of $\ahlbl$. 
One type of modification concerns the kernel
$\kernel_{[\rho,\sigma];\mu\nu\lambda}(x,y)$.  Various subtraction
terms to the kernel have been proposed to beneficially change the
shape of the integrand \cite{Blum:2016lnc,Asmussen:2019act} without
changing the resulting integral. The importance of performing such
subtractions cannot be understated as the unsubtracted kernel is
poorly suited for practical lattice simulations due to being too
peaked at short distances.

Here we make extensive use of a new subtraction scheme for the QED kernel~\cite{Asmussen:2019act},
\begin{equation}\label{eq:lamsub}
\begin{aligned}
\kernel^{(\Lambda)}_{[\rho,\sigma];\mu\nu\lambda}(x,y) = &\kernel_{[\rho,\sigma];\mu\nu\lambda}(x,y)\\ 
	&-\partial_\mu^{(x)} (x_\alpha e^{-\Lambda m_\mu^2 x^2/2}) \kernel_{[\rho,\sigma];\alpha\nu\lambda}(0,y) \\&- \partial_\nu^{(y)} (y_\alpha e^{-\Lambda m_\mu^2 y^2/2})\kernel_{[\rho,\sigma];\mu\alpha\lambda}(x,0),
\end{aligned}
\end{equation}
where $\Lambda$ is an arbitrary, tuneable, dimensionless free
parameter. When $\Lambda=0$ we have the $\kernel^{(2)}$ kernel
of~\cite{mainzHLbL1} and as $\Lambda\rightarrow\infty$ we
recover the unsubtracted $\kernel^{(0)}$. The benefit of such a choice
of kernel is that we are able to tune the shape of the integrand to
reduce the long-distance effects while still preserving the beneficial
properties of short-distance subtractions. An investigation of this kernel with the
infinite-volume lepton loop is presented in Appendix~\ref{app:leploop}.

In this section we will write the formulae for obtaining $\ahlbl$ in terms of continuous integrals. The lattice, however, is discrete so we can only approximate these integrals with finite sums\footnote{If using open boundary conditions (as we mostly do in this work) some care in truncating the temporal sum is required as to not incorporate boundary effects.},
\begin{equation}
\int d^4x \approx a^4\sum_{t/a=-N_t/2}^{N_t/2-1}\:\:\sum_{z/a=-N_z/2}^{N_z/2-1}\:\:\sum_{y/a=-N_y/2}^{N_y/2-1}\:\:\sum_{x/a=-N_x/2}^{N_x/2-1},
\end{equation}
where $a$ is the lattice spacing and $N_\mu=L_\mu/a$. In Eq.~\eqref{eq:pihat},
we set $z_\rho=L/2\rightarrow 0$ to accommodate the discontinuity when it changes sign. 
The integral of $f(|y|)$ with respect to $|y|$ can be performed with the trapezoid rule and in practice we will average over equivalent values of $f(|y|)$ to both increase statistical precision and reduce computational cost. Later on in the analysis we will show results for the partially-integrated value of $a_\mu$,
\begin{equation}
\amu(|y|_\Max) = \int_0^{|y|_\Max} d|y| f(|y|)
\end{equation}
with the hope that we will see a plateau at large-enough $|y|_\Max$, indicating that our integral has saturated.

In the following three subsections we will discuss two methods to compute the connected contribution: the first, Method 1, is a direct calculation of the three pairs of connected Wick contractions; the second, Method 2, uses rearrangements of the integrand expressed in terms of only one pair of Wick contraction to make the calculation cheaper. We will then give the methodology used for the calculation of the disconnected contribution, which also uses some rearrangements in the integrand.

\subsection{Connected contribution (Method 1)}

\begin{figure}[t]
	\includegraphics*[width=0.18\linewidth]{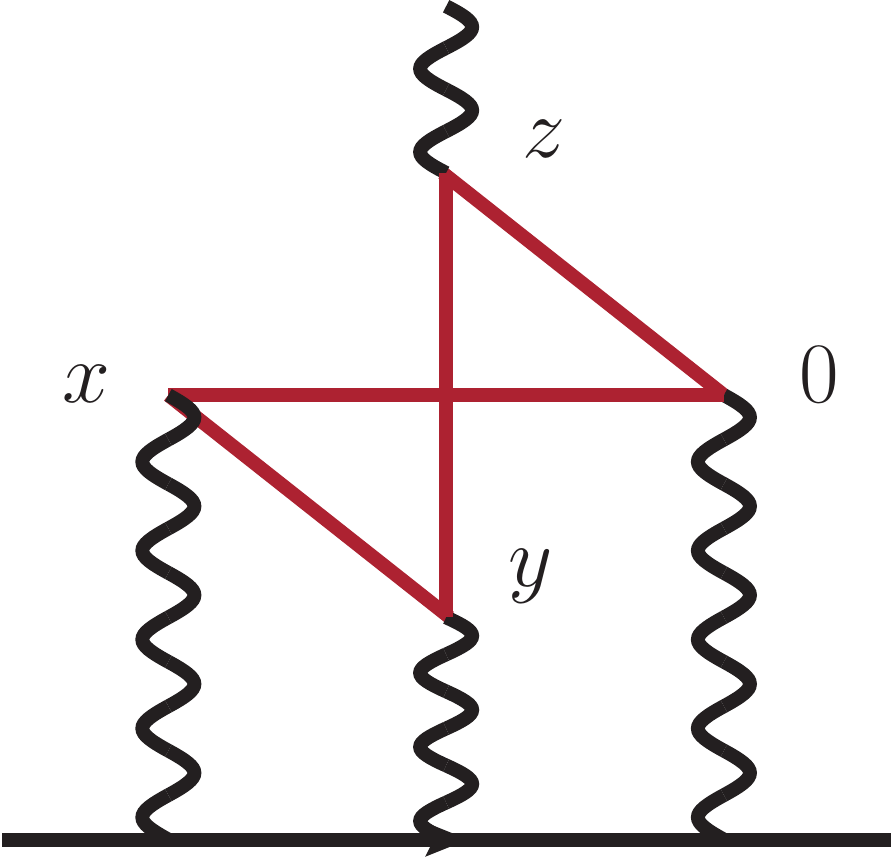} \hspace{1cm}
	\includegraphics*[width=0.18\linewidth]{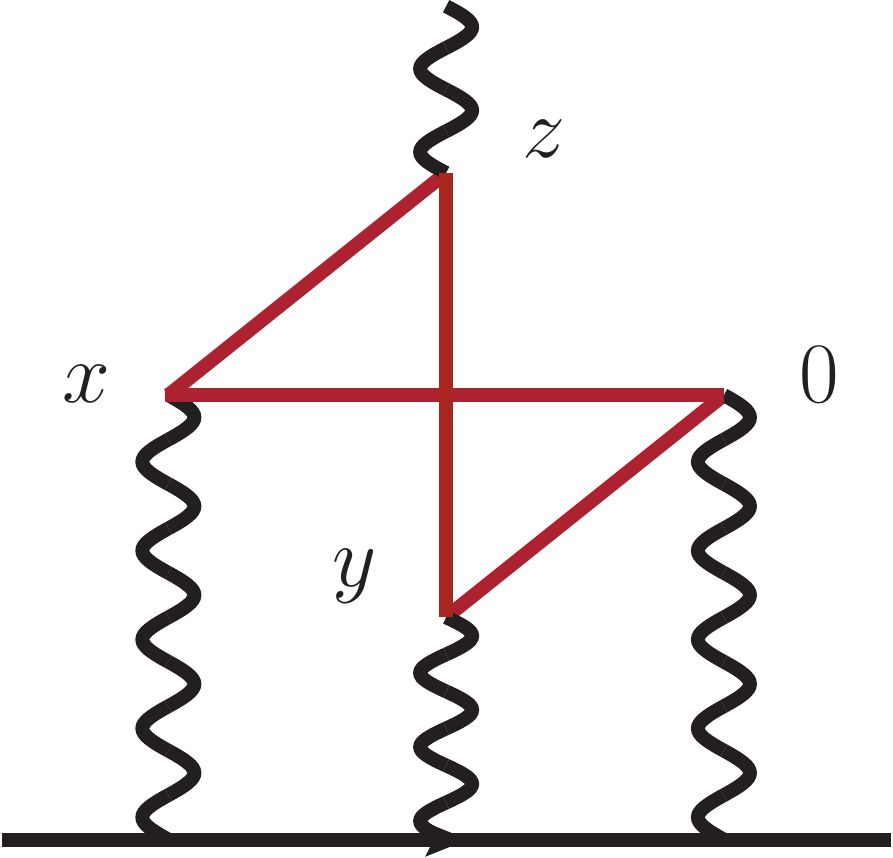} \hspace{1cm}
	\includegraphics*[width=0.18\linewidth]{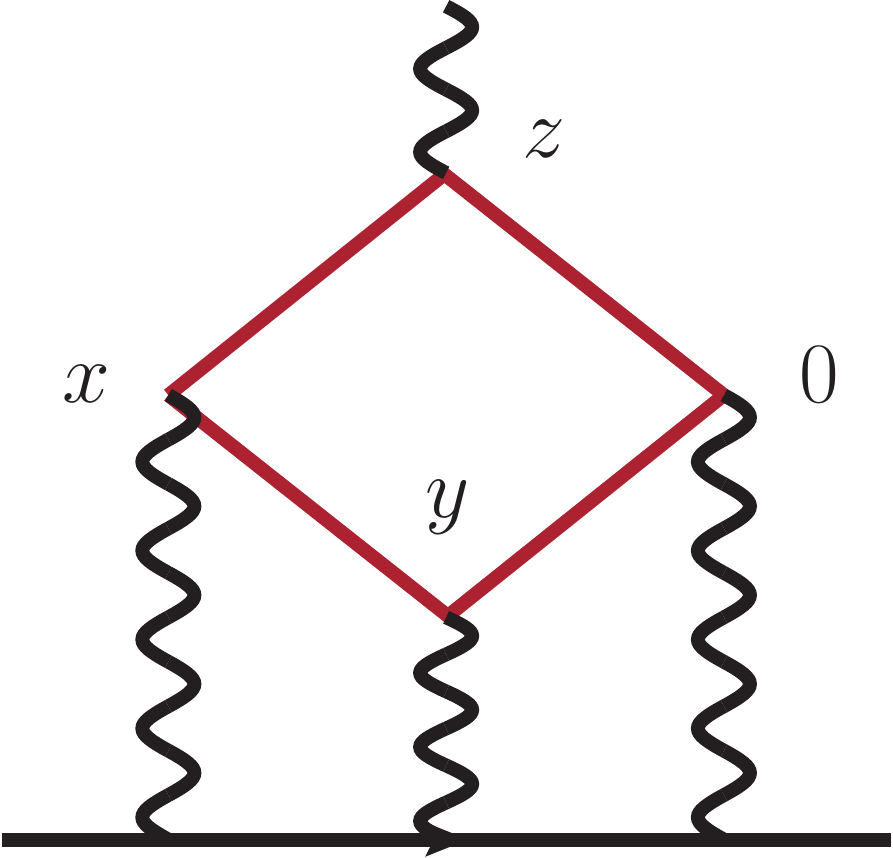}
			
	\caption{Wick contractions for the connected contribution. Each diagram represents two contractions with quark flow in opposite directions.}
	\label{fig:wickM1}
\end{figure}

The Wick contractions for the connected contribution are shown in Fig.~\ref{fig:wickM1}. With four local vector currents, the four-point correlation function can be written as
\begin{equation}\label{eq:piconn}
\widetilde\Pi^{\rm conn}_{\mu\nu\sigma\lambda}(x,y,z) = \frac{18}{81} Z_V^4 \left( \widetilde\Pi_{\mu\nu\sigma\lambda}^{(1)}(x,y,z) + \widetilde\Pi_{\mu\nu\sigma\lambda}^{(2)}(x,y,z) + \widetilde\Pi_{\mu\nu\sigma\lambda}^{(3)}(x,y,z) \right) \,,
\end{equation}
where $Z_V$ is the renormalization factor of the vector current\footnote{Note that the fourth power in $Z_V$ in Eq.~\eqref{eq:piconn} comes from the use of four local currents. A conserved current does not require such a factor. This should thus be adapted correspondingly in some particular cases where the effect of conserved currents is studied in this work.} and each term represents one diagram in Fig.~\ref{fig:wickM1}, i.e.\ a pair of Wick contractions.
We will use the $Z_V$ values from \cite{Gerardin:2018kpy} without the $O(a)$-improvement terms proportional to quark mass combinations. The value $18/81$ is the necessary charge factor for the degenerate light and strange quarks. Writing the $\widetilde\Pi^{(1,2,3)}$s in terms of propagators yields
\begin{equation}\label{eq:diagrams_conn}
\begin{aligned}
\widetilde\Pi_{\mu\nu\sigma\lambda}^{(1)}(x,y,z) = & -2\text{Re}\left\langle\Tr\left[ S(0,x) \gamma_{\mu} S(x,y) \gamma_{\nu} S(y,z) \gamma_{\sigma} S(z,0) \gamma_{\lambda} \right]\right\rangle_U,\\
\widetilde\Pi_{\mu\nu\sigma\lambda}^{(2)}(x,y,z) = & -2\text{Re}\left\langle\Tr\left[ S(0,y) \gamma_{\nu} S(y,z) \gamma_{\sigma} S(z,x) \gamma_{\mu} S(x,0) \gamma_{\lambda} \right]\right\rangle_U, \\
\widetilde\Pi_{\mu\nu\sigma\lambda}^{(3)}(x,y,z) = & -2\text{Re}\left\langle\Tr\left[ S(0,y) \gamma_{\nu} S(y,x) \gamma_{\mu} S(x,z) \gamma_{\sigma} S(z,0) \gamma_{\lambda} \right]\right\rangle_U,
\end{aligned}
\end{equation}
where $S(x,y)$ is the quark propagator with sink at $x$ and source at $y$ and $\langle\cdots\rangle_U$ is the expectation value over gauge configurations. The trace is over both Dirac and color indices. In Method 1, the six contractions are computed explicitly and we choose the direction $y/a \propto (1,1,1,1)$\footnote{We use the notation ($x$,$y$,$z$,$t$) where time is the last component}. In this method, we first compute the point-to-all propagator $S(\cdot,0)$ and the six sequential propagators using the fields $z_{[\rho} \gamma_{\sigma]} S(z,0)$ as sources (the anti-symmetrization of the indices $\rho$ and $\sigma$ is imposed by the symmetries of the QED kernel). Then, for each value of $|y|$ used to sample the integrand in Eq.~(\ref{eq:masterM1}), we compute the point-to-all propagator $S(\cdot, y)$ and the six sequential propagators using the fields $z_{[\rho} \gamma_{\sigma]} S(z,y)$ as sources. Therefore, for $N$ evaluations of the integrand, we need $7(N+1)$ propagator inversions. 
In our set-up all currents are local except for the one at $x$, which will be either local or the point-split, the latter implying a suitable modification of Eq.~\eqref{eq:diagrams_conn}.

In practice, since we are (mostly) using open-boundary conditions in the time direction, the origin is located somewhere near the middle time-slice and randomly distributed in the spatial volume. We also average over the 16 combinations $y/a = (\pm n, \pm n, \pm n, \pm n)$ to increase statistics. 

\subsection{Connected contribution (Method 2)}

The idea for Method 2 is simple: we pick a reference diagram that is easiest to compute and use a change of variables in the integrals to relate the other diagrams to this reference. Here, we pick the diagram that does not have a propagator from the origin to $y$, or from $z$ to $x$ (the leftmost of Fig.~\ref{fig:wickM1}). Such a choice avoids the extra inversions required for the sequential sources as we discussed in Method 1. Simplistically, for two samples of a single $|y|$ (i.e.\ $+y$ and $-y$) we only need two point-to-all propagators. In fact, we can do much better than this if we keep propagators in memory so that they can be reused.

The diagrams $\widetilde\Pi^{(2)}$ and $\widetilde\Pi^{(3)}$ are
related to $\widetilde\Pi^{(1)}$ by applying symmetries to
Eq.~\eqref{eq:diagrams_conn}: Euclidean-space translations and
inversions, along with cyclicity of the trace~\cite{Asmussen:2019act,
 mainzHLbL1}. This leads to our master formula for Method 2:
\begin{equation}\label{eq:Method2}
\begin{gathered}
a_\mu^{\rm conn} = -\frac{18}{81}Z_{\rm V}^4 \frac{m_\mu e^6}{3}2\pi^2\int d|y| |y|^3 \;\int d^4x \\
 \bigg((\kernel^{(\Lambda)}_{[\rho,\sigma];\mu\nu\lambda}(x,y)+\kernel^{(\Lambda)}_{[\rho,\sigma];\nu\mu\lambda}(y,x)-\kernel^{(\Lambda)}_{[\rho,\sigma];\lambda\nu\mu}(x,x-y))\int d^4z \:z_\rho \widetilde\Pi^{(1)}_{\mu\nu\sigma\lambda}(x,y,z)\\
+\kernel^{(\Lambda)}_{[\rho,\sigma];\lambda\nu\mu}(x,x-y)x_\rho \int d^4z \:\widetilde\Pi^{(1)}_{\mu\nu\sigma\lambda}(x,y,z)\bigg).
\end{gathered}
\end{equation}
In infinite volume, the integral is equal to the result of Method 1,
but the integrand $f(|y|)$ will generally be different. As a result,
the systematic effects in Method 2 will be different from Method 1.

We invert point-source propagators along the line $(n,n,n,3n+t_\text{min}/a)$, and what we call $|y|$ will be the difference between these points; here the integer $n$ ranges from 0 to $N_i/2$. The closest source position in time ($t_\text{min}$) is chosen to be suitably away from our (usually) open temporal boundary, ideally $m_\pi t_{\text{min}}>4$ and $m_\pi(L_t-t_\text{max})>4$.  This line of propagator sources was chosen as we typically have a large anisotropy in the temporal direction, allowing us to achieve sizeable values of $|y|$.

In our implementation, we keep all $N_i/2$ of the propagators in memory and perform the integrals for every possible $y$ and origin, so for $N$ propagator inversions we have $N(N-1)$ samples distributed among the different values of non-zero $|y|$. To further boost statistics we will also average other directions that give the same $|y|$, e.g. $y/a=(\pm n,\pm n,\pm n,t_\text{max}/a-3n)$.

\subsection{Disconnected contribution}

The disconnected contribution can be computed from the two-point contraction
\begin{equation}\label{eq:2p2_Pi}
\Pi_{\mu\nu}(x,y)= -\text{Re}\left(\text{Tr}[S(y,x)\gamma_\mu S(x,y)\gamma_\nu]\right).
\end{equation}
An important point to note is that $\Pi_{\mu\nu}$ has a vacuum expectation value (VeV) that must be subtracted to ensure that the two ``disconnected'' quark loops are still connected by gluons. To this end, we define
\begin{equation}
\hat{\Pi}_{\mu\nu}(x,y) = \Pi_{\mu\nu}(x,y) - \langle \Pi_{\mu\nu}(x,y) \rangle_U,
\end{equation}
and use this to compute the $2+2$ quark-disconnected contribution to $\ahlbl$,
\begin{equation}
\begin{aligned}
a_\mu^{\rm disc} =
&-\frac{36}{81}Z_{\rm V}^4\frac{m_\mu e^6}{3}2\pi^2\int d|y| |y|^3 \int d^4x \\
\biggl\langle
&(\kernel^{(\Lambda)}_{[\rho,\sigma];\mu\nu\lambda}(x,y)+\kernel^{(\Lambda)}_{[\rho,\sigma];\nu\mu\lambda}(y,x)) 
\hat{\Pi}_{\mu\lambda}(x,0)\int d^4z \:z_\rho
\hat{\Pi}_{\sigma\nu}(z,y)\\
&+\kernel^{(\Lambda)}_{[\rho,\sigma];\mu\nu\lambda}(x,y)\hat{\Pi}_{\mu\nu}(x,y)
\int d^4z \: z_\rho \hat{\Pi}_{\sigma\lambda}(z,0)
\biggr\rangle_U.
\end{aligned}
\end{equation}

In our implementation we compute a grid of point-source propagators from some $t_{\text{min}}$ to $t_{\text{max}}$ wrapping completely around the spatial directions alternating between $(2n,2n,2n,6m+t_\text{min}/a)$ and $(2n+1,2n+1,2n+1,6m+3+t_\text{min}/a)$ where $n$ and $m$ take values between $0$ and $N_i/2-1$, and $0$ and $(t_\text{max}-t_{\text{min}})/(6a)$ respectively.

\begin{figure}[h!]
\centering
\includegraphics[trim={5cm 14cm 5cm 4cm},clip,scale=0.7]{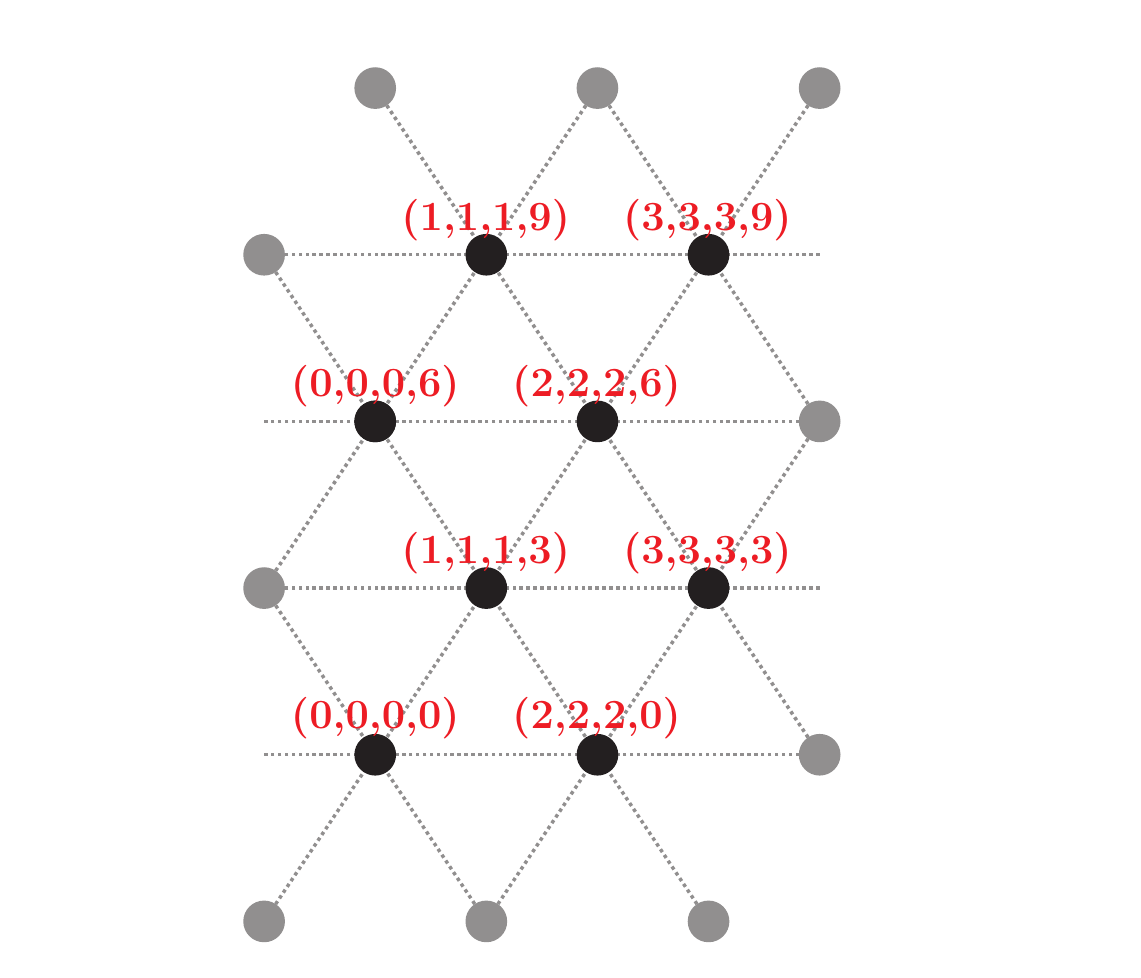}
\caption{Illustration of the grid of propagators used in the toy example given in the text. Filled circles indicate sites that lie on the lattice, gray circles are ones accessible through periodicity and the dashed lines indicate lines along our preferred directions $(1,1,1,3)$ and $(2,2,2,0)$.}\label{fig:gridex}
\end{figure}

Giving a toy example; for a $4^3\times 12$ periodic lattice we would invert point-source propagators at $(0,0,0,0),(2,2,2,0),(1,1,1,3),(3,3,3,3),(0,0,0,6),(2,2,2,6),(1,1,1,9),$ and $(3,3,3,9)$. This is illustrated in Fig.~\ref{fig:gridex}. We then compute the two-point contraction of Eq.~(\ref{eq:2p2_Pi}) for each of these sources and keep it in memory. We then perform a doubly-nested loop over all of the source positions, where we only integrate combinations of two-point contractions where our $y$ lies in a direction we like, e.g. along the directions $(1,1,1,3)$ and $(2,2,2,0)$. The results of the $z$ and $x$ integrations are saved to disk, the VeV is subtracted, and the final contraction of indices is performed offline in the analysis.

The benefit of such a method over calculating each separate $|y|$ individually is the quadratic growth of equivalent combinations of $|y|$ that are available as the number of source fields grows. For our $4^3\times 12$ example we have $6$ values of equivalent $|y|$ per source position. It would be impractical to perform this calculation without exploiting this fact. We find it is beneficial to truncate the $x$-integral in our set-up to avoid negligible, noisy contributions at large distances. As our kernels are computed on-the-fly this is also useful as a computer-time saving measure as it reduces the number of QED kernel calls. We truncate the integral over $x$ to points that lie within a maximum distance [$(r/a)^2=81,81,121,169$ for our coarsest to finest ensembles respectively] from the origin or from $y$, while we do not truncate the $z$-integral. Although these truncations differ in physical volume, the ensembles H101 and H200 were tested for smaller and larger values of $(r/a)^2$ on a subset of our data and even smaller values of $(r/a)^2$ than used here, for example $49$ and $64$ for the ensemble H101, were found to be consistent.

To reiterate, in our full calculation we do not perform the integrals for all possible $y$-vectors, as there are many that we expect to have bad finite-volume or discretization effects. For instance, in the toy example we could calculate $f(|y|)$ for the $y$-vector $y/a=(0,0,0,6)$, but this would have significant cut-off effects. Typically, we filter-out about $\approx 80\%$ of the possible $y$-vectors and keep only (the modulus of) those parallel to $(1,1,1,3)$, $(2,2,2,0)$, and occasionally $(1,1,1,1)$.
	
\section{Lattice parameters and properties of \texorpdfstring{SU$(3)_{\rm f}$}{SU(3)-flavor}-symmetric QCD \la{sec:latparms}}

\begin{figure}
  \centering
  \includegraphics[width=0.5\textwidth]{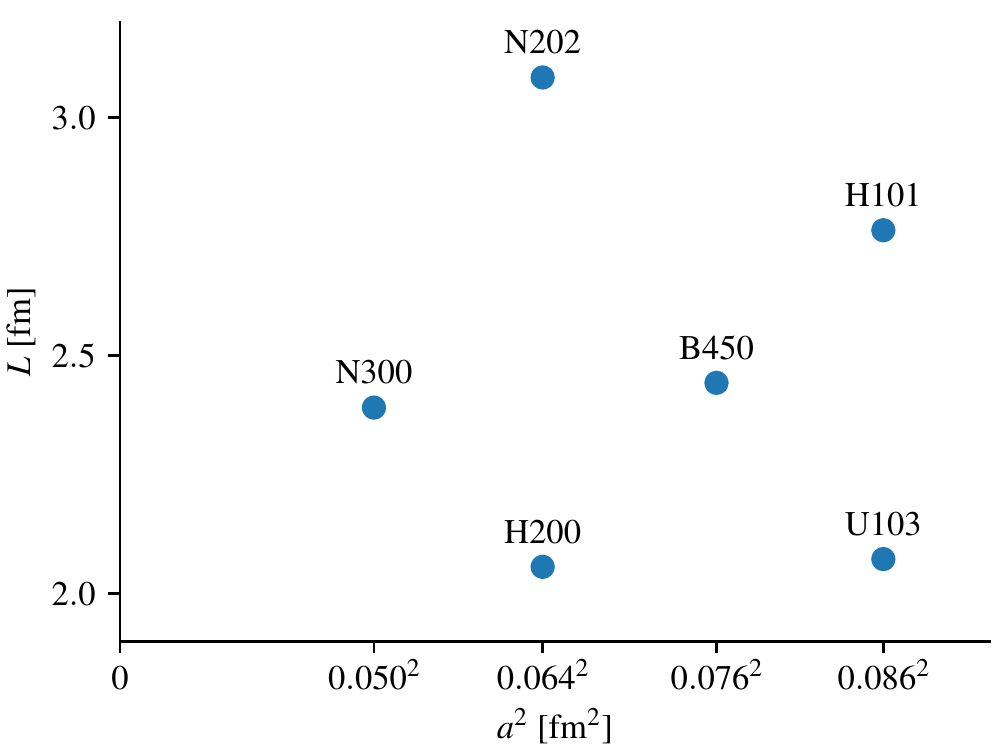}
  \caption{Spatial extent $L$ and lattice spacing $a$ for the 
    ensembles with SU$(3)_{\rm f}$-symmetry used in this work.}
  \label{fig:landscape}
\end{figure}

Calculations have been performed on lattice ensembles provided by the
Coordinated Lattice Simulations (CLS) initiative~\cite{Bruno:2014jqa},
which have been generated using three flavors of non-perturbatively
$O(a)$-improved Wilson-clover fermions and with the tree-level
$O(a^2)$-improved Lüscher-Weisz (Symanzik) gauge action. In particular, we consider only those with SU$(3)_{\rm f}$-symmetry.
On these ensembles, the mass of
the octet of light pseudoscalar mesons is approximately 420~MeV. These
ensembles are summarized in Table~\ref{tab:ensembles} and in
Fig.~\ref{fig:landscape}: there are four lattice spacings, as well as
two pairs of ensembles that differ only by their volume.

\begin{table}
  \centering
  \begin{tabular}{c|cllc|lcc}
  \toprule
Label & $\beta$ & $\kappa$ & size & bdy.\ cond. & $a$ (fm) & $m_{\pi,K,\eta}$ (MeV) & $m_\pi L$ \\
\hline
U103 & 3.40 & 0.13675962 & $24^3\times 128$ & open & 0.08636(98)(40) & 415(5) & 4.4 \\
H101 & 3.40 & 0.13675962 & $32^3\times 96$  & open & 0.08636(98)(40) & 418(5) & 5.9 \\
B450 & 3.46 & 0.13689 & $32^3\times 64$ & periodic & 0.07634(92)(31) & 417(5) & 5.2 \\
H200 & 3.55 & 0.137      & $32^3\times 96$  & open & 0.06426(74)(17) & 421(5) & 4.4 \\
N202 & 3.55 & 0.137      & $48^3\times 128$ & open & 0.06426(74)(17) & 412(5) & 6.4 \\
N300 & 3.70 & 0.137      & $48^3\times 128$ & open & 0.04981(56)(10) & 421(5) & 5.1 \\
\botrule
  \end{tabular}
  \caption{Lattice ensembles with SU$(3)_{\rm f}$-symmetry used in this work.
    Each ensemble is parametrized by the gauge coupling parameter $\beta$, 
    the quark hopping parameter $\kappa$, the lattice size, and the temporal
    boundary condition. The lattice spacing $a$ was determined in
    Ref.~\cite{Bruno:2016plf}.}
  \label{tab:ensembles}
\end{table}

\subsection{The \texorpdfstring{$\text{SU}(3)_{\rm f}$}{SU(3)-flavor} meson spectrum}\label{subsec:spectroscopy}

At long distances, the leading contributions to the four-point
correlator emanate from the lowest-lying mesons. Clearly, the degenerate
$\pi^0$ and $\eta$-pole contributions dominate at the longest
distances. However, at intermediate distances the contributions from heavier states,
including resonances in the two-pseudoscalar channel, may also play a 
significant r\^ole. Especially since in our calculation at the SU$(3)_{\rm f}$-point the pseudoscalar mesons are not appreciably light compared to the other mesons. Therefore, we present results from a limited spectroscopic study of the low-lying meson spectrum in all relevant channels with total angular momentum $J\leq 1$.

Only charge-conjugation-even ($C$-even) mesons can couple to two
electromagnetic currents and thus contribute in exchange diagrams to
the four-point function. However, from the Wick-contraction structure,
one can view Method 2 as using charged vector currents (see
Appendix~\ref{app:pi0matching}) and it is possible to couple a $C$-odd
state to two of them. Therefore, the integrand for Method~2 can
receive contributions from $C$-odd mesons such as the 
rho~\cite{Asmussen:2019act}. For that reason, we also inspect 
the spectrum of $C$-odd mesons, even if their contributions to $\ahlbl$  must vanish in the infinite-volume limit once all integrals are performed.

Our dedicated spectroscopy calculation is performed on
the ensemble H101.
This made use of the distillation framework~\cite{Peardon:2009gh}
(for the connected hadron two-point functions) and its stochastic
formulation~\cite{Morningstar:2011ka}, which was essential
for efficiently computing the
disconnected hadron two-point functions needed in the singlet sector.
However, we have only used smeared quark bilinears as interpolating
operators, and the lack of non-local multimeson-like interpolators
means that this calculation should not generally be considered as a robust
determination of the spectrum. This analysis precludes the use of finite-volume
quantization conditions; therefore, only the approximate locations of
resonances can be found, provided that they are narrow.
Nevertheless, since we compute diagonal correlators, the effective masses
taken at any Euclidean time provide an upper bound on the ground-state energy in a given channel. This observation is particularly useful in the flavor-singlet $0^{++}$ channel, which turns out to admit a stable ground-state meson. Such a stable scalar meson has been found previously~\cite{Briceno:2016mjc} in a lattice calculation at a similar pion mass, though not at an SU(3)$_{\rm f}$-symmetric point.

\begin{figure*}
  \centering
  \includegraphics[width=\textwidth]{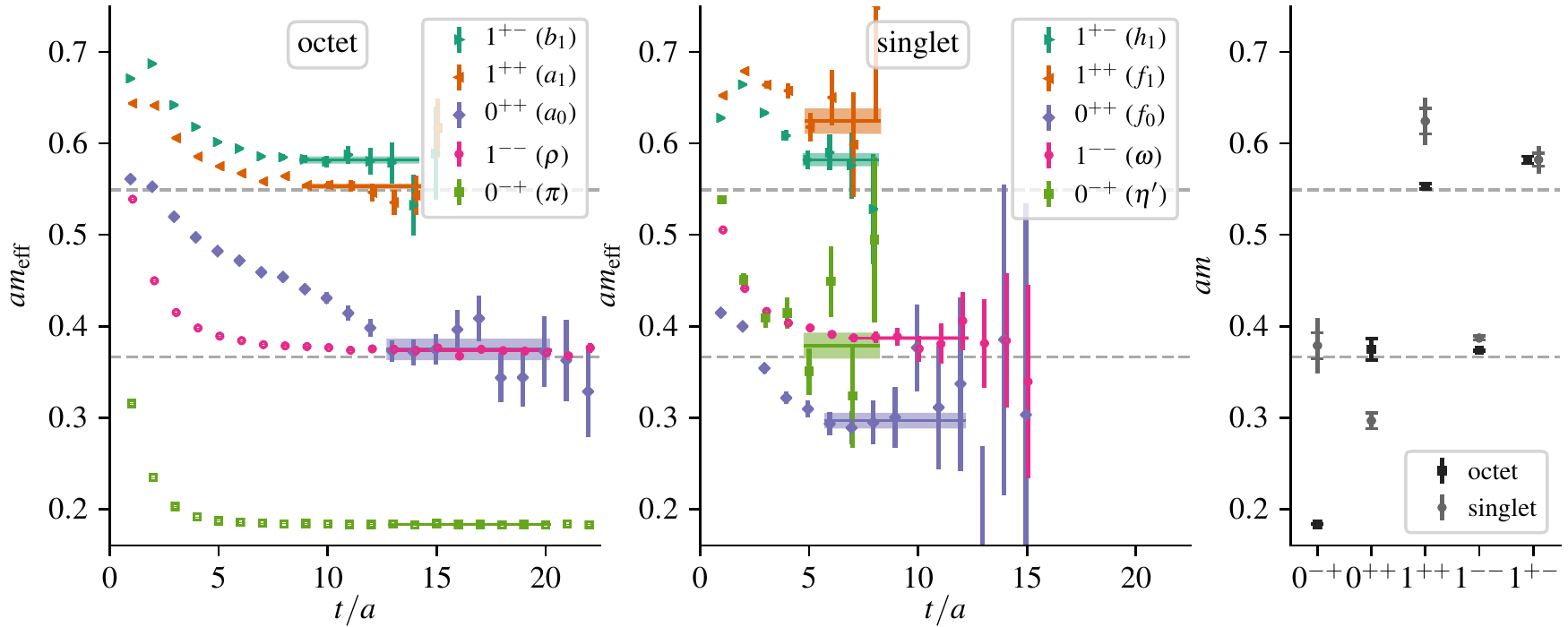}
  \caption{The meson spectrum of ensemble H101. Effective
    masses are shown for each channel in the flavor octet (left) and
    singlet (center) sectors, along with bands indicating the plateau
    fits. Note that in many cases (in particular the octet
    pseudoscalar and vector channels), the error bar is smaller than
    the plotted symbol. Fitted masses are shown in the right panel;
    the outer error bars include an estimate of systematic uncertainty
    obtained by shifting the plateau fit range. Horizontal dashed
    lines indicate thresholds at twice and three times the octet
    pseudoscalar mass.}
  \label{fig:spectrum}
\end{figure*}

\begin{table}
  \centering
  \begin{tabular}{c|ll|ll}
  \toprule
  $J^{PC}$ & \multicolumn{2}{c|}{octet} & \multicolumn{2}{c}{singlet} \\\hline
  $0^{-+}$ & \ $\pi$ \ & \hphantom{0}418(1)(0) & \ $\eta'$ \ & \hphantom{0}865(32)(61) \\
  $0^{++}$ & \ $a_0$ \ & \hphantom{0}856(27)(12) & \ $f_0$ \ & \hphantom{0}678(20)(10) \\
  $1^{++}$ & \ $a_1$ \ & 1264(7)(9) & \ $f_1$ \ & 1427(32)(50) \\
  $1^{--}$ & \ $\rho$ \ & \hphantom{0}853(2)(1) & \ $\omega$ \ & \hphantom{0}884(5)(3) \\
  $1^{+-}$ & \ $b_1$ \ & 1329(9)(3) & \ $h_1$ \ & 1330(16)(30) \\
  \botrule    
  \end{tabular}
  \caption{Estimated meson masses in MeV from ensemble H101. The first
    uncertainty is statistical and the second was determined by varying
    the plateau fit range. The scale-setting uncertainty has not been
    included.}
  \label{tab:spectrum}
\end{table}

Results are shown in Fig.~\ref{fig:spectrum} and summarized in
Table~\ref{tab:spectrum}. Note that since the
signal in the flavor-singlet sector is much worse, the plateau fits
had to be done at relatively short time separations, so that the shown
uncertainty on the mass may be an underestimate.
In addition, the plateau for the flavor-octet scalar is relatively poor, 
which might be due to its coupling to
two octet pseudoscalars in S-wave and the plateau's proximity to the
corresponding threshold. In the case of a mis-identified plateau, the
true ground-state energy would be lower, since the effective masses
shown here are expected to approach the ground-state energy from above.

For the flavor-octet sector, after the $\pi^0$, the
next-longest-distance meson-exchange contributions in the integrand
for $\ahlbl$ come from the $a_0$ and (for the Method~2 integrand) the
$\rho$, both of which sit near $2m_\pi$. For the flavor-singlet
sector, the three corresponding mesons are also the lightest, although
the ordering is different, with the $f_0$ (or $\sigma$) being a bound
state with mass near 680~MeV and the $\eta'$ mass sitting higher,
somewhere near $2m_\pi$. For the disconnected diagrams, which
receive the difference between contributions from the exchanges of
singlet and octet mesons (see section \ref{sec:intgnd_predn}), the
pseudoscalars will provide the longest-distance contribution but we
should also expect a significant contribution from scalars. The
difference between the octet and singlet vector meson masses is very
small; if the same holds for their form factors, then their combined
contribution to the integrand will be negligible.

\subsection{Pion-pole and \texorpdfstring{$\eta$}{eta}-pole contributions to \texorpdfstring{$\ahlbl$}{a-mu HLBL}}\label{sec:pipole_tff}

In Ref.~\cite{Gerardin:2019vio} a model independent parametrization of the pion TFF has been obtained on the same set of lattice ensembles as used in this work. These results can be used to compute the pion-pole contribution, $\ahlblpi$, at the SU$(3)_{\rm f}$-symmetric point in the continuum limit.
This result will be used in Section~\ref{sec:physpt} to obtain a rough estimate of $\ahlbl$ at the physical pion mass. 
More importantly, in Sections~\ref{sec:meth1},~\ref{sec:meth2}, and~\ref{sec:disc}, a vector meson dominance (VMD) parametrization of the TFF will be used to estimate the finite-size corrections to our lattice data at the symmetric point. 
For this purpose, we have performed a new fit to the data from \cite{Gerardin:2019vio} assuming a VMD parametrization, where we have restricted the fit to the singly-virtual case where the model provides a good description of our data. The fit parameters used in this work are collected in Table~\ref{tab:vmdparam}.
 
\begin{figure}
  \centering
  \includegraphics[width=0.64\textwidth]{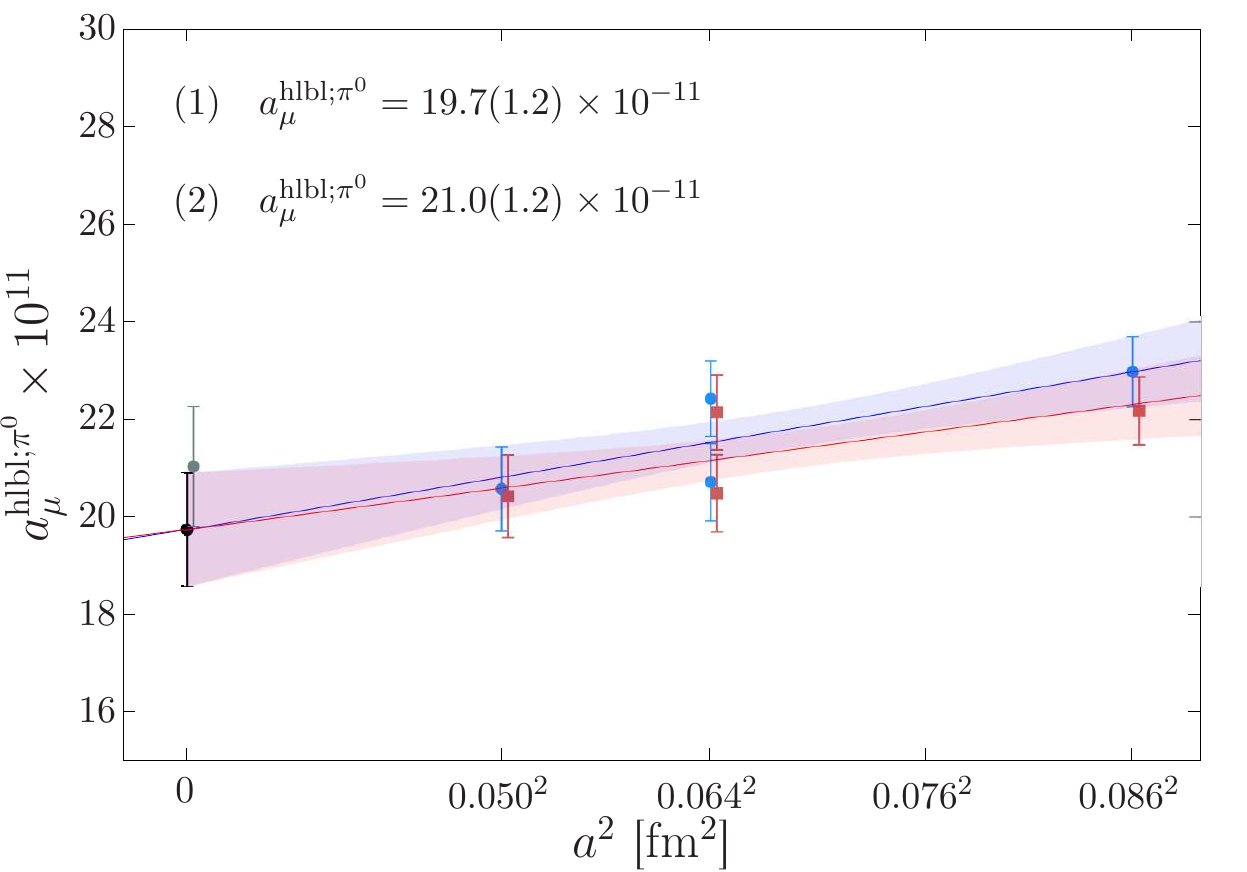}
  \caption{The pion-pole contribution to $\ahlbl$ at the SU(3)$_{\rm f}$ symmetric point with a pion mass of 420~MeV. Blue and red points correspond to two different discretizations of the 3-pt correlation function. The results (1) and (2) are explained in the main text.}
  \label{fig:pi0lattice}
\end{figure}

In~\cite{Gerardin:2019vio}, two strategies have been used to extract the pion TFF, and the corresponding results for the pion-pole contribution are displayed in Fig.~\ref{fig:pi0lattice} for two different discretizations of the 3-point correlation function (see~\cite{Gerardin:2019vio} for details).
In the first strategy, the TFF was computed on each lattice ensemble separately. This allowed us to determine the pion-pole contribution for different values of the pion mass and lattice spacing. The physical value was then obtained by a combined chiral-continuum extrapolation of $\ahlblpi$. We have repeated this analysis but now restrict the fit to only the ensembles included in this study. Here we use the pion mass of the given ensemble (instead of the physical one as done in~\cite{Gerardin:2019vio}) in the weight functions that appear in Eq.~(74) of~\cite{Gerardin:2019vio}. This leads to the result (1) of Fig.~\ref{fig:pi0lattice}.

In the second strategy, the pion TFF was directly extrapolated to the physical point using a global fit that includes several ensembles including, and away from, the SU$(3)_{\rm f}$-point. From the resulting fit parameters, we can extract the pion TFF in the continuum limit at a pion mass of 420~MeV. Using this result in Eq.~(74) of~\cite{Gerardin:2019vio}, we obtain the grey point of Fig.~\ref{fig:pi0lattice}, which we find to be in very good agreement with the first estimate.

\begin{table}
\begin{tabular}{l@{\quad}c@{\quad}r}
\toprule
Label & ${\cal F}_{\pi^0\gamma\gamma}(0,0)$$\,[{\rm GeV}^{-1}]$ & $M_V$\,[MeV] \\ 
\hline
H101 & 0.237(5) &  921(13) \\
 B450 & 0.230(5) &  942(25) \\
 H200 & 0.219(6) &  979(26) \\
 N202 & 0.227(5) &  952(15) \\
N300 & 0.214(5) &  1001(23) \\
\botrule
\end{tabular}
\caption{\label{tab:VMDparams}The VMD fit parameters of the $\pi^0$ transition form factor.}
\label{tab:vmdparam}
\end{table}

The second strategy has the advantage of using an expanded set of ensembles (15 in total) to determine the TFF,  
\be\la{eq:amupi0su3}
\ahlblpi=21.0(1.2) \times 10^{-11} \qquad (\textrm{SU$(3)_{\rm f}$ point})
\ee
at the SU$(3)_{\rm f}$-point which is to be compared to the physical-pion value 
\be\la{eq:amupi0phys}
\ahlblpi = 59.7(3.6)\times 10^{-11} \qquad (\textrm{physical point}).
\ee
Unsurprisingly, we observe a strong dependence on the pion mass. 
The smaller pion contribution in Eq.~\eqref{eq:amupi0su3} compared with \eqref{eq:amupi0phys} is due roughly in equal parts to the heavier pion mass and to the reduced coupling to photons, 
as can be seen by comparing the entries in Table~\ref{tab:vmdparam} to the physical value, ${\cal F}_{\pi^0\gamma\gamma}(0,0)\approx 0.274\,{\rm GeV}^{-1}$.

At the SU$(3)_{\rm f}$-symmetric point, the $\eta$ is mass-degenerate with the pion, $m_\eta \simeq 420\,$MeV, and contributes exactly $1/3$ of the $\pi^0$ exchange to $\ahlbl$, i.e. 
\be\la{eq:amuetasu3}
a_\mu^{\rm hlbl;\eta}=(7.0\pm0.4)\times 10^{-11} \qquad (\textrm{SU$(3)_{\rm f}$-point}).
\ee
A lattice calculation at the physical point for this contribution is not yet available, but the most recent estimate\footnote{An older estimate is $a_\mu^{\rm hlbl,\eta,{\rm phys}}=14.5\times 10^{-11}$ using the VMD model~\cite{Nyffeler:2016gnb}.}, 
which comes from Canterbury approximants~\cite{Masjuan:2017tvw,Aoyama:2020ynm}, is $a_\mu^{\rm hlbl,\eta,{\rm phys}}=16.3\times 10^{-11}$.
This comparison shows that at the physical point the $\eta$
gives a much larger contribution to $\ahlbl$ than at the SU$(3)_{\rm f}$-symmetric point,
in spite of being heavier.
This can be traced back to its much larger coupling to two photons,
\be
{\cal F}_{\eta\gamma\gamma}(0,0)[{\rm GeV}^{-1}]  \simeq \left\{\begin{array}{ll}  0.12 & \textrm{SU$(3)_{\rm f}$ point}, \\ 0.27 & \textrm{physical point} .
 \end{array}\right.
\ee
 
\section{Hadronic models vs.\  the lattice integrand \texorpdfstring{$f(|y|)$}{f(|y|)} \la{sec:intgnd}}

In subsection \ref{sec:intgnd_predn}, we state the theoretical
predictions for the integrand $f(|y|)$ corresponding to the
quark-connected diagrams in Method 1 and Method 2, as well as for the
disconnected diagram.  Then, in subsection \ref{sec:conn_intgnd}, we compare the lattice
$|y|$-integrand obtained by Method~1 for the quark-connected contribution to hadronic model predictions.
For this purpose, we will focus on the ensembles N202 and H200. N202 has the largest physical volume ($L=3.08$\,fm) of all ensembles considered in this work, and H200 (with $L=2.06$\,fm) differs only by its comparatively smaller volume.  Since these only differ by their volume,
they allow us to test our understanding of finite-volume
effects. Finally, a comparison of the theory predictions for the
integrand of the quark-disconnected contributions to the corresponding
lattice data is made in subsection \ref{sec:disc_intgnd}.

\subsection{Predictions for the  integrand}\la{sec:intgnd_predn}

In order to gain some insight into the various contributions to the
quantity $\ahlbl$, we will compare predictions for the pseudoscalar
exchanges as well as the contribution from a constituent-quark loop, and a charged-pseudoscalar loop, to the lattice data.
Here, we provide some details as to how these predictions are obtained.
Expressions for the amplitudes $i\widehat\Pi_{\rho;\mu\nu\lambda\sigma}$
calculated with a fermion loop, a charged-pion loop, or with the pseudoscalar exchange 
can be found in~\cite{Asmussen:2017bup,mainzHLbL1}.
In QCD with exact SU(3)$_{\rm f}$-symmetry, the $\eta$ contribution to
$\ahlbl$ is always one third of the contribution of the $\pi^0$.

The flavor structure of single-meson exchanges in the different
Wick-contraction topologies of four-point functions was discussed in
detail in Ref.~\cite{Gerardin:2017ryf}, including the case of $N_{\rm f}=3$
QCD. 
The quark-connected diagrams receive contributions only from
flavor-octet mesons, enhanced\footnote{The enhancement of the
  $\pi^0$-exchange contribution in quark-connected diagrams was first
  pointed out in~Refs.~\cite{Bijnens:2015jqa, Bijnens:2016hgx}.} by a
factor of three. This is compensated by the quark-disconnected
diagrams, which contain the differences between the flavor-singlet and twice the flavor-octet meson contributions.
In the large-$N_c$ limit, the singlet and
octet contributions to the disconnected diagrams will cancel
as their spectra becomes the same. For QCD, this
degeneracy is most strongly broken in the pseudoscalar sector, where
the octet's mass is far below the singlet's.

To interpret the integrand in Method 2, one needs a mapping of
individual quark-level Wick contractions onto meson-exchange
diagrams~\cite{Asmussen:2019act} (see Appendix~\ref{app:pi0matching}
for a derivation based on partially quenched chiral perturbation
theory).  It turns out that
$\widetilde{\Pi}^{(1)}_{\mu\nu\sigma\lambda}(x,y,z)$ does not contain
the meson-exchange diagram in which the $\pi^0$ and $\eta$ propagate
between the pair of vertices $(0,y)$ and $(x,z)$. Also,
the normalization of the two other $(\pi^0,\eta)$-exchange diagrams is
such that $\widetilde\Pi^{\rm conn}_{\mu\nu\sigma\lambda}$ contains
the same $(\pi^0,\eta)$ contribution as
$\widetilde\Pi_{\mu\nu\sigma\lambda}$, enhanced (in the present
SU(3)$_{\rm f}$ case) by the charge factor three. 

In addition, we need the mapping of individual quark-level disconnected diagrams onto meson-exchange diagrams.
Here it turns out that there is a one-to-one mapping between a given quark-level diagram and a meson-exchange diagram.
For instance, take a quark-level diagram, consisting of two quark loops each containing two vectorial vertices; each quark loop
thus defines a pair of vertices. Such a diagram is in one-to-one correspondence with the diagram 
in which the meson is exchanged between the two pairs of vertices. For octet mesons, the latter diagram
has a weight of $-2$ relative to its normalization in the full HLbL amplitude. For singlet mesons, this relative weight factor is simply unity.

The short-distance contribution to $\ahlbl$ is sometimes modelled by a
constituent-quark loop. This corresponds to an effective degree of
freedom, and the mass assigned to the `constituent quark' is typically
on the order of 300\,MeV~\cite{Bijnens:1995xf}. In the following sub-section, we will address the question ``to what extent does such a contribution describe the
short-distance part of the $|y|$ integrand?''. In this case, the Wick
contractions and weight factors of the constituent quark simply
correspond to those of the fundamental quark degrees of freedom.

We have computed the charged pion and kaon loop contributions in the
framework of scalar QED~\cite{Asmussen:2019act,mainzHLbL1}.  The
contribution of these loops to the set of quark-connected contractions
and their contribution to the set of quark-disconnected contractions
add up coherently, the latter being twice as large as the former. This
result can again be derived in partially-quenched perturbation theory.
While the $(\pi^0,\eta)$ exchange contribution to the full $\ahlbl$ 
is three times smaller than its contribution to $a_\mu^{\rm conn}$,
the charged-pseudoscalar loop contribution is three times larger.
The charged-pseudoscalar loop's contribution might seem negligible in comparison to the integrand of the quark-connected, but it need not be negligible in the full integrand.

\subsection{Lattice connected contribution}\la{sec:conn_intgnd}

\begin{figure}[t]
        \centering
        \includegraphics*[width=0.64\linewidth]{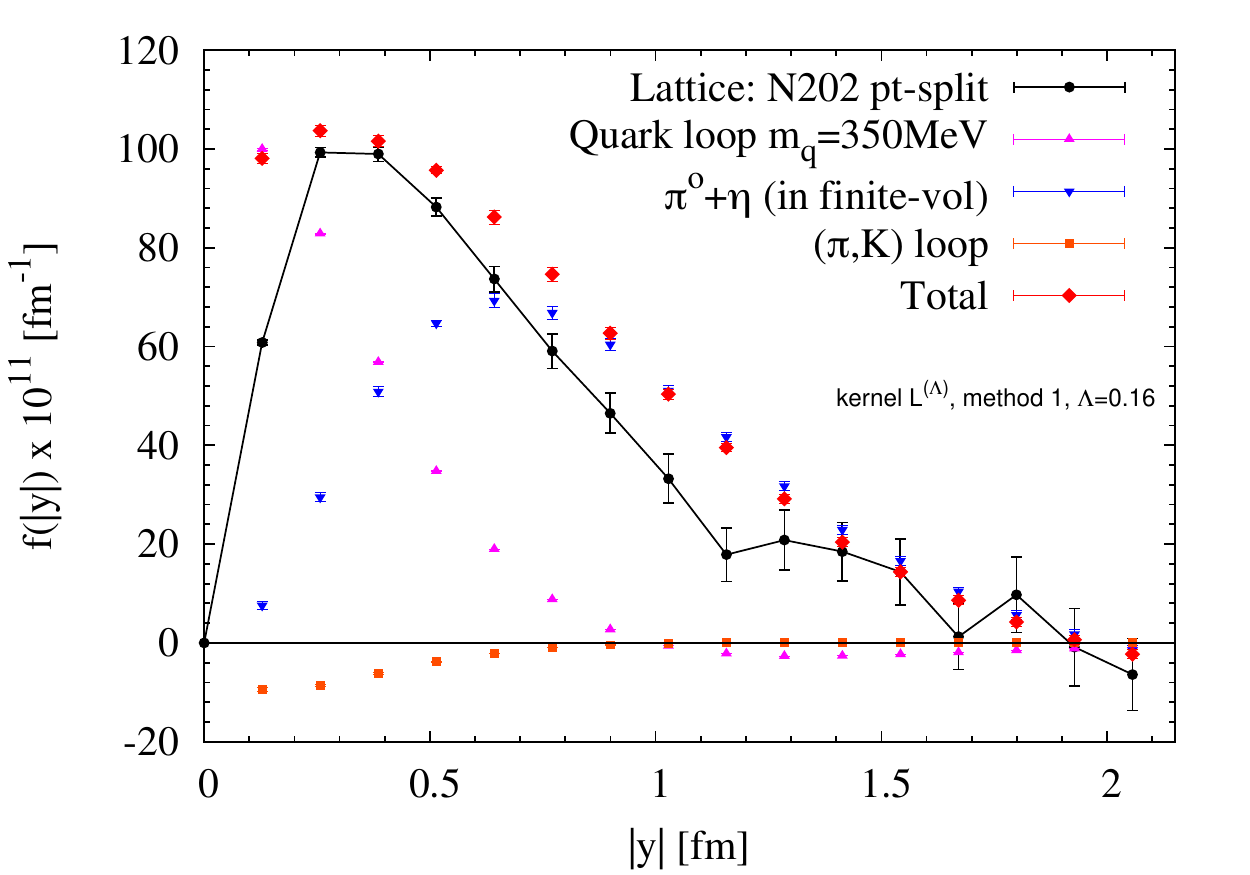} 
                        
        \caption{Integrand for the connected contribution on ensemble N202 using Method 1 with $\Lambda = 0.16$. The lattice data use a point-split current at $x$. The integrand is compared to the prediction for the exchange of the $\pi^0$ and $\eta$
mesons with a VMD transition form factor, which is expected to provide a good approximation to the tail. In addition, an attempt to describe the
 short-distance contribution with  a constituent-quark loop with a quark mass of 350\,MeV is made.}  
        \label{fig:N202_CQM}
\end{figure}

Fig.~\ref{fig:N202_CQM} displays the integrand obtained with
Method~1 and $\Lambda=0.16$. It is compared to the integrand for the
exchange of the $(\pi^0,\eta)$ mesons with a VMD TFF and using the parameters of Table~\ref{tab:vmdparam}. 
Beyond 1.5\,fm, the prediction is
consistent with the lattice data, albeit within large relative
errors. Between 0.8\,fm and 1.4\,fm, the lattice data lie noticeably
below the pseudoscalar-octet exchange prediction. 

At distances up to 0.8\,fm, one would certainly not expect the
pseudoscalar-octet exchange to saturate the integrand. We have
attempted to model the higher-energy contributions using a
constituent-quark loop, displayed in Fig.~\ref{fig:N202_CQM}. For a
constituent-quark mass of 350\,MeV, the sum of this contribution and
the pseudoscalar-octet exchange provides a good description of the
maximum height of the integrand.  At distances $|y|\lesssim0.4$\,fm,
one must expect large cutoff effects on the lattice integrand, as the  separation of sources becomes comparable to our lattice spacing.
This interpretation is confirmed by comparing the integrand to the one
obtained using exclusively the local current, as in the left panel of
Fig.~\ref{fig:N202_M1}, where a sizeable difference is visible up to
about 0.4\,fm.  
An interesting observation is that upon adding the
constituent-quark loop to the pseudoscalar-octet exchange the result does not
improve the agreement with the lattice data in the region
$0.8{\rm\,fm}<|y|<1.4{\rm\,fm}$.  A clear excess remains in the
prediction; we currently do not have a compelling explanation for this
excess. The exchange of the lightest scalar-octet mesons ($a_0$ type)
would have the right sign to explain the effect, since scalar-meson
exchanges are known to contribute negatively to $\ahlbl$. In the
future, a calculation of the scalar contribution along the lines
of~\cite{Knecht:2018sci} would be worthwhile to find out whether it
accounts for this missing effect. The charged-pion loop is also
expected to contribute negatively and we have studied this contribution in the
scalar-QED framework in~\cite{Asmussen:2019act}. We found the
integrand to be negligible compared with the $\pi^0$ contribution beyond
$|y|=0.6\,$fm, and introducing a vector form factor for the charged
pion would likely only reduce the integrand further. It thus remains
an open problem to understand the physics underlying the integrand for
$|y|$ around 1\,fm.

The effect of the finite volume on the lattice integrand is
illustrated in Fig.~\ref{fig:FSE_M1}.  A clear effect is seen
between the comparatively  `small' ensemble H200 and the `large' N202, the former
integrand lying below the latter. This finite-volume dependence
matches in sign and typical size the volume dependence of the
pseudoscalar-octet exchange contribution, as seen in the figure.  

\subsection{Lattice disconnected contribution}
\la{sec:disc_intgnd}
Fig.~\ref{fig:disc_intgnd} (later in the text) displays the disconnected integrand for
several gauge ensembles.  It is compared to the prediction of the
$(\pi^0,\eta)$ exchange, including its appropriate weight factor of
$-2$ for the disconnected contribution, both in finite and
infinite volume. The finite-volume effect predicted from the
$(\pi^0,\eta)$ exchange calculation is very small, and the lattice
data from ensembles N202 and H200 confirms this expectation, at least
at distances $|y|<0.9\,$fm, where the statistical errors allow for a
meaningful comparison.

The main difference between the disconnected and the connected contributions is that for the disconnected the 
$(\pi^0,\eta)$ exchange already provides a decent description of the
lattice data for $|y|$ below 1\,fm; in other words, we do not observe a
large short-distance contribution.  The $\eta'$ and the $\sigma$
mesons being singlets, they contribute to the disconnected diagrams
with the same weight factor as they would to the full HLbL
amplitude~\cite{Bijnens:2016hgx,Gerardin:2017ryf}.  In order to
estimate the typical size of these heavy-meson contributions, we have
computed the $\eta'$ contribution under the following assumptions: The
$\eta'$ mass was set to 982\,MeV, a value close to the result of our
calculation on ensemble H101. Its coupling to two photons, ${\cal
  F}_{\eta'\gamma\gamma}$, was assumed to be equal to its value at
physical quark masses. The virtuality dependence of the transition
form factor can be modelled with a VMD ansatz, with vector mass
952\,MeV. Under these assumptions, the contribution is positive and
sizeable (compared to the $(\pi^0,\eta)$ exchange) up to
$|y|=1.5\,$fm. In addition, we expect a significant contribution from the
stable $\sigma$ meson, whose mass we have found to lie well below the $\pi\pi$
threshold. As a scalar, the $\sigma$ would contribute negatively and
thus compensate to some extent the $\eta'$ contribution. Again, a
dedicated calculation in the framework of~\cite{Knecht:2018sci} would certainly be worthwhile. 
The estimate for the $\eta'$ contribution displayed in Fig.~\ref{fig:disc_intgnd}
is only meant to be representative of one particular meson-exchange
contribution, with other cancellations being expected.

\section{Results for the quark-connected contribution\la{sec:res_conn}}

\subsection{Results from Method 1 \la{sec:meth1}}

\begin{figure}[t]
	\centering
	\includegraphics*[width=0.495\linewidth]{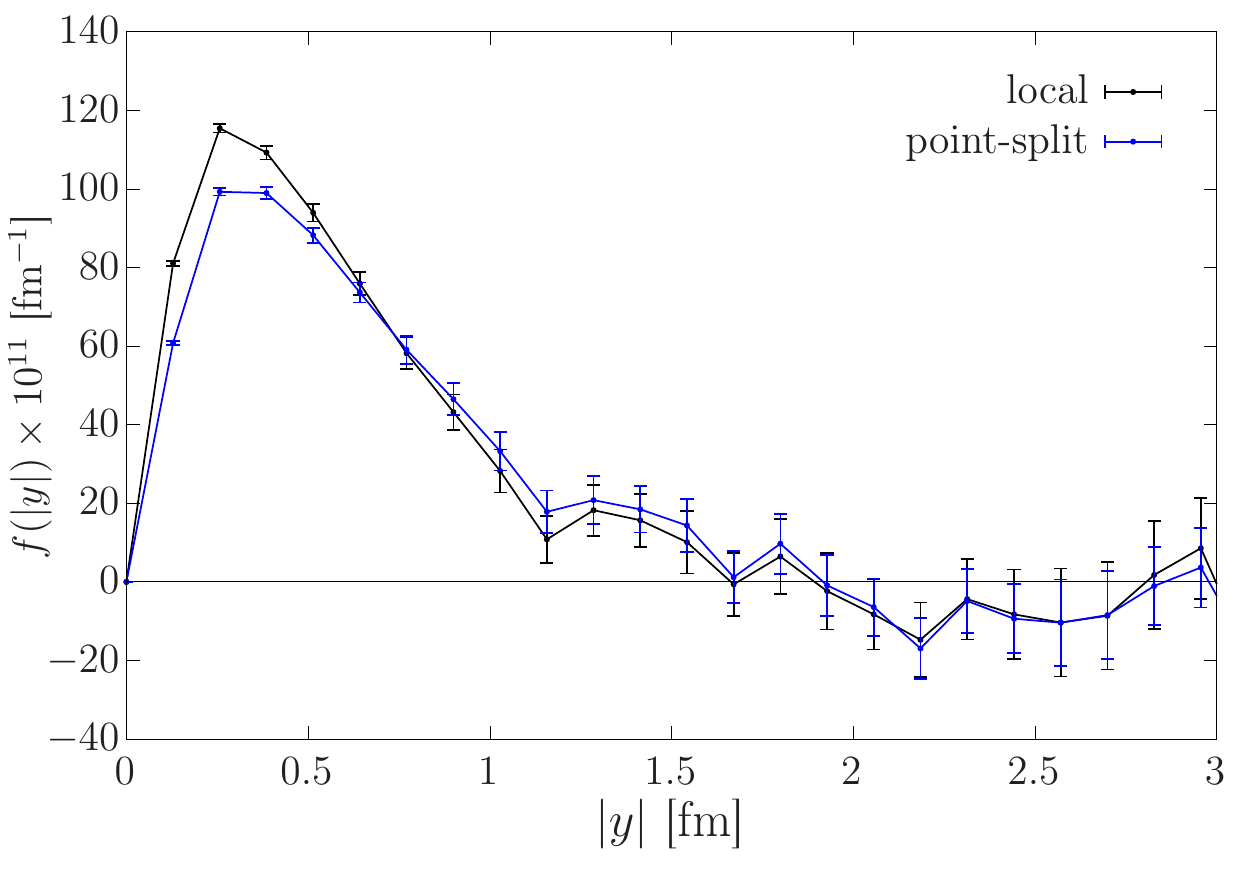}
	\includegraphics*[width=0.495\linewidth]{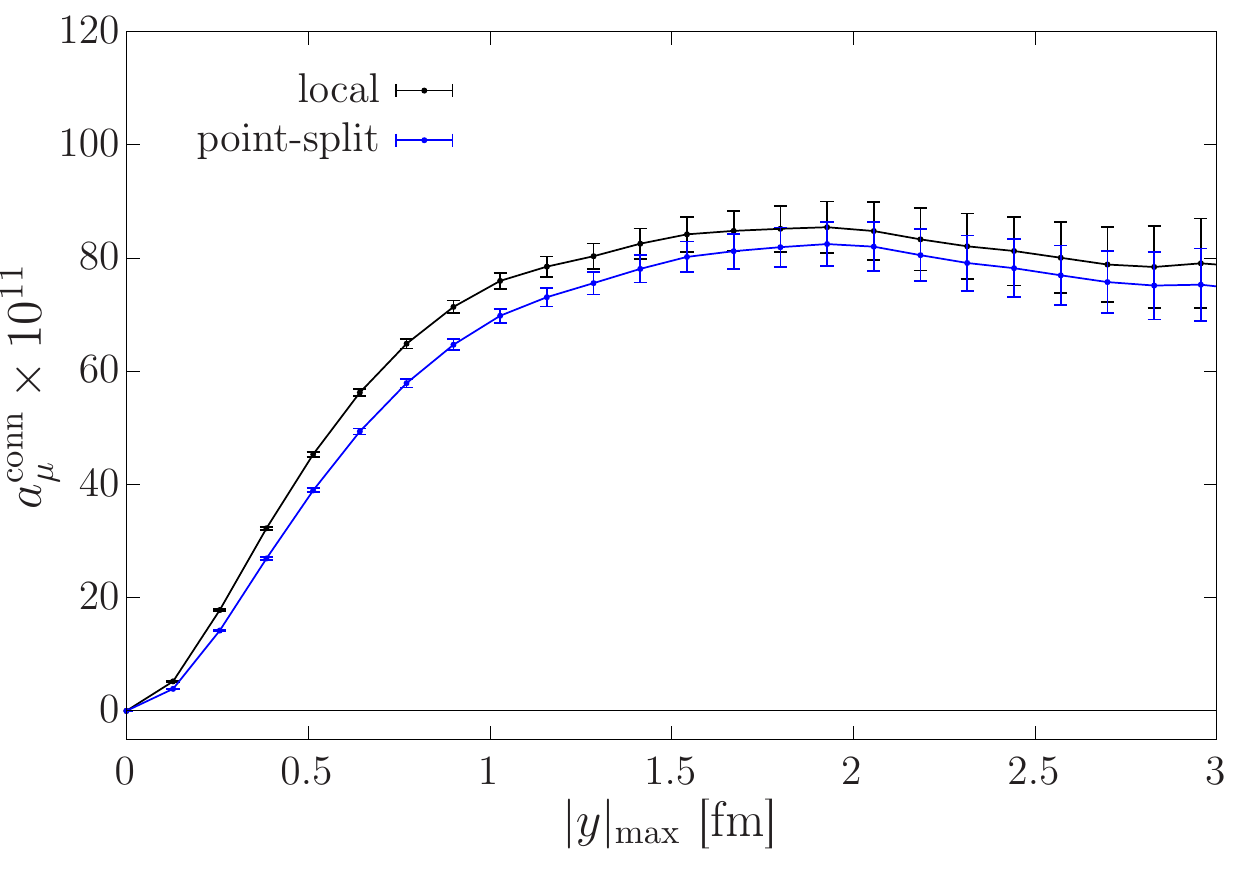}
	\caption{Results for the ensemble N202 using Method 1 with $\Lambda = 0.16$. Black points use a local current at the site $x$ in Eq.~\eqref{eq:integrand} (which is our default approach) and blue points use a conserved current at $x$. Left: integrand $f(|y|)$ as a function of $|y|$. Right: integrated value of $a_{\mu}^{\rm conn}$ as a function of $|y|_{\rm max}$.}
	\label{fig:N202_M1}
\end{figure}

The lattice results for the quark-connected contribution to $\ahlbl$ using Method 1 have been generated along the direction $y/a \propto (1,1,1,1)$. We have used two different discretizations of the four-point correlation function: the vector current located at the site $x$ is either local or point-split (conserved), while the three other currents are always local. Our results are summarized in Table~\ref{tab:resM1} and the integrand for the ensemble N202 (also presented in Sec.~\ref{sec:conn_intgnd}, Fig.~\ref{fig:N202_CQM}) is shown on Fig.~\ref{fig:N202_M1}. The signal-to-noise ratio clearly deteriorates rapidly at large distances and we observe slightly better statistical precision when using a conserved vector current at $x$. Both discretizations give similar results, suggesting that there are small discretization effects present. In the end we will quote a final result with the fully-local discretization for a direct comparison with the results obtained using Method 2.

\begin{figure}[t]
	\centering
	\includegraphics*[width=0.495\linewidth]{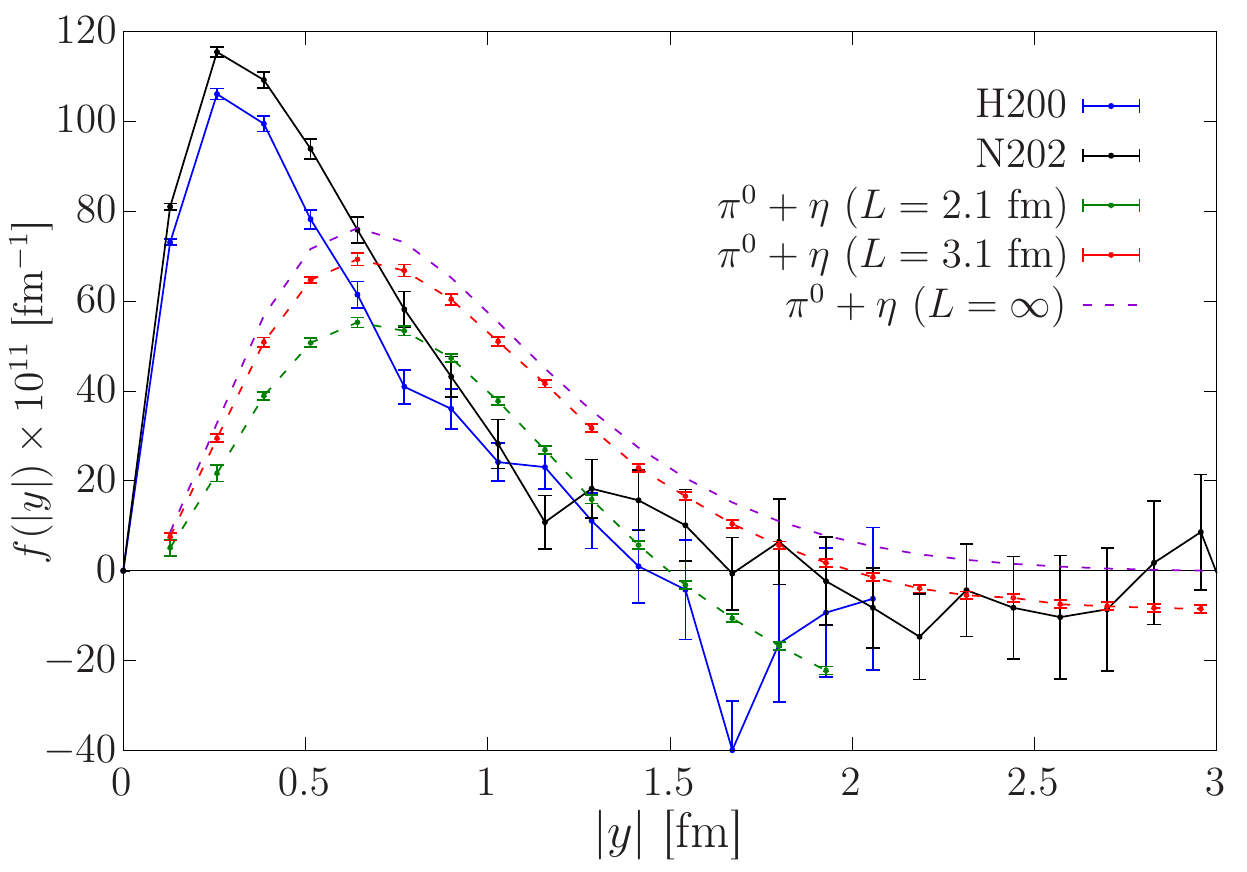}
	\includegraphics*[width=0.495\linewidth]{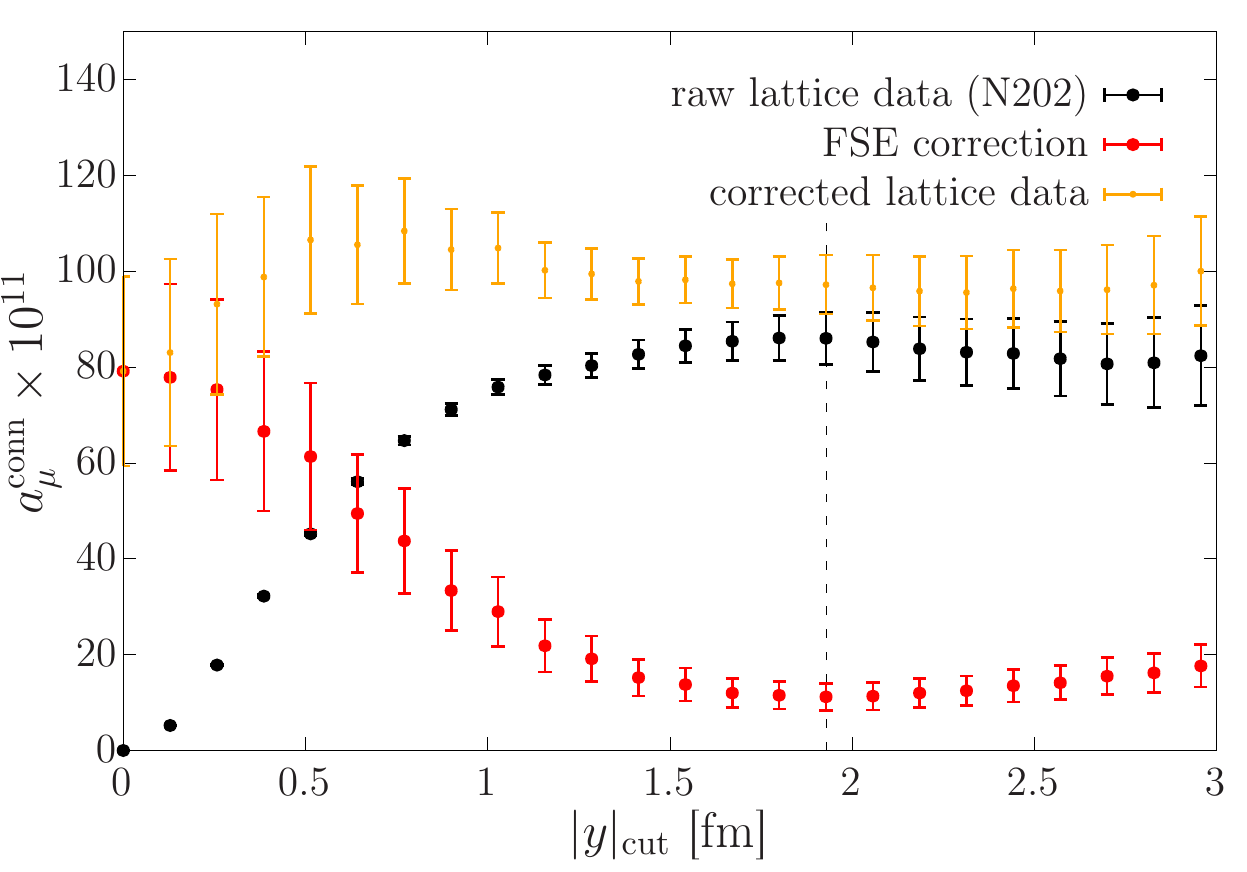}
			
	\caption{Study of FSEs for Method 1. Left: Integrand for the ensembles H200 and N202 with $m_{\pi} L = 4.4$ and 6.4 respectively. Right: Value of $a_{\mu}^{\rm conn}$ for the ensemble N202 using the FSE correction prescription described in the text, as a function of $\ycut$}	
	\label{fig:FSE_M1}
\end{figure}

Although all the ensembles used in this work satisfy $m_{\pi} L > 4$, we still expect significant finite size effects (FSEs) due to the  pseudoscalar-pole contribution, which is the dominant contribution in model estimates of hadronic light-by-light scattering~\cite{Prades:2009tw}, being a long-range phenomenon. Even with heavy pions $m_{\pi} \approx 400~$MeV, the tail extends beyond $|y|=2.5~\text{fm}$~\cite{mainzHLbL1}. A comparison of the integrand for the ensembles H200 and N202, which only differ by their physical volumes, is depicted on the left panel of Fig.~\ref{fig:FSE_M1}.
Assuming a VMD model for the TFF, the pseudoscalar contribution can be computed in both finite and infinite volume. For more information on the calculation of the TFF we refer the reader back to Sec.~\ref{sec:pipole_tff}; the parameters we use for modelling our data are summarized in Table~\ref{tab:vmdparam} in that section.

To obtain our final estimate, the lattice data are integrated up to $\ycut$ where the integrand is compatible with zero. For $|y|>\ycut$, the tail is aproximated by the pseudoscalar-pole contribution in infinite volume. Finally, for $|y|<\ycut$, the FSEs are estimated as the difference between the pseudoscalar-pole's contribution computed in finite and infinite volume:
\begin{equation}
a_{\mu}^{\rm conn} = a_{\mu}^{\rm data}  + a_{\mu}^{\rm tail}  + a_{\mu}^{\rm FSE} \,.
\label{eq:decomposition}
\end{equation}
A systematic error of 25\% is attributed to both the tail extension and the FSE correction. Since the same VMD model is used in both cases, we treat these corrections as being fully correlated.

The value of $a_{\mu}^{\rm conn}$ as a function of $\ycut$ is shown in the right panel of Fig.~\ref{fig:FSE_M1}; we observe a nice plateau for values $\ycut > 1.2~$fm, suggesting that our systematic error estimate is quite conservative. Estimates of the finite-size corrections for each ensemble are summarized in Table~\ref{tab:resM1} with their corresponding values of $\ycut$. In particular, we note that the systematic error on the FSEs is always larger than the statistical precision, except for our largest ensemble N202.

\begin{table}[b]
\begin{tabular}{c@{\hskip 02em}c@{\hskip 02em}c@{\hskip 02em}c@{\hskip 02em}c@{\hskip 02em}c}
\toprule
Label	&	$\ycut$~[fm] 	& $\amu^{\rm data}$ 	& $\amu^{\rm FSE}$   & $\amu^{\rm tail}$  &        $a_\mu^{\text{conn}}$	\\ 
\hline
U103    &	1.55 		& 69.0(2.2)   	& 	31.0		& 	4.3			& 104.3(2.2)(8.8)	\\
H101    & 	1.73		& 71.6(2.9) 	&	13.1		& 	3.8			& 88.5(2.9)(4.2)		\\
B450    & 	1.68		& 72.9(3.8)  	& 	18.3		& 	4.0			& 95.2(3.8)(5.6)		\\
H200    & 	1.54		& 71.1(2.8)  	& 	27.3		& 	6.1			&104.4(2.8)(8.4)	\\
N202    & 	1.93		& 85.5(4.5)  	& 	9.0		& 	1.7			& 96.3(4.5)(2.7)		\\
N300    & 	1.59		& 79.2(4.2) 	& 	15.6		& 	4.1			& 99.0(4.2)(4.9) 	\\
\botrule
\end{tabular}
\caption{Results for the connected contribution using Method 1 with four local vector currents using the decomposition given by Eq.~(\ref{eq:decomposition}). A $25\%$ systematic to the total  correction is used.
  \label{tab:resM1}}
\end{table}

\begin{figure}[h!]
	\centering
	\includegraphics*[width=0.6\linewidth]{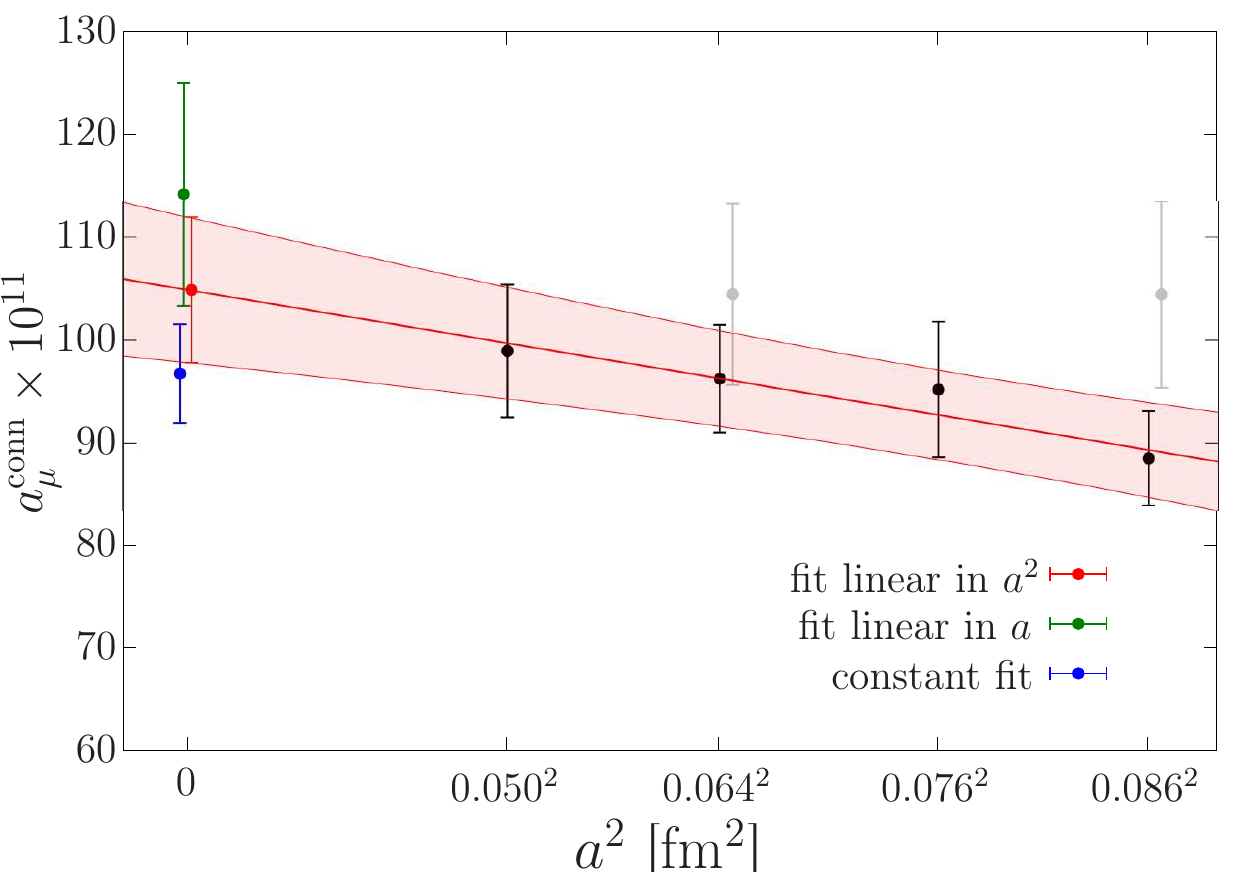}
			
	\caption{Continuum extrapolation of the connected contribution using Method 1. We perform a linear fit in $a^2$ (red), a linear fit in $a$ (green) or a constant fit (blue).  The coarsest lattice spacing is excluded from the constant fit. Ensembles in grey have $m_{\pi}L<5$ and are not included in the fits and they have been shifted for clarity.}		
	\label{fig:extrap_M1}
\end{figure}

For the ensembles U103 and H200, with $m_{\pi} L < 5$, we see very large FSE corrections, of the order of $50~\%$. After these corrections the values of $a_{\mu}^{\rm conn}$ do become compatible with the results at $m_\pi L>5$ within about 1.5~$\sigma$. We also observe a systematic over-estimate in comparison to the larger-volume results, and when it comes to our final continuum extrapolation we will omit these results.

$O(a)$-improvement is not implemented for the vector currents used in this work, but our experience with other observables involving electromagnetic currents, such as the LO HVP~\cite{Gerardin:2019rua} and the pion TFF~\cite{Gerardin:2019vio}, suggests the remaining $O(a)$ terms are small compared to the quadratic contribution. A linear fit in $a^2$ leads to $a_{\mu}^{\rm conn, M1} = 104.1(6.9) \times 10^{-11}$ with $\chi^2/\mathrm{d.o.f.} = 0.4$. To estimate the systematic error associated with this continuum extrapolation, we perform a constant fit which excludes the coarsest lattice spacing and obtain the slightly smaller value $a_{\mu}^{\rm conn, M1} = 96.7(4.7) \times 10^{-11}$ with $\chi^2/\mathrm{d.o.f.} = 0.3$. Finally, we also tried a linear fit in the lattice spacing which leads to $a_{\mu}^{\rm conn, M1} = 113.8(10.5) \times 10^{-11}$. The results of all these fits are shown in Fig.~\ref{fig:extrap_M1}.

We quote our continuum-extrapolated value for the quark-connected contribution to $\ahlbl$ using Method 1 at the SU(3)$_{\rm f}$-symmetric point as
\begin{equation}
a_{\mu}^{\rm conn, M1} = 104.1(6.9)(3.7) \times 10^{-11} ,
\end{equation}
where the first error includes both the statistical error and the systematic from the finite-size correction. The second error is an estimate of the continuum-limit extrapolation systematic error, taken as half the difference between the linear in $a^2$ and constant fit ans\"atze.

\subsection{Results from Method 2 \la{sec:meth2}}

In our measurement of the quark-connected contribution to $\ahlbl$ using Method 2 we focus on $\Lambda=0.4$; this value was already indicated as being beneficial for the lepton loop as discussed in Appendix~\ref{app:leploop}. We performed measurements on all ensembles with $\Lambda=0.0,0.4,0.8,$ and $1.0$ and found that $\Lambda=0.0$ approached plateau too slowly and $\Lambda=1.0$ had a significant peak in the integrand at short distances but a more-pronounced negative-valued tail. It appears that $\Lambda=0.4$ is a near-optimal choice for our calculation.

Although not presented here, we also performed the contractions with conserved currents at $x$ and/or $z$. We found that putting a conserved current at $z$ yields a result roughly comparable (point-by-point) to the determination with just 4 local currents. Having a conserved current at $x$ appears to introduce large, unwanted discretization effects. From a computational standpoint the calculation with four local currents is simpler and has no apparent downside, so that is what we will present from here on.

\begin{table}[h!]
\begin{tabular}{c@{\hskip 02em}c@{\hskip 02em}c@{\hskip 02em}c@{\hskip 02em}c@{\hskip 02em}c}
\toprule
Label	&	$\ycut$~[fm] 	& $\amu^{\rm data}$ 	& $\amu^{\rm FSE}$   & $\amu^{\rm tail}$  &        $a_\mu^{\text{conn}}$ \\
\hline
U103    &	1.85 		& 58.9(1.8)  	& 14.3 	& 13.3 	& 86.5(1.8)(6.9) \\
H101    & 	2.55		& 75.7(2.9)  	& 6.9		& 2.6		& 85.2(2.9)(2.4) \\
B450    & 	2.00		& 73.6(3.3)  	& 7.6 	& 9.4		& 90.3(3.3)(4.3) \\
H200    & 	1.75		& 68.6(1.8)  	& 13.7 	& 13.6 	& 95.8(1.8)(6.8) \\
N202    & 	2.60		& 91.0(2.5)  	& 3.8 	& 1.9		& 96.7(2.5)(1.5) \\
N300    & 	2.10		& 79.0(1.8)  	& 7.1 	& 6.2		& 92.3(1.8)(3.4) \\
\botrule
\end{tabular}
 \caption{Results for the connected contribution using Method 2. Here $a_{\mu}^{\rm data}$ corresponds to the value using lattice data up to some linearly-interpolated value of $y = \ycut$, chosen to minimize the total error of $a_\mu^{\rm conn}$. Again, a $25\%$ systematic to the total  correction is used. In the last column we give the infinite-volume result.}
\label{tab:resM2}
\end{table}

\begin{figure}[t]
	\centering
	\includegraphics*[width=0.495\linewidth]{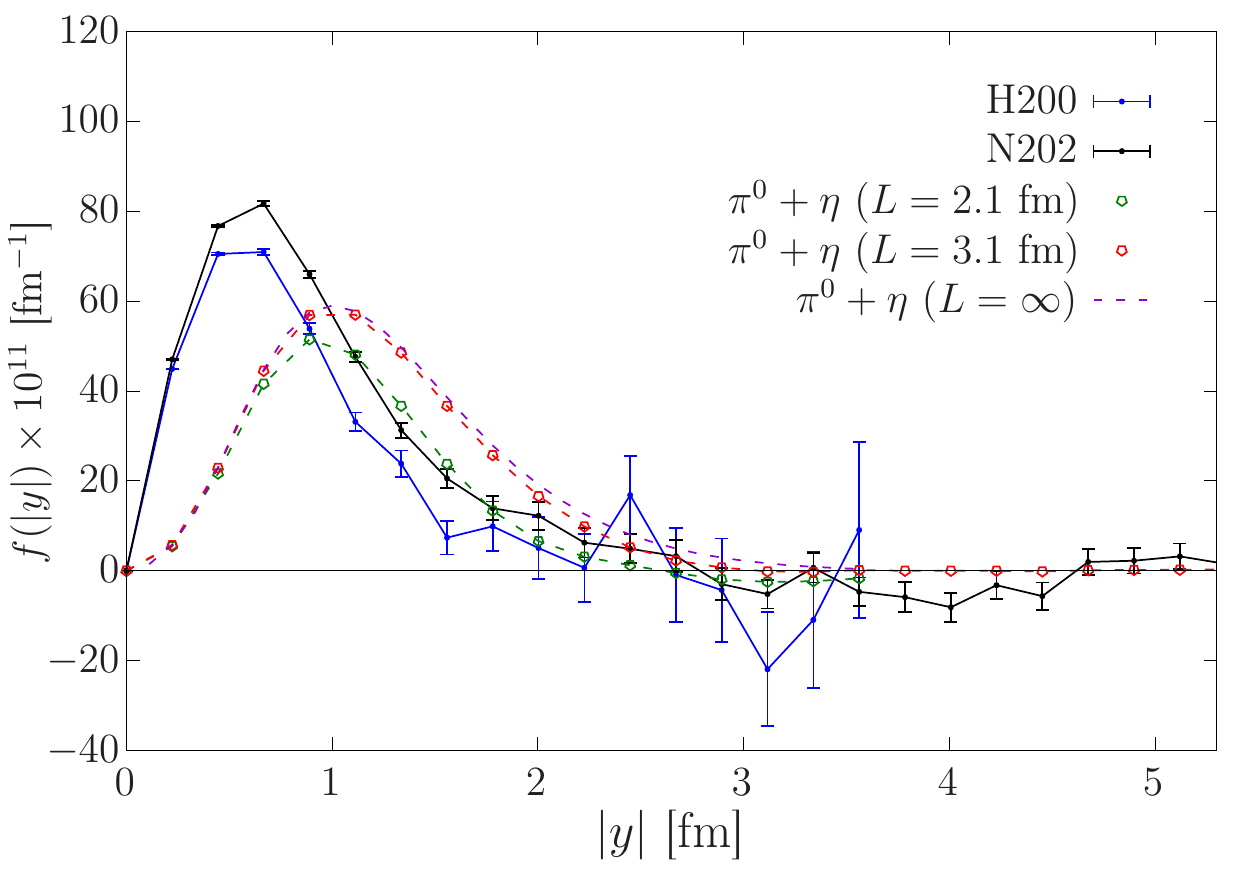}
	\includegraphics*[width=0.495\linewidth]{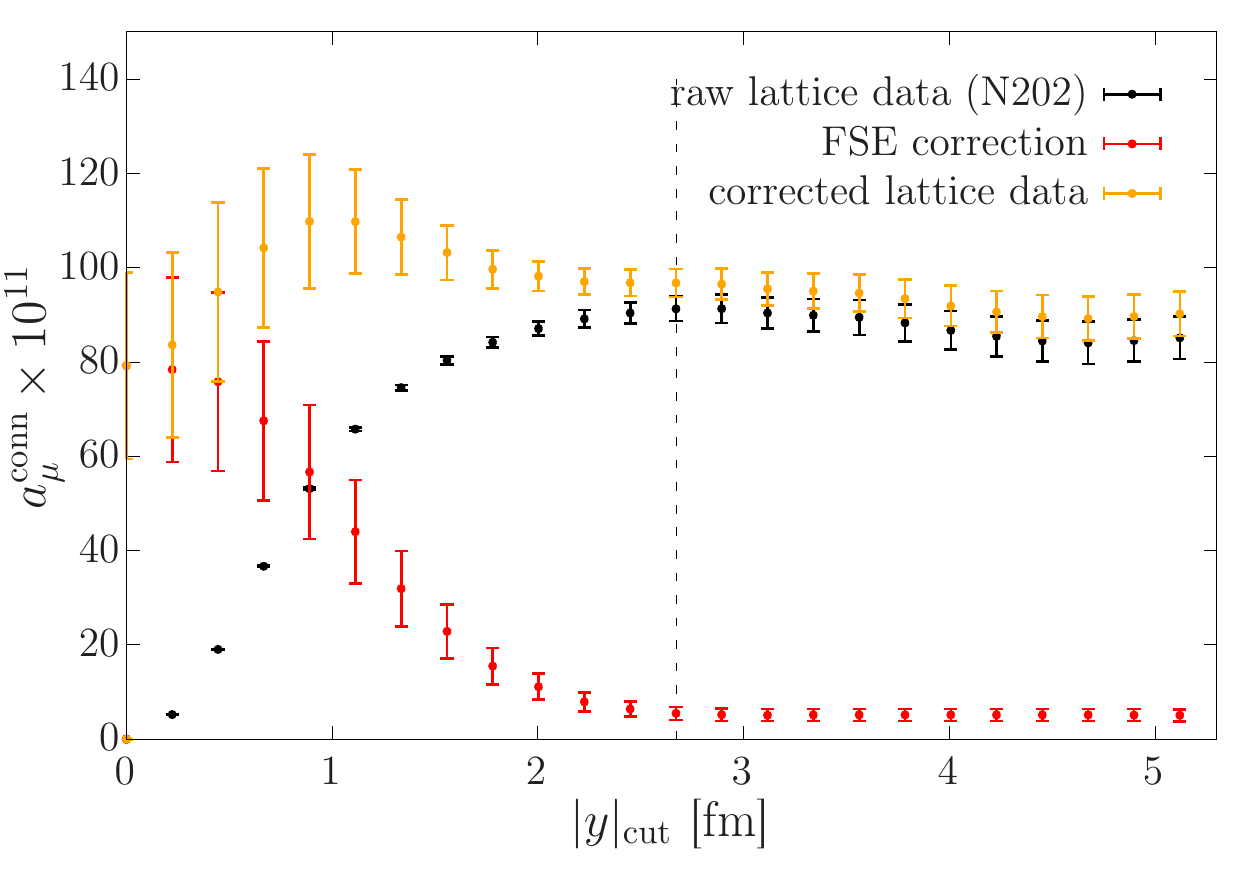}
	\caption{Study of FSEs for Method 2. Left: Integrand for the ensembles H200 and N202 with $m_\pi L=4.4$ and $6.4$ respectively. Right: Value of $a_{\mu}^{\rm conn}$ for the ensemble  N202 using the FSE correction prescription described in the text, as a function of $|y|_{\rm cut}$.}\label{fig:N202_M2}
\end{figure}

The left plot of Fig.~\ref{fig:N202_M2} illustrates the finite-size effect between ensembles N202 and H200, and the discrepancy between these ensembles that differ only by volume is significant. The pion-pole prediction describes the tails of both data-sets reasonably well at large-enough values of $|y|$, although it completely under-estimates the position and height of the short-distance peak of the integrand. It is also worth noting how less statistically precise the result of H200 is compared to N202 for a comparable number of measurements; this gives some indication that the statistical precision is linked to either $m_\pi L$ or the physical volume. It is clear that much like for Method 1 there is a significant signal-to-noise problem for large values of $|y|$.

We perform the same finite-size correction procedure for Method 2 as we did for Method 1 above. On the right of Fig.~\ref{fig:N202_M2} we show the stability of performing the FSE correction with varied $\ycut$ matching point on ensemble N202. We find excellent stability for many different values of $\ycut$ and the matching point of $2.6$ fm  was chosen in an attempt to minimize the total error.

If we consider the results of Table~\ref{tab:resM2} we see good agreement after finite-size correction between ensembles that only differ by volume (compare U103 with H101, and H200 with N202), which suggests that our finite-size correction procedure is sensible. Unlike for Method~1 we see no reason to exclude these smaller volumes from our final extrapolation. An unusual result in our determination in Method 2 comes from the ensemble N300, which lies below the trend of all our other data points. Since it is the finest ensemble, we have no reason to exclude it, even though this point will reduce the quality of our final extrapolations.

We hold off on presenting the continuum-limit extrapolation here as we will employ a combined extrapolation after the next section (See Sec.~\ref{sec:finalres}, Fig.~\ref{fig:M2conn_disc}). However, we will quote the result of the connected continuum extrapolation,
\begin{equation}
a_\mu^{\text{conn,M2}} = 98.9(2.5)\times 10^{-11}.
\end{equation}
The quoted error is a combination of the statistical and 100\%-correlated finite-size systematic. We observe that within error this value is in complete agreement with the determination of Method 1.

\subsection{Comparison of the two methods}

\begin{table}
  \centering
  \begin{tabular}{c|cc|c}
    \toprule
    Calculation & $N_\text{Conf}$ & $N_\text{Src}$ & $N_\text{Prop}$ \\
    \hline
    Method 1 & $800$ & $48$ & $280,800$ \\
    Method 2 & $901$ & $25$ & $22,525$ \\
    Disconnected & $225$ & $480$ & $108,000$ \\
    \botrule
  \end{tabular}
  \caption{Resources used for the calculations on ensemble N202. $N_\text{Conf}$ gives the number of gauge
    configurations used, $N_\text{Src}$ the number of source positions per gauge
    configuration, and $N_\text{Prop}$ the total number of propagator solves.}
  \label{tab:costs}
\end{table}

The appeal of Method 2 for computing the connected contribution is mostly practical: it is computationally far less expensive than Method 1. The saving between the two is roughly an order of magnitude, see Tab.~\ref{tab:costs} for a exemplary comparison of computational cost for one particular ensemble, N202. This is because Method 2 effectively replaces sequential propagator solves by additional, much cheaper, QED kernel evaluations. The downside of using these additional kernels is that their combination tends to broaden the integrand $f(|y|)$. This behavior was seen in the lepton loop study of Appendix~\ref{app:leploop} and is also clearly the case with all the lattice QCD data in this work. We can use the parameter $\Lambda$ of Eq.~\eqref{eq:lamsub} to partially ameliorate this broadening; the use of such a subtraction kernel appears to be very important specifically for Method 2, as this regulator offers little to no advantage for Method~1.

If we compare the results for the integrand of H200 and N202 using Method 1 with those of Method 2 (Figs.~\ref{fig:N202_M1} and \ref{fig:N202_M2} respectively) we see that the integrand for Method 2 is in general less-peaked at short distances and extends further in $|y|$. For example, the integrand for N202 using Method 1 is effectively zero around 2 fm, whereas for Method 2 it becomes zero closer to 3 fm. This behavior is reflected in Tables~\ref{tab:resM1} and \ref{tab:resM2} by larger choices of $|y|_\text{cut}$ for Method 2 compared to Method 1.

Again comparing Tables~\ref{tab:resM1} and \ref{tab:resM2}, we see that for both methods the smaller boxes (those with $m_\pi L<5$) require a significant finite volume correction. As the data for Method 2 uses the direction $(1,1,1,3)$ we approach the boundary of the lattice ($L/2$) slowly for increasing $|y|$, and so the finite volume effect is smaller in comparison to the direction used in Method~1. We do, however, see somewhat larger discretization effects for Method 2 compared to those found in Method 1, perhaps $O(15\%)$ opposed to $O(10\%)$ at our coarsest lattice spacing respectively. It is quite possible this is due to the $\Lambda$-regulator enhancing the integrand at shorter distances. Nevertheless, this is not a significant problem as we have several fine lattice spacings to help determine the continuum limit.

If we were to use $\Lambda=0.0$ with Method 2 the tail would extend even further into the region where finite-volume effects become significant. This is likely still controllable for the symmetric-point ensembles used here as they have large volumes and $m_\pi L$, but this would become much more problematic for lighter-pion-mass ensembles where the signal is expected to degrade quickly at large distances and the integrand is expected to be even broader.

In the following sections, when we combine the results for the connected and disconnected contributions, we will use the results from Method 2 as our connected contribution. This is because they are statistically more precise while still being consistent with those of Method~1.

\section{Results for the quark-disconnected contribution\la{sec:disc}}

\begin{figure}[h!]
        \centering
        \includegraphics*[width=0.495\linewidth]{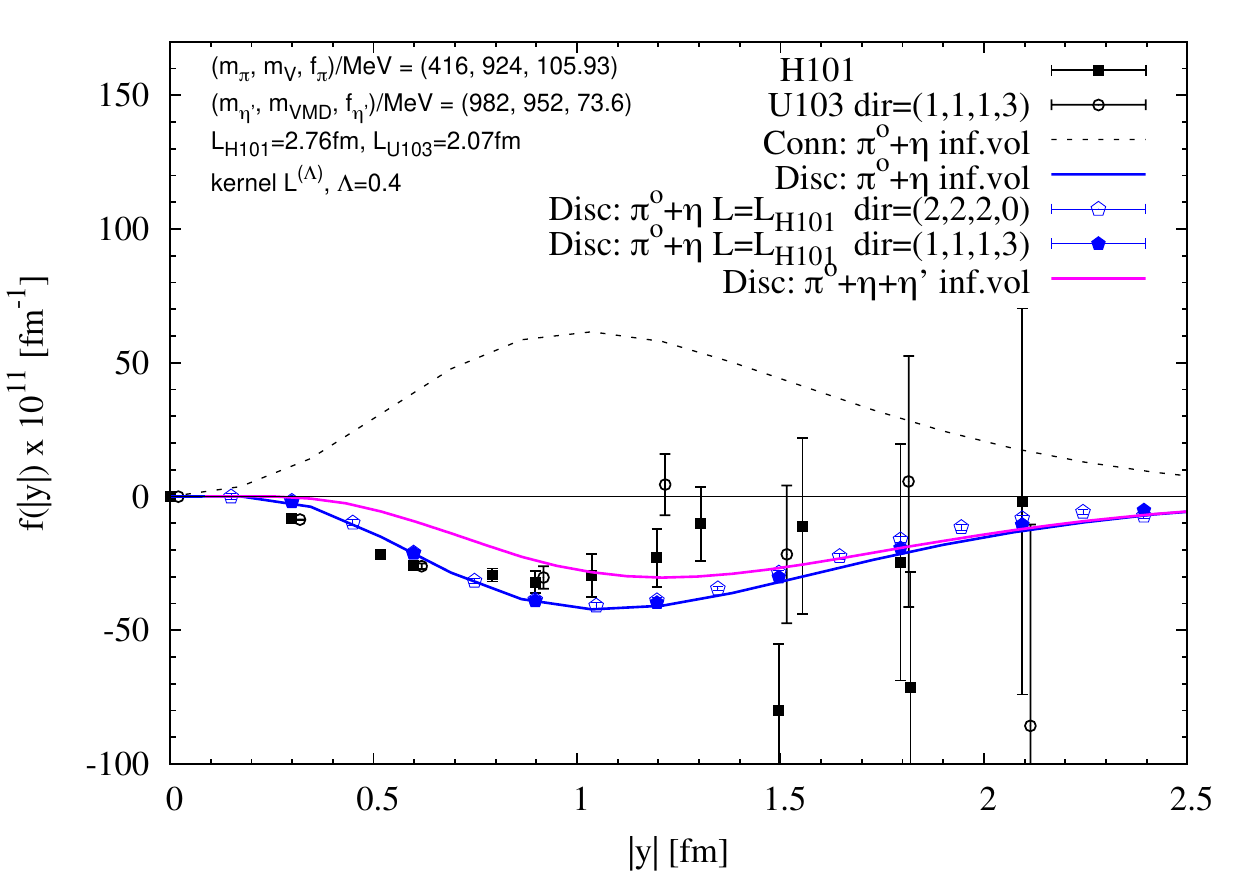}
        \includegraphics*[width=0.495\linewidth]{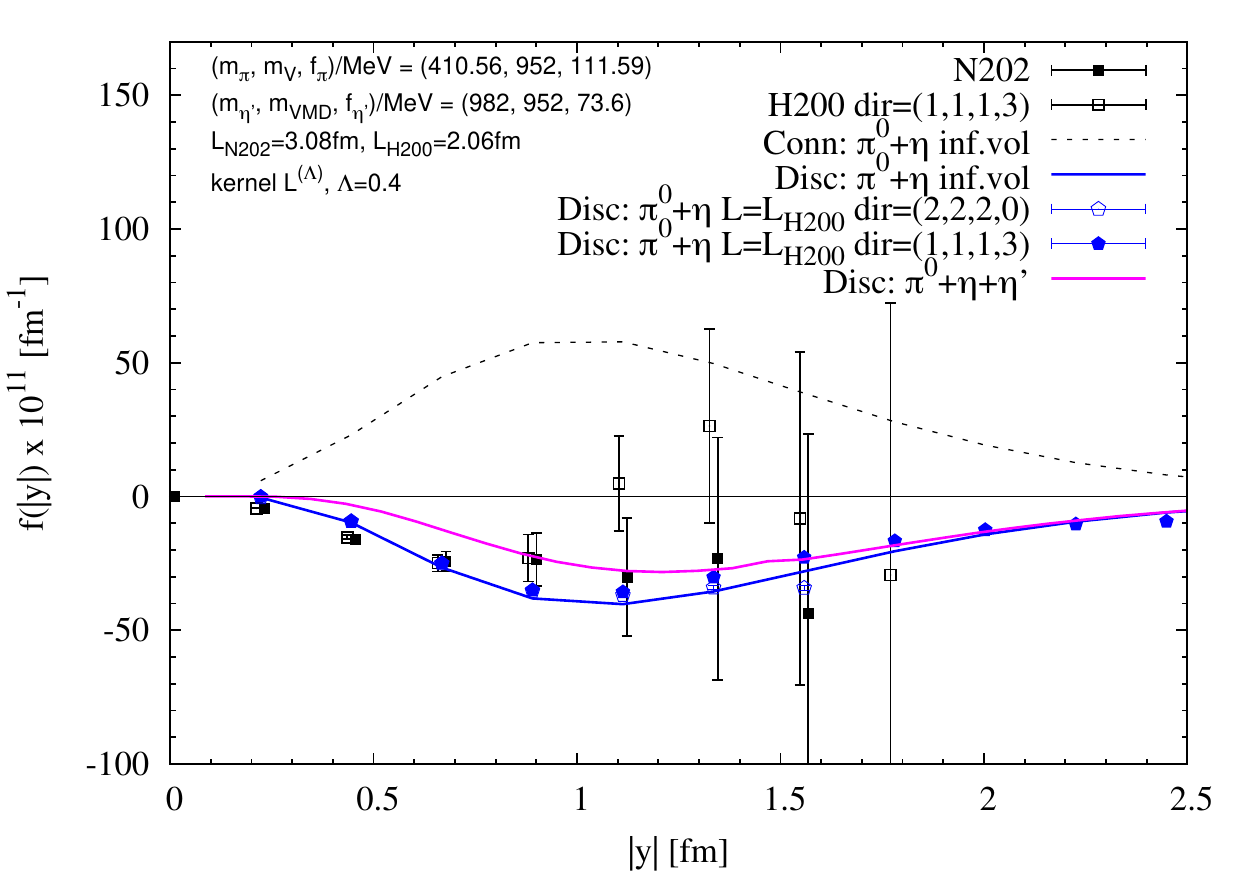}
    \caption{\label{fig:disc_intgnd} Integrand for the disconnected contribution on ensembles H101 and U103 (left), as well as 
N202 and H200 (right). The lattice data are shown as black points. The black dashed line shows the fully-connected model prediction, 
the blue line the $\pi^0 + \eta$ contribution for the disconnected and the magenta line gives the $\pi^0+\eta + \eta^\prime$. 
The blue points show the $\pi^0+\eta$ contribution in finite volume.
}
\end{figure}

Table~\ref{tab:disconn} lists the ensembles and statistics used for the computation of
the quark-disconnected contribution. As the smaller ensembles (U103, H101, B450, H200)
were considerably cheaper to perform inversions on, their statistics is greatly
enhanced. As the lattice volume increases, the cost of propagator
inversions increases with some power $V^n$, with $n>1$, and this
quickly becomes the dominant cost of the computation. The column
$N_\text{Src}$ indicates the number of point-source propagators
inverted per gauge configuration to build the grid and the final
column indicates the
maximum and minimum number of equivalent values of $|y|$ available
from the set-up. For the ensembles with open boundary conditions, the
number of self-averages, $N_{\rm{Equiv}}$, for a given $|y|$ decreases as $|y|/a$ increases. Therefore open temporal boundaries make this calculation much more difficult as the signal degrades rapidly with large $|y|/a$.

\begin{table}[t!]
\begin{tabular}{c|ccc|c}
\toprule
Label & $(t_\text{min},t_\text{max})/a$ & $N_\text{Src}$ & $N_{\rm Conf}$ & $N_{\rm{Equiv}}$(max,min) \\
\hline
U103 & $(20,107)$ & 360 & 1030 & (2112,432) \\
H101 & $(24,72)$ & 272 & 1008 & (1568,128) \\
B450$^*$ & $(0,64)$ & 128 & 1611 & (512,128) \\
H200 & $(24,72)$ & 272 & 500 & (1568,128) \\
N202 & $(36,93)$ & 480 & 225 & (2784,672) \\
N300 & $(27,99)$ & 600 & 271 & (3504,1152) \\
\botrule
\end{tabular}
\caption{The setup used for each ensemble in the computation of the quark-disconnected contribution. $N_\text{Src}$ gives the number of propagator inversions per configuration, $N_{\rm{Conf}}$ gives the number of gauge configurations used and $N_{\rm{Equiv}}$ gives the maximum and minimum number of equivalent values of $|y|$ averaged per configuration. Shorter separations have larger numbers of self-averages, whereas larger values of $|y|$ have smaller $N_{\rm{Equiv}}$. The ensemble B450 has a periodic temporal boundary and temporal length of $2\times$ that of the the spatial, so values of $y/a=n(2,2,2,4)$ could be used, as well as the full periodicity in time.}\label{tab:disconn}
\end{table}

\begin{figure}[t!]
        \centering
        \includegraphics*[width=0.495\linewidth]{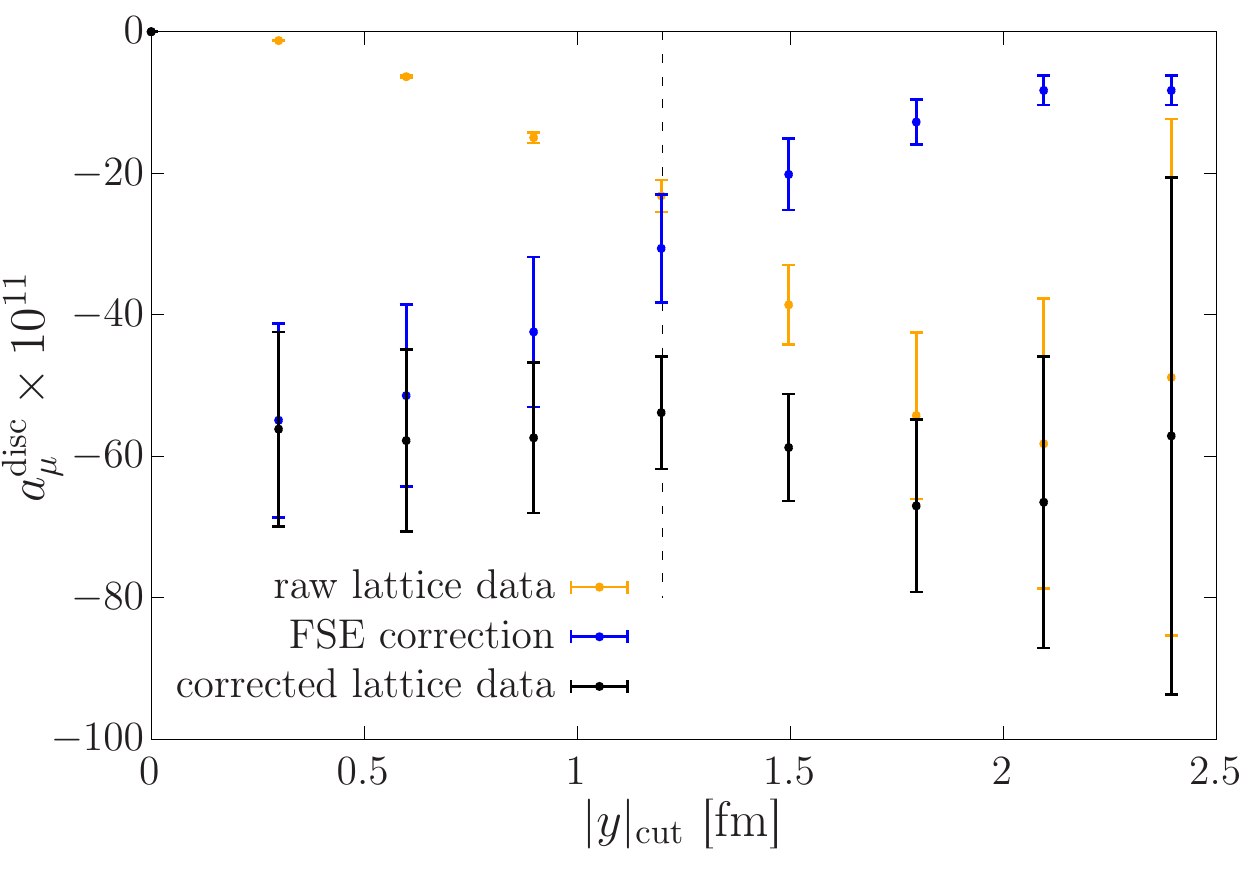}
        \includegraphics*[width=0.495\linewidth]{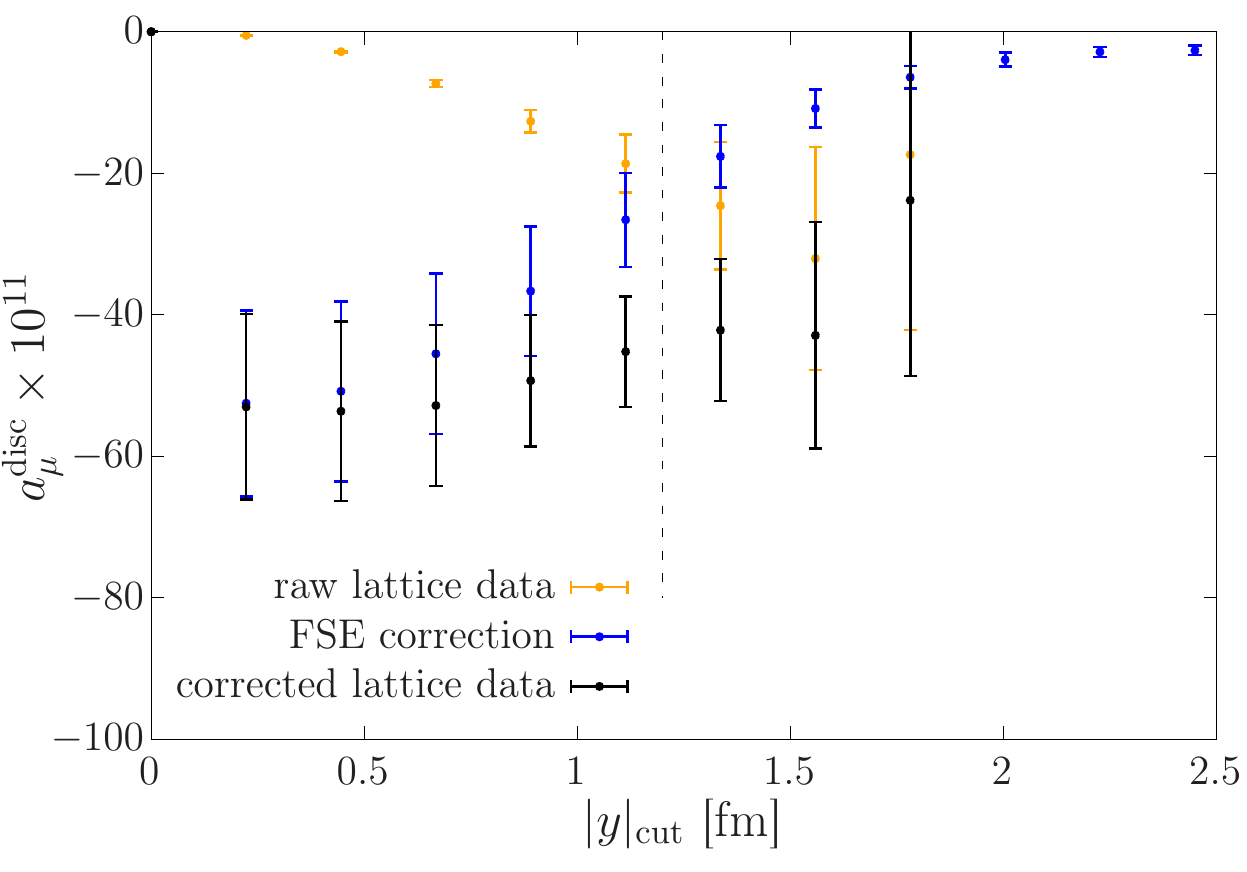}
        \caption{\label{fig:disc_summed} Value of $\amu^{\rm disc}$ as a function of $\ycut$ for ensembles H101 (left) and N202 (right). }
\end{figure}

The ensemble B450 has lattice volume $32^3\times 64$ and periodic boundary condition in time, so a fully-periodic grid built of the two basis vectors $(4,4,4,0)$ and $(2,2,2,4)$ was used. All of the other determinations had multiples of $(1,1,1,3)$ and $(2,2,2,0)$ $|y|$ directions, so it is possible that B450 might have noticeably different discretization and finite-volume effects as this direction lies in a different lattice irreducible representation of the broken rotation group $O(4)$. However, as we see in Figs.~\ref{fig:disc_intgnd} and \ref{fig:disc_summed} the short-distance contribution to the integral is small, and so we can assume the same is true of the discretization effects. As we appropriately correct for FSEs with this choice of direction, we do not expect a significant discrepancy compared to the open-boundary data. Upon continuum extrapolation (Fig.~\ref{fig:M2conn_disc}) it does seem that this ensemble is consistent with the others, indicating that rotation-breaking artifacts are not the main source of discretization effects.

The integrand for the disconnected contribution is displayed in
Fig.~\ref{fig:disc_intgnd} for the value $\Lambda=0.4$; much like for Method 2 we find this value to be preferable.  The integrand is
compared to the prediction for the exchange of the $\pi^0$ and $\eta$
mesons with a VMD TFF. In addition, the same
prediction including an estimate of the $\eta'$
contribution, based on the assumptions in Section~\ref{sec:disc_intgnd},
is indicated.

We do not see any statistically significant finite-volume effects in the integrands between the ensembles U103 and H101, and H200 and N202, for $|y|<1\,$fm. This observation is consistent with the predictions for the $(\pi^0,\eta)$ exchange in finite volume. The central values of the integrand obtained on ensembles with different volumes differ substantially at some larger values of $|y|$, however this is also in the regime where the signal is rapidly deteriorating, if not lost already. There is a trend in the tail for the larger-volume results to enhance the magnitude of the disconnected contribution, much like what we saw in the connected contribution. We see some enhancement of the integrand compared to the $\pi^0+\eta+\eta^\prime$ prediction at short distances; the likely cause of this is the contribution from scalar mesons.

Clearly, the $\pi^0$ and $\eta$ exchanges already provide a rather
good description of the shape of the integrand, unlike in the case of
the connected contribution (e.g. Figs.~\ref{fig:N202_M1} and~\ref{fig:N202_M2}). At the same time, the loss of the signal
beyond 1.2\,fm means that we cannot, at this point, confirm the validity
of the $(\pi^0,\eta)$ exchange at long distances. Given the rapid degradation in signal of the disconnected data, probing large distances of the disconnected contribution will be a very challenging undertaking. There are however
good reasons to believe that this description should apply in that
regime, and in the following we assume this to be the case.

\begin{table}[h!]
\begin{tabular}{c@{\hskip 01em}c@{\hskip 01em}c@{\hskip 01em}c@{\hskip 01em}c@{\hskip 01em}c}
\toprule
Label 	& $\amu^{\rm data}$ & $\amu^{\rm FSE}$ & $\amu^{\mathrm{tail}; \pi^0+\eta}$     & $\amu^{\mathrm{tail}; \eta^{\prime}}$ &       
$a_\mu^{\text{disc}}$ \\
\hline
U103   &	$-17.7(1.9)$  & $-2.4$ & $-30.0$ & 4.11 & $-45.9(1.9)(7.1)$ \\
H101   &	 $-23.4(2.3)$  & $-0.6$ & $-30.0$ &4.11 & $-49.8(2.3)(6.6)$ \\
B450  &	$-26.7(3.4)$  & $-0.9$ & $-28.4$ & 4.11 & $-51.8(3.4)(6.3)$ \\
H200  &	$-12.6(3.7)$  & $-2.2$ & $-28.2$ & 4.14 & $-38.9(3.7)(6.6)$ \\
N202  &	$-21.0(6.0)$  & $-0.2$ & $-28.2$ & 4.14 & $-45.3(6.0)(6.1)$ \\
N300  &	$-20.6(5.4)$  & $-0.7$ & $-24.7$ & 4.01 & $-42.0(5.4)(5.4)$ \\
\botrule
\end{tabular}
\caption{Finite-volume corrected disconnected contributions to $a_\mu^{\text{hlbl}}$ at the SU$(3)_\text{\rm f}$-symmetric point. A breakdown of the FSE contributions to the total results are shown, again a $25\%$ systematic to the total finite-size correction is used. The value of $|y|_\text{cut}$ was $1.2\text{ fm}$ for each ensemble.}\label{tab:conndisc}
\end{table}

Table~\ref{tab:conndisc} summarizes our results. We perform the FSE
matching at a single linearly-interpolated point of
$\ycut=1.2\:\text{fm}$. This point appears to be where we start losing
signal for most of our ensembles, and so a significant proportion of
the tail of the integrand has to be modelled. We take solace in the
fact that the model appears to describe our data well even at
distances far shorter than 1.2 fm, as can be seen in
Figs.~\ref{fig:disc_intgnd} and \ref{fig:disc_summed}. 
By far the largest part of the correction comes from modelling the tail
with the $(\pi^0,\eta)$ exchange. This correction 
is of the order of $100\%$ of the lattice-determined
contribution. 
We have also computed an estimate of the $\eta'$ exchange contribution 
as described in section \ref{sec:disc_intgnd}. 
The values in Table \ref{tab:conndisc} show that its contribution to the tail, 
$|y|>1.2$\,fm, is much smaller. Its magnitude is covered 
by the systematic uncertainty we assign to the modelling of the tail.
We have chosen to include the estimated contribution of the $\eta'$ to the tail 
in the central value of $a_\mu^{\rm disc}$.

Anticipating the combined analysis presented in section \ref{sec:finalres}, 
the continuum-extrapolated disconnected value we obtain from a constrained-slope fit 
to both the connected Method~2 data and the disconnected data is
\be
\amu^{\rm disc} = -33.5(4.2)\times10^{-11}.
\ee
We observe that this result amounts to be about $(-1/3)$ of the connected contribution.

\section{The total \texorpdfstring{$\ahlbl$}{a-mu HLBL}}

\la{sec:combined}
The main purpose of this section is to describe how we arrive at our final result for the total $\ahlbl$ (subsection \ref{sec:finalres}), 
in which the systematics of the continuum extrapolation are discussed in detail. The following sub-section \ref{sec:m_mu_depd} presents a study of the dependence of our result on the muon mass, and finally, subsection \ref{sec:physpt} discusses what outcome one may expect for $\ahlbl$ at physical quark masses based on our findings.

\subsection{Final result for \texorpdfstring{$\ahlbl$}{a-mu HLBL} in \texorpdfstring{SU$(3)_{\rm f}$}{SU(3) flavor}-symmetric QCD}
\la{sec:finalres}
\subsubsection{Combining the connected and disconnected contributions}

\begin{figure}[t!]
        \centering
        \includegraphics*[scale=0.4]{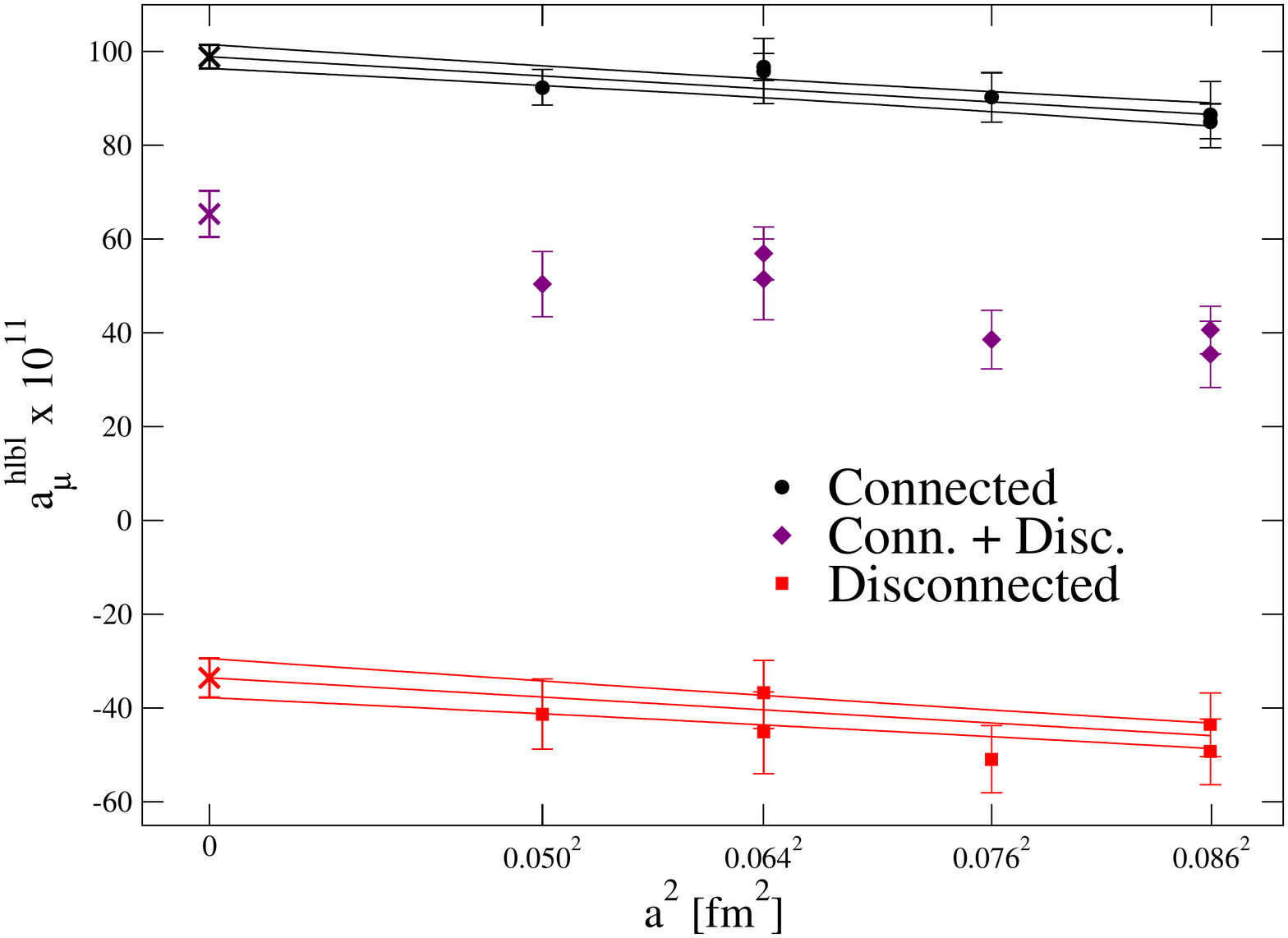}
        \caption{Combined continuum-extrapolation analysis for the connected and disconnected data. The fits were constrained to have the same slope and the final sum of the two was performed on the constant fit parameters. Also shown is the individual sum of the connected and disconnected pieces. The purple cross represents the addition of the continuum-extrapolated results for the connected and disconnected contributions.}\label{fig:M2conn_disc}
\end{figure}

In order to obtain the final result for $\ahlbl=\amu^{\rm conn}+\amu^{\rm disc}$,
we combine the disconnected data with the connected data of Method~2, as it is consistent with, yet statistically more precise than that of Method~1. We then perform two analyses on this combined data set.
The full set of data for the two contributions and their sum is shown
in Fig.~\ref{fig:M2conn_disc}, along with fits and results from the
first analysis whose description follows. Both the connected and
disconnected data show a negative slope in $a^2$ despite the two
contributions having opposite signs. This indicates that the leading
discretization effects do not arise from a common multiplicative
effect such as the renormalization factor.

The first analysis consists in simultaneously extrapolating both the connected and the
disconnected contributions to the continuum with a slope in $a^2$ constrained to be equal for the two contributions, before summing the two constant parameters of the fit to obtain $\ahlbl$.  In this procedure, the results for the connected contribution and the disconnected contribution are both corrected for 
finite-size effects and for the extension of the $|y|$-integrand. The statistical errors are obtained under the bootstrap, and a correlated gaussian sample with appropriate width is propagated for the systematic error due to the correction.
The systematic
error of the correction, which is set to $25\%$ of its size, is
treated as being fully anti-correlated between the connected and the
disconnected contribution (recall that the two contributions have
opposite sign). The fit ansatz for the two continuum extrapolations is 
a polynomial of degree one in $a^2$. For this analysis we obtain the result,
\begin{equation}
\ahlbl = 65.4(4.9)\times 10^{-11}.
\end{equation}

In the second analysis method, the connected and disconnected
contributions are summed prior to performing the continuum
extrapolation. This extrapolation is then performed linearly in $a^2$.
In this analysis, the correction applied to the lattice data is split
into two parts, (a) $a_{\mu}^{\rm FSE}$, the pure finite-size correction on the
integrand, and (b) $a_{\mu}^{\rm tail}$, the extension of the integrand beyond some
$|y|_{\rm cut}$.  In combining the connected and disconnected
contributions, each part is treated as being fully anti-correlated
between the connected and the disconnected contribution; however, in
contrast with the first analysis, the
correlation between the systematic errors of the two parts is
considered to be zero.  The alternative point of view adopted here is that
predicting the tail of the infinite-volume integrand is a rather
different exercise from predicting the finite-size effect on the
finite-volume integrand; one of the two predictions could be
successful while the other is not. The result of this procedure is
\begin{equation}
\ahlbl = 64.5(6.7)\times 10^{-11}.
\end{equation}

The two analyses produce compatible values for $\ahlbl$ but the first
is more precise and allows more flexibility since the two
contributions are treated separately. Therefore, we will use the first
analysis for the final result and study variations on it to estimate
the systematic uncertainty.

\subsubsection{Continuum extrapolation systematics}

\begin{figure}[h!]
\centering
\includegraphics[scale=0.4]{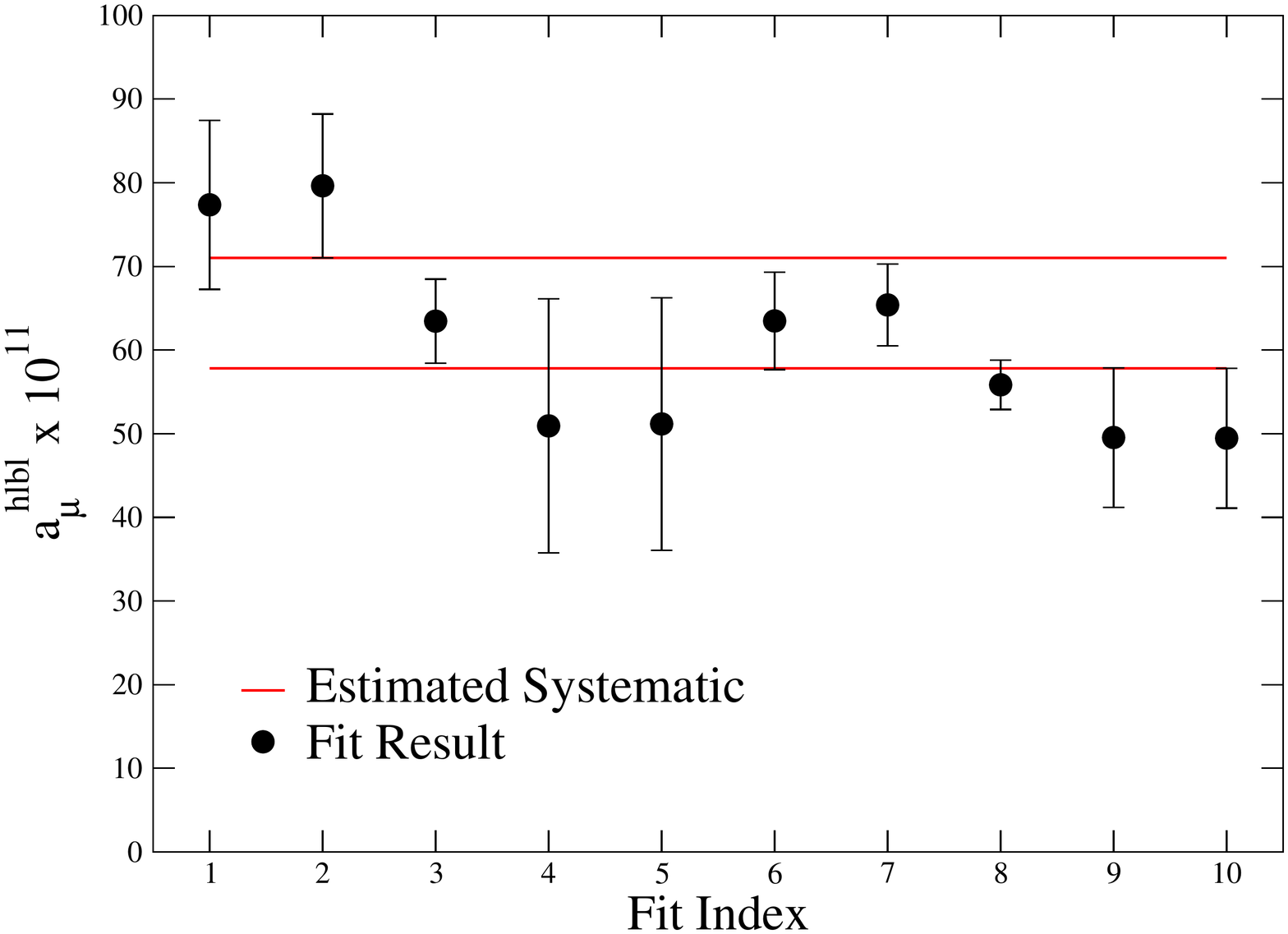}
\caption{An estimate of our continuum extrapolation systematic on the full $\ahlbl$ for the combination of Method 2 and the disconnected data.}
\label{fig:cont_spread}
\end{figure}

Our data are noisy and it is difficult to find a fit that describes them perfectly, so we identify several different choices of continuum extrapolation to investigate the spread and provide an associated continuum-extrapolation systematic. We note that (also expressed in Sec.~\ref{sec:meth1}) as we are not using the $O(a)$-improved vector currents it is possible that we have a term linear in the lattice spacing. We consider the two fit forms:
\begin{equation}
a_\mu^{\text{conn}}(a)=a_\mu^{\text{conn}}(0)+A a^n, \qquad a_\mu^{\text{disc}}(a)=a_\mu^{\text{disc}}(0)+B a^m,
\end{equation}
with various cuts to the data, constraints on $A$ and $B$, and choices of $n$ and $m$ as listed in Table~\ref{tab:sysfits}. We then add the distributions $\ahlbl = a_\mu^{\text{conn}}(0) + a_\mu^{\text{disc}}(0)$ to obtain the results shown in Fig.~\ref{fig:cont_spread}.

\begin{table}[h!]
\begin{tabular}{c|cccc|c|cccc}
\toprule
Index & $a$-Cut [fm] & Constraint & $n$ & $m$ & Index & $a$-Cut [fm] & Constraint & $n$ & $m$ \\ 
\hline
1 & None & None & 1 & 1 & 6 & None & None & 2 & 2 \\
2 & None & $A=B$ & 1 & 1 & 7 & None & $A=B$ & 2 & 2 \\
3 & None & $B=0$ & 1 & - & 8 & None & $B=0$ & 2 & - \\
4 & $< 0.0864$ & None & 1 & 1 & 9 & $< 0.0864$ & None & 2 & 2 \\ 
5 & $< 0.0864$& $A=B$ & 1 & 1 & 10 & $< 0.0864$ & $A=B$ & 2 & 2 \\
\botrule
\end{tabular}
\caption{Different fit forms used to estimate the continuum extrapolation systematic.}\label{tab:sysfits}
\end{table}

It is clear that any time we omit the coarse data the fit wants to flatten the slope. This is due to the anomalously low result of N300. When we do not include the coarse ensembles the error increases substantially, which is fairly indicative that the fit is struggling to accurately model the data.

If we perform a fit linear in $a$ the central value moves up, as was also the case for the Method 1 continuum extrapolation. We choose to quote the linear in $a^2$ fit to all the data with a constrained slope as our final result as it has the best $\chi^2/\mathrm{d.o.f.}=2$. For the continuum-extrapolation systematic, we use the lower error bound of the largest fit result and the upper bound of our lowest fit result. It is clear that constraining the slope or letting it vary does little to the position of the central value apart from reducing the error, this suggests that the fit is having a hard time accurately determining the slope with the quality of the data we have at present.

\subsection{The dependence of our results on the muon mass}
\la{sec:m_mu_depd}

Since the mesons at the SU(3)$_{\rm f}$-symmetric point can be viewed as `heavy' degrees of freedom relative to the muon, we would expect $\ahlbl$ to be roughly proportional to $m_\mu^2$.
Here, we will study what happens if we re-scale the muon mass on one of our ensembles. We are motivated to do so by our experience on related projects~\cite{Gerardin:2019rua} whereby adjusting the muon mass by some dimensionless ratio can flatten the approach to the chiral limit. Here we investigate this idea by defining two quantities,
\begin{equation}
\overline{a}^{\text{hlbl}}_\mu = 
\left(\frac{f_\pi^{\text{Latt.}}}{f_\pi^{\text{Phys.}}}\right)^2 a_\mu^{\rm hlbl}(m_\mu^{\text{Phys.}}),
\end{equation}
and,
\begin{equation}
\quad \tilde{a}^{\text{hlbl}}_\mu = a_\mu^{\text{hlbl}}\left(\frac{f_\pi^{\text{Latt.}}}{f_\pi^{\text{Phys.}}}m_\mu^{\text{Phys.}}\right).
\end{equation}
The first quantity rescales the integrated result by the lattice-determined pion decay constant divided by the value in continuum, squared. The second quantity re-scales the muon mass used as input in our determination by this ratio. These two definitions would be comparable if $\ahlbl$ scales as $m_\mu^2$, which is to be expected in the heavy-quark limit.

\begin{figure}[t]
\includegraphics[scale=0.4]{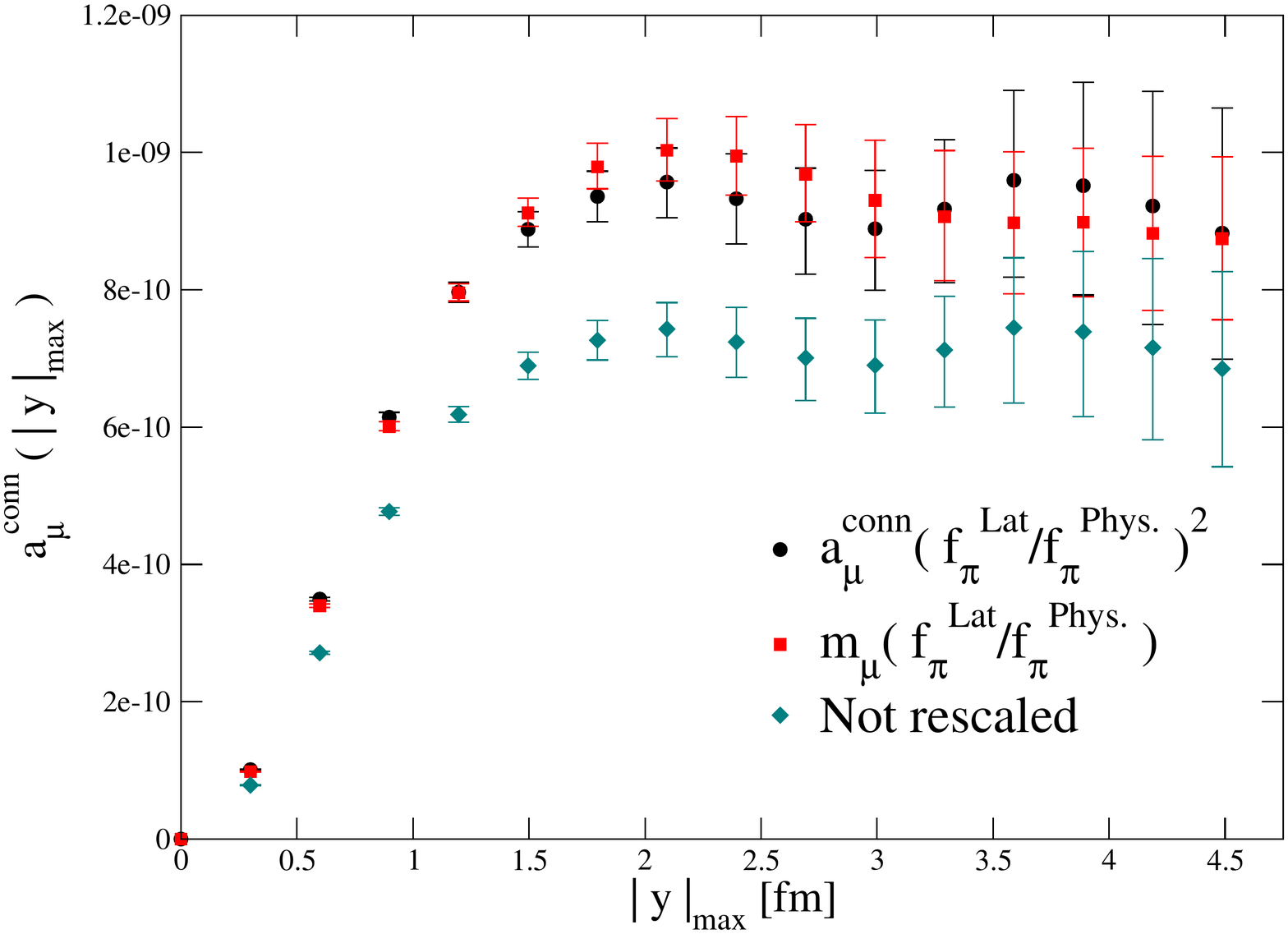}
\caption{Partially-integrated lattice results for the ensemble H101 with and without re-scaling of the integral.}\label{fig:H101PartInt}
\end{figure}

Fig.~\ref{fig:H101PartInt} illustrates the effect of these re-scaling procedures on the connected contribution to $\ahlbl$ on one of our coarsest and largest boxes, H101 (results are from Method 2 with $\Lambda=0.4$). It is clear that these two prescriptions are equivalent within error, which suggests that any change in the muon mass leads to a quadratic change in the integrated result. We can also use this analysis to \textit{very naively} infer how much we expect the result to grow as we approach the chiral limit, and it appears that for the connected contribution this could be of the order of a $25\%$ increase.

\subsection{Expectations for \texorpdfstring{$a_\mu^{\rm hlbl}$}{a-mu HLBL} in QCD with physical quark masses}
\la{sec:physpt}

In this subsection, we first quantify the contributions to $a_\mu^{\rm hlbl}$ at the SU$(3)_{\rm f}$-symmetric point 
not coming from the light pseudoscalars. These contributions are expected to have only a modest quark-mass dependence, 
and they are also the hardest to determine quantitatively by (experimental-)data-driven methods. Therefore 
any information on these contributions is worthwhile collecting. By using our determination of this contribution together with
our previous calculation of the $\pi^0$ exchange, we can arrive at an estimate for $a_\mu^{\rm hlbl}$ at physical quark masses.

We begin by noting that, subtracting the $\pi^0$ and $\eta$ contributions (respectively Eqs.\ (\ref{eq:amupi0su3}) and (\ref{eq:amuetasu3})) from our final SU(3)-point result (\ref{eq:final_amu}), 
the contribution of heavier intermediate states amounts to
\be\la{eq:amu_heavierstates}
 a_\mu^{\rm hlbl,SU(3)_{\rm f}} - a_\mu^{\rm hlbl,\pi^0+\eta,SU(3)_{\rm f}}   = (37.4 \pm 8.3)\times 10^{-11}.
\ee
In particular, this contribution accounts for 57\% of the total $a_\mu^{\rm hlbl}$, and we have added all statistical and systematic errors
in quadrature.

Next, we may try to roughly estimate  $a_\mu^{\rm hlbl}$ at the physical point.
As the $(u,d)$ quark masses are lowered to their physical values at
fixed trace of the quark mass matrix, it is the pion whose mass
changes by the largest factor: it becomes a factor of three lighter. Since we have an evaluation of the $\pi^0$ exchange
contribution at the SU$(3)_{\rm f}$-symmetric point and at the physical point (see Eqs.\ (\ref{eq:amupi0su3}--\ref{eq:amupi0phys})), we can 
correct for this effect,
\be\la{eq:amuphys_pi0corrected}
a_\mu^{\rm hlbl,SU(3)_{\rm f}} - a_\mu^{\rm hlbl,\pi^0,SU(3)_{\rm f}} + a_\mu^{\rm hlbl,\pi^0,{\rm phys}} 
= (104.1\pm9.1)\times 10^{-11}.
\ee
One can think of Eq.\ (\ref{eq:amuphys_pi0corrected}) as a  rough estimate of $a_\mu^{\rm hlbl}$ at the physical point,
purely based on our lattice QCD results and the assumption of a
negligible quark-mass dependence of the non-$\pi^0$-exchange contributions. 
As argued in the next paragraph, we expect such an estimate to hold at the 20\% level.
It is well in line with the most recent evaluations~\cite{Aoyama:2020ynm}. 

In order to assess the systematic uncertainty of such an estimate,
we perform a slightly more sophisticated method to correct for the quark-mass dependence of the $\eta$ contribution and the charged pion loop.
Within the scalar QED framework, we find $-6.3\times10^{-11}$ for the pion loop, to be doubled to include the kaon loop,
and we expect a factor of two to three reduction if one includes an electromagnetic form factor for the pseudoscalar.
Therefore, further subtracting $a_\mu^{{\rm hlbl},(\pi^\pm,K^\pm),{\rm SU}(3)_{\rm f}}\approx 2\times(-3.0)\times10^{-11}$ 
from Eq.\ (\ref{eq:amu_heavierstates}), we obtain $(43.4\pm9.3)\times10^{-11}$.
This number represents our estimate of the non-pseudo-Goldstone contributions at the SU$(3)_{\rm f}$ -symmetric point\footnote{Note that since the lightest $f_0$ meson is stable, there is no reason to treat its contribution as part of the $\pi\pi$ rescattering effect.}.
We observe that by neglecting the quark-mass dependence of this contribution, and using the dispersive $\pi^0$-exchange~\cite{Hoferichter:2018dmo,Aoyama:2020ynm} and the Canberbury-approximant
$\eta$-exchange~\cite{Masjuan:2017tvw,Aoyama:2020ynm} results for 
$a_\mu^{\rm hlbl,\pi^0+\eta,{\rm phys}}=79.3\times 10^{-11}$ and the box contribution~\cite{Colangelo:2017fiz,Aoyama:2020ynm} 
to $ a_\mu^{{\rm hlbl},(\pi^\pm,K^\pm),{\rm phys}}=-16.4\times 10^{-11}$, we arrive 
at the estimate $10^{11}a_\mu^{\rm hlbl,phys}=79.3(3.0)-16.4(2)+43.4(9.3)=106.3(9.8)$. It is only slightly different from the more naive estimate 
of Eq.\ (\ref{eq:amuphys_pi0corrected}). Therefore we consider it safe to assign a systematic error of 20\% to Eq.\ (\ref{eq:amuphys_pi0corrected}) as an 
estimate of $a_\mu^{\rm hlbl}$ at the physical point. This uncertainty estimate also generously covers the $a_\mu^{\rm hlbl,phys}$ value obtained by assuming that the 
non-pseudo-Goldstone contributions increase from the SU(3)$_{\rm f}$ to the physical point by a factor $(f_\pi^{\rm SU(3)_{\rm f}}/f_\pi^{\rm phys})^2$ to account approximately 
for the quark-mass dependence of the QCD resonances; see the previous subsection concerning this point.

\section{Conclusions\la{sec:concl}}

In this work we have computed the hadronic light-by-light contribution
to the $g-2$ of the muon using lattice QCD at the SU$(3)_\text{\rm f}$-symmetric point with $m_{\pi} = m_{K} \approx 420$~MeV. We
chose to initially work at the symmetric point for several reasons:
Due to the significantly reduced computational cost (as compared to
simulations at physical quark masses), we are able to control all
known sources of systematic error, in particular the finite-size and
finite lattice spacing effects.  Second, the SU(3)$_{\rm f}$-symmetry
implies that only two out of five quark-contraction topologies
contribute, and it simplifies the hadronic models with which the
integrand can be compared and interpreted.  For instance, the
$\eta$-exchange contribution simply amounts to one third of the
pion-exchange. Third, the overall contribution from states beyond the
light pseudoscalars is not expected to be strongly quark-mass
dependent, so that the present calculation already constrains its
size.

In order to help interpret our results for $\ahlbl$, we have performed
an exploratory study of the low-lying meson spectrum at the
SU(3)$_{\rm f}$ point.  The most remarkable feature is the existence
of a stable singlet $J^{PC}=0^{++}$ meson with a mass of about
680\,MeV. Also, our previous calculation of the pion transition form
factor~\cite{Gerardin:2019vio} allows us to quantify the contributions
of the $\pi^0$ and $\eta$ exchanges.

Our strategy to calculate $\ahlbl$ relies on coordinate-space
perturbation theory, for which muon and photon propagators are
computed in infinite volume.  We have presented the integrand for the
final, one-dimensional integral over the distance $|y|$ of a
quark-photon vertex from one of the other two internal vertices, since
it contains more information than the final $\ahlbl$ value.  For the
quark-connected contribution we have identified two methods of
calculation that we call Method~1 and Method~2. The former amounts to a
direct computation of the three connected diagrams, but it is a
numerically costly approach, as it involves many sequential-propagator
calculations. To make this computation much cheaper, we have utilized
several changes of variables and translational invariance to rewrite
the integral in terms of an easy-to-calculate single diagram and a
combination of different kernels; this we call Method~2.

For a single $|y|$-value, Method~2 requires
only two propagator inversions, at the meagre cost of a
more-complicated QED-kernel calculation. Method~2 has another
computational advantage over Method~1 in that it allows one to store a
set of propagators in memory and perform their integrals, redefining
the origin to be each propagator source; thereby, for $N$ propagators we
can compute $N(N-1)$ non-zero samples of $f(|y|)$. If the source points
are evenly spaced and the volume is periodic this amounts to $N$
self-averages per $|y|$. Such self-averaging is crucial in reducing
the cost of the calculation.

We note that the combination of kernels needed for Method~2 broadens the
integrand in $|y|$ significantly. To counteract this effect, we have introduced a
new class of subtracted kernels with a gaussian-regulator
$\Lambda$. This parameter $\Lambda$ effectively allows us to tune the
shape of the integrand to peak at shorter-distances. We find that a
value of $\Lambda=0.4$ suits our purposes quite well and allows for
the integral of the connected lattice data to saturate at reasonably
short distances of about $2\text{ fm}$.

We have handled the disconnected contribution by introducing a sparse
sub-grid of equi-distant point sources, the idea being that we obtain
a large number of self-averages from treating our origin as each point
on the grid. Such a technique is necessary as this contribution is
extremely noisy and suffers from a significant signal-to-noise problem
at large distances. We note that potentially millions of averages are
needed in order to get good control over errors at
distances of order $2\text{ fm}$.

In our calculation, we have seen that finite-size effects are
significant, and having a good theoretical understanding of the tail
of the integrand is very important. For the quark-connected
contribution, we have approximated the $\pi^0$ and $\eta$ meson
exchange contribution to the integrand using a vector-dominance
transition form factor and compared it to the lattice data; only at
fairly large $|y|$ does the prediction quantitatively represent the
integrand. We therefore attempt to conservatively incorporate as much
lattice data as possible before making contact with the model. For the
disconnected contribution, the model does a satisfactory job of
describing the data over the entire range of $|y|$ where we have
signal, and we therefore match on to the prediction at shorter
distances, where we still have control over the statistical errors of
the lattice data.

We have shown that both the disconnected and connected contributions
have a non-negligible discretization effect within our measured
precision. It appears to be of the same sign and comparable magnitude
for both contributions, but ultimately this extrapolation appears well
under control. We decide to quote a result from a fit linear in $a^2$
with a constrained slope to the data of Method~2 and the disconnected
as,
\begin{equation}\la{eq:final_amu}
\ahlbl = (65.4\pm 4.9\pm 6.6)\times 10^{-11},
\end{equation}
at the SU(3) flavor-symmetric point, 
where the first error results from the uncertainties on the individual gauge ensembles, 
and the second is the systematic error of the continuum extrapolation.

We have discussed in subsection \ref{sec:physpt} how we expect our
result for $\ahlbl$ to evolve as the up and down quark masses are
lowered towards their physical values at fixed trace of the quark mass
matrix. Correcting for the increase in the $\pi^0$ exchange
contribution\footnote{For this purpose, a model-independent parametrization of the pion transition form factor is used, 
as opposed to the simple vector-meson dominance parametrization.} using our previous lattice
calculation~\cite{Gerardin:2019vio}, we arrive at a value
(Eq.\ (\ref{eq:amuphys_pi0corrected})) which is very well in line with
the most recent phenomenological~\cite{Aoyama:2020ynm} and lattice QCD
results~\cite{Blum:2019ugy}. This value is quite stable under varying
the assumptions about the quark-mass dependence of heavier-state
contributions.

In order to reduce the systematic uncertainty of $\ahlbl$ at physical
quark masses using lattice QCD, obviously simulations at lighter quark
masses are needed. The methods we have developed to correct for
finite-size effects and to extend the tail of the $|y|$-integrand
based on the $\pi^0$ exchange will be extremely valuable in this
endeavor.  While we have reached a semi-quantitative understanding of
the integrand in terms of hadronic models, further work is needed on
the theory side to bring this description to a fully quantitative
level.

\acknowledgments{This work is supported by the European Research Council (ERC) under the
  European Union's Horizon 2020 research and innovation programme
  through grant agreement 771971-SIMDAMA, as well as by  the Deutsche
  Forschungsgemeinschaft (DFG) through the Collaborative Research
  Centre 1044 and through
 the Cluster of Excellence \emph{Precision Physics, Fundamental Interactions, and Structure of Matter} (PRISMA+ EXC 2118/1)  within the German Excellence Strategy (Project ID 39083149).
 The project leading to this publication has also received funding from the Excellence Initiative of Aix-Marseille University - A*MIDEX, a French “Investissements d’Avenir” programme, AMX-18-ACE-005.
Calculations for this project were partly performed on the HPC clusters ``Clover'' and ``HIMster II'' at the Helmholtz-Institut Mainz and ``Mogon II'' at JGU Mainz. 
Our programs use the deflated SAP+GCR solver from the openQCD package~\cite{Luscher:2012av}, as well as the QDP++ library
\cite{Edwards:2004sx}.
The spectroscopy calculation also made use of the PRIMME library~\cite{PRIMME}
and was partly performed on the supercomputer JUQUEEN~\cite{juqueen} at
Jülich Supercomputing Centre (JSC); we acknowledge the Gauss Centre for
Supercomputing e.V.\ (\url{http://www.gauss-centre.eu}) for providing
computer time.
We are grateful to our colleagues in the CLS initiative for sharing ensembles.}

\appendix
\section{Pseudoscalar-meson-exchange-channel matching in Method 2 and the quark-disconnected contribution}
\label{app:pi0matching}

In this work, when we compute the quark-connected diagrams using Method 2 and the quark-disconnected diagrams, only a subset of Feynman diagrams are considered due to the contributing Wick-contractions at QCD level.
In order to study the FSE in an effective field theory framework, one has to map these contractions to the corresponding Feynman diagrams in the effective theory. 
Our study of the FSE is based on pseudoscalar-meson (PS) exchange from Chiral Perturbation Theory.
The matching between the QCD Feynman diagrams and the PS-meson-exchange channels can be done in various ways.
Here, we present a determination of the matching using Partially-Quenched QCD (PQQCD) and Partially-Quenched Chiral Perturbation Theory (PQChPT). 
There have been many applications of PQQCD in the Lattice QCD community (see e.g. \cite{Bernard:1992mk,Sharpe:2000bc,Giusti:2008vb}).
In particular, it can serve as a tool to give an estimate of the size of the contribution coming from quark-disconnected diagrams compared to the connected ones for some observables~\cite{DellaMorte:2010aq}.
Here we do not intend to detail the formalism, but only to give some arguments to explain how we reach the mappings explained in our methodology.

PQQCD is a theory with a graded Lie-group SU$(N|M)$ as symmetry group.
In this theory there are $N-M$ sea quarks and $M$ valence (quenched) quarks with their ghost counterparts. 
The presence of these quenched quarks  
does not change the partition function from the un-quenched theory with $N-M$ dynamical quarks.
Therefore, one can formulate each of the different Wick contractions needed for computing an $n$-point function (in an un-quenched theory) in a partially-quenched theory by adding certain number of quenched quarks.
Then, one can study the long-distance behavior of the observable using PQChPT, which is the corresponding effective field theory to PQQCD.

With only three flavors (as is the case in this work), one cannot build a four-point function that requires only quark-connected diagrams, because each quark line would require a different flavor in order to avoid disconnected diagrams.
However, such a four-point function can be constructed with an additional quenched quark flavor, i.e. SU$(4|1)$ PQQCD.
For instance, if we introduce a quenched quark $r$ and its ghost $\tilde{r}$ with the same mass as the ones originally present in the SU$(3)_{\rm f}$ theory, namely $u$, $d$ and $s$, in the computation of the four-point function
\begin{equation}\label{eq:app-pi0m-4p0m2}
\langle \bar{u}\gamma_\mu d (x)
\bar{d}\gamma_\nu s(y)
\bar{s}\gamma_\sigma r(z)
\bar{r}\gamma_\sigma u(0)
+ h.c.
\rangle,
\end{equation}
only quark-connected diagrams are involved. 
Eq.~(\ref{eq:app-pi0m-4p0m2}) is exactly the four-point function $\widetilde\Pi^{(1)}_{\mu\nu\sigma\lambda}(x,y,z,0)$ that is used for the connected contribution in Method 2.

Moreover, one can study the quark-disconnected diagrams that we consider in this work within the same partially quenched theory. For instance, 
\begin{equation}
\langle \bar{u}\gamma_\mu d (x)
\bar{d}\gamma_\nu u(y)
\bar{s}\gamma_\sigma r(z)
\bar{r}\gamma_\sigma s(0)
+ h.c.
\rangle
\end{equation}
only contains the disconnected diagram where $(x,y)$ and $(z,0)$ are connected by quark lines.

\begin{figure}[t]
	\centering
	\includegraphics*[width=0.6\linewidth]{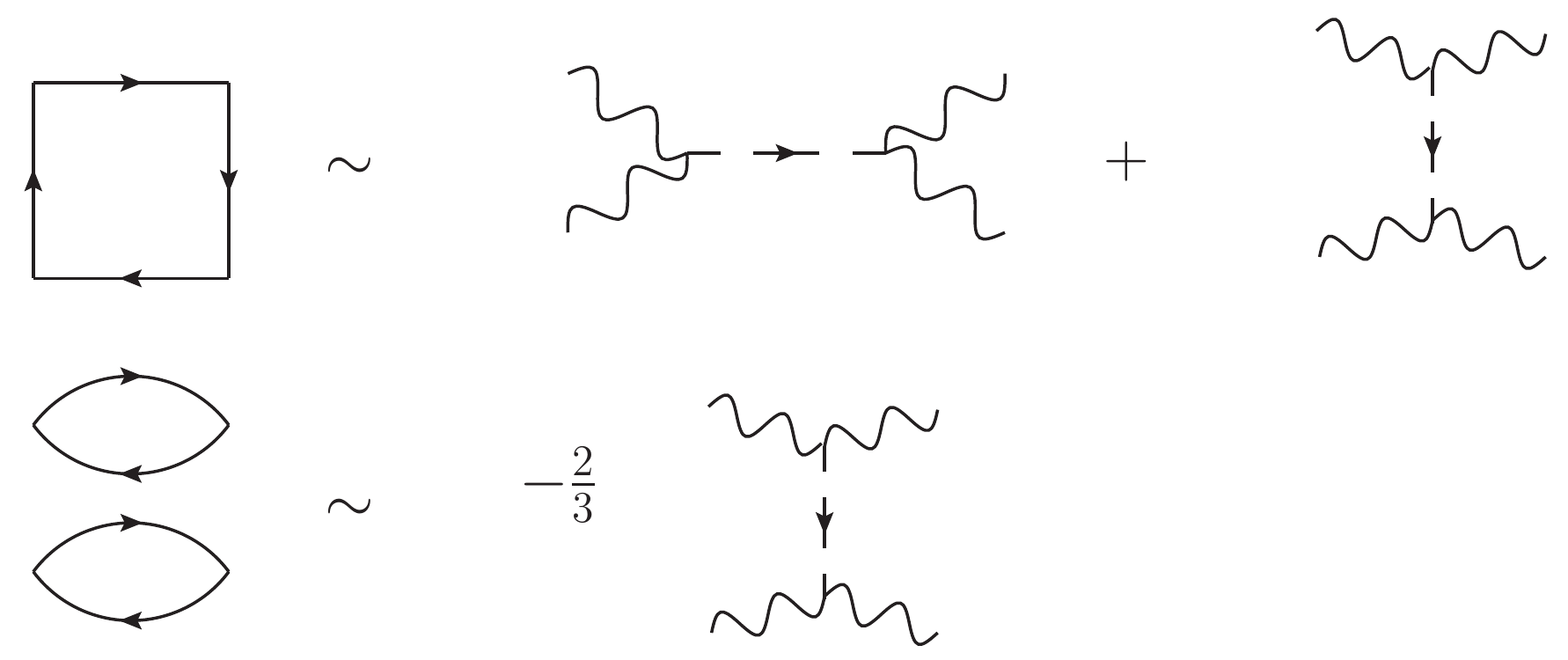} \hspace{1cm}
	\caption{Correspondence between connected (Method 2) and disconnected diagrams (left) and PS-meson exchange channels (right) in position space of the four-point functions described in the text. The top-left figure should be understood as containing quark flow in both orientations, representing exactly all the relevent diagrams for Method 2. Note that the charge factors are not included here.}	
	\label{fig:app-pi0m-matching}
\end{figure}

To determine the relevent PS-meson-exchange channel, one can consider the Wess-Zumino-Witten term~\cite{Wess:1971yu, Witten:1983tw} in PQChPT at leading order in the decay constant.
Eq.~(\ref{eq:app-pi0m-4p0m2}) can be understood as a four-point vector current correlation function. 
A general procedure is to include all the possible \textit{super} traces of operators~\cite{Detmold:2006vu}.
However, in the case of the coupling to external charged vector currents, which are the Noether currents of the symmetry group SU$(N|M)$ whose generators are super-traceless, the only allowed non-vanishing term is proportional to
\begin{equation}
\textrm{str}(T^aT^b T^c)\int d^4 x \phi^c \epsilon^{\mu\nu\rho\sigma} F^{a}_{\mu\nu} F^{b}_{\rho\sigma} ,
\end{equation}
where $\phi$ is the Goldstone boson/ghost field, $F_{\mu\nu}$ is the field strength of the external vector current and the $T^a$'s are the generators of the group. 
As a result, in momentum-space, the PS-meson exchange in the $s$-channel of the two-vector-currents-to-two-vector-currents process $v^{a_1}v^{b_1} \rightarrow v^{a_2}v^{b_2}$ is proportional to
\begin{equation}
\textrm{str}(\{ T^{a_1},T^{b_1}\}T^{c_1}) \textrm{str}(\{T^{a_2},T^{b_2}\}T^{c_2})g^{c_1c_2}G(p^2),
\end{equation}
where $G$ is the scalar field propagator ,
\begin{equation}
G(p^2) = \frac{1}{p^2 + m_\pi^2},
\end{equation}
and 
\begin{equation}
g^{ab}  = 2 \textrm{str}(T^a T^b).
\end{equation}
The matching between different contractions and the PS-exchange channels based on this computation is shown in Fig.~\ref{fig:app-pi0m-matching}.
With the charge factors included, one will recover the ratio between the total quark-disconnected contribution and the total quark-connected given in Ref.~\cite{Gerardin:2017ryf}, based on charge factor arguments.

\section{The new subtraction with the lepton-loop}\label{app:leploop}

In this section we investigate the new subtraction for both Methods 1 and 2 using the infinite volume lepton-loop. A comparison of previous results with our kernel to the lepton-loop can be found in \cite{Asmussen:2017bup,mainzHLbL1}. 
The lepton-loop result can be determined easily by standard numerical integration techniques, the results of which can be used to inform our choices of kernel parameters in the full QCD calculation. 

\begin{figure}[h!]
\begin{minipage}{0.5\textwidth}
\centering \footnotesize
\includegraphics[scale=0.28]{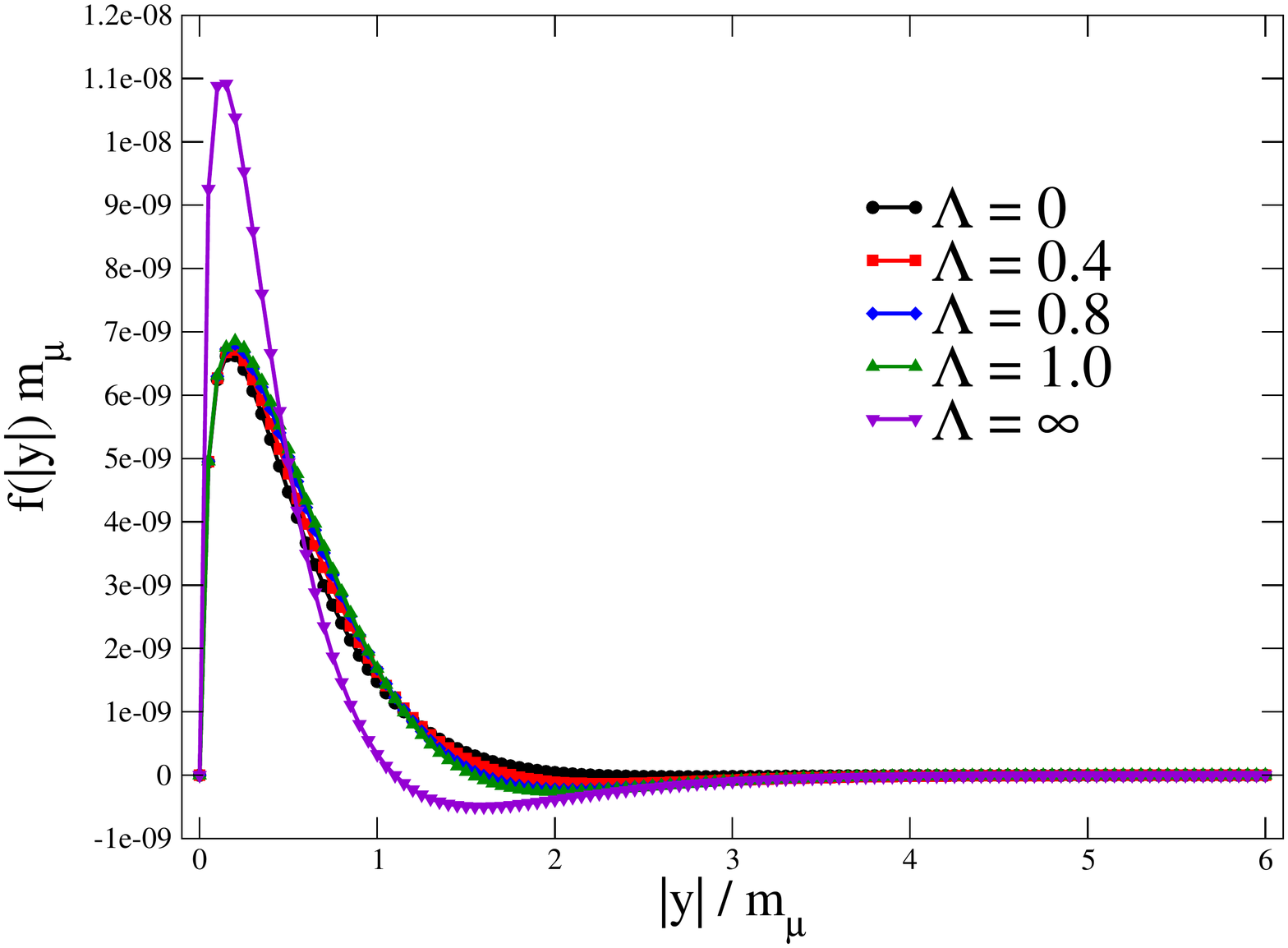}
(a) Method 1 integrands
\end{minipage}
\begin{minipage}{0.5\textwidth}
\centering \footnotesize
\includegraphics[scale=0.28]{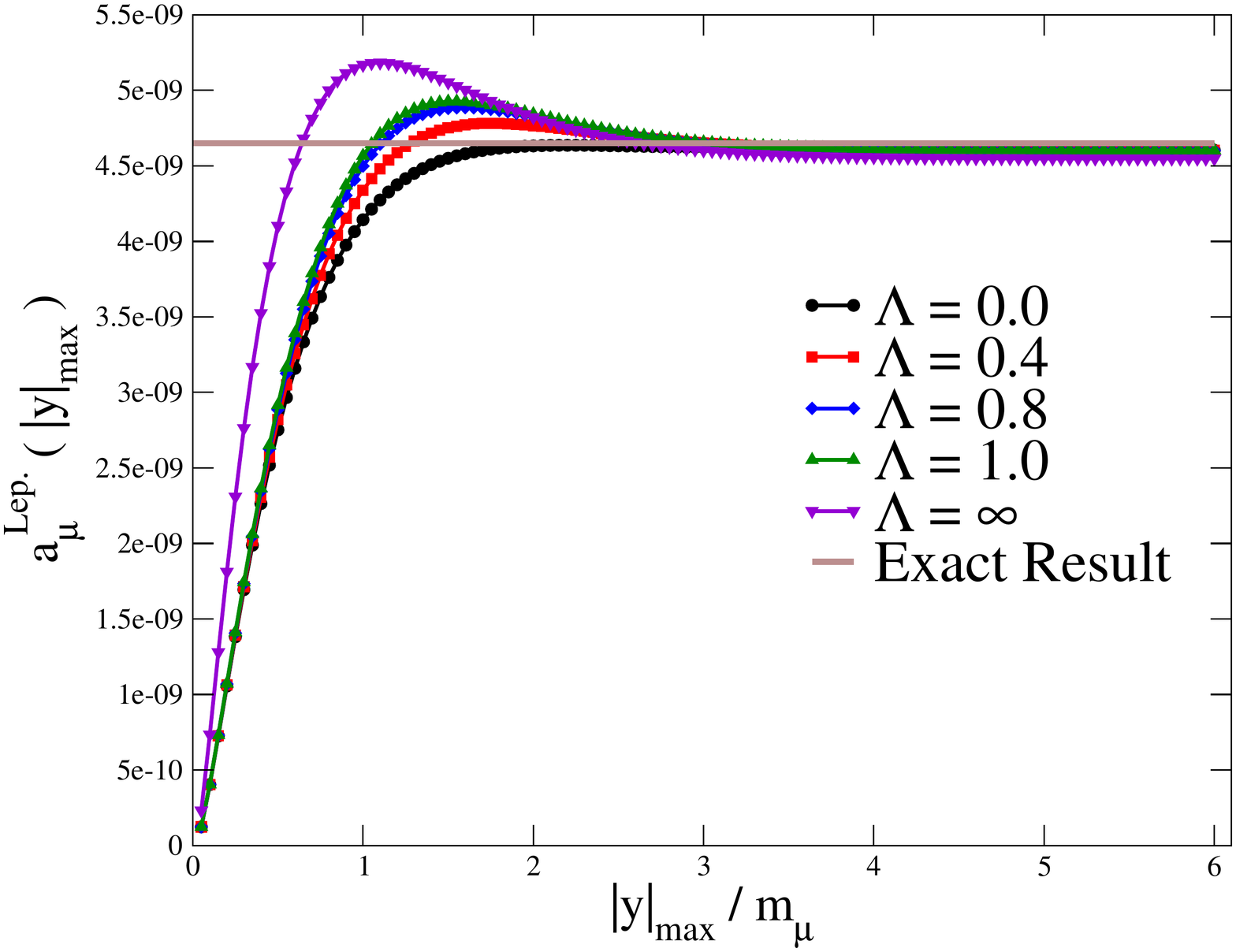}
(b) Method 1 integrals
\end{minipage}
\caption{Left: the integrands in Method 1 of several $\Lambda$ choices of the new subtracted kernel. Right: the results of the partial integration of the integrands up to some $|y|_{\Max}$. These results were obtained with lepton mass equal to the muon mass and the exact result used was $4.6497\times 10^{-9}$ \cite{Laporta:1992pa}.}\label{fig:leploopM1}
\end{figure}

\begin{figure}[h!]
\centering
\begin{minipage}{0.5\textwidth}
\centering \footnotesize
\includegraphics[scale=0.28]{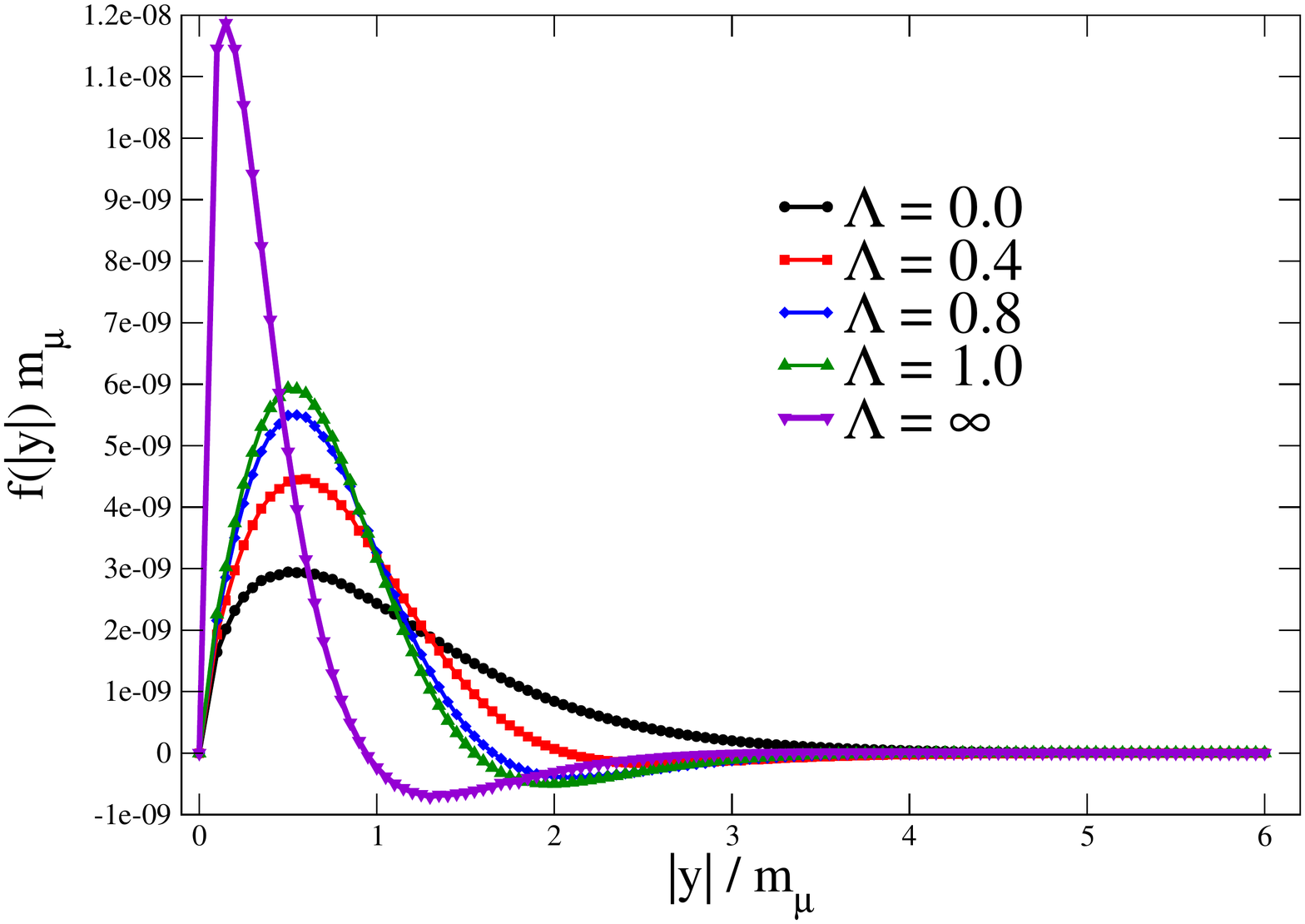}
(a) Method 2 integrands
\end{minipage}
\begin{minipage}{0.5\textwidth}
\centering \footnotesize
\includegraphics[scale=0.28]{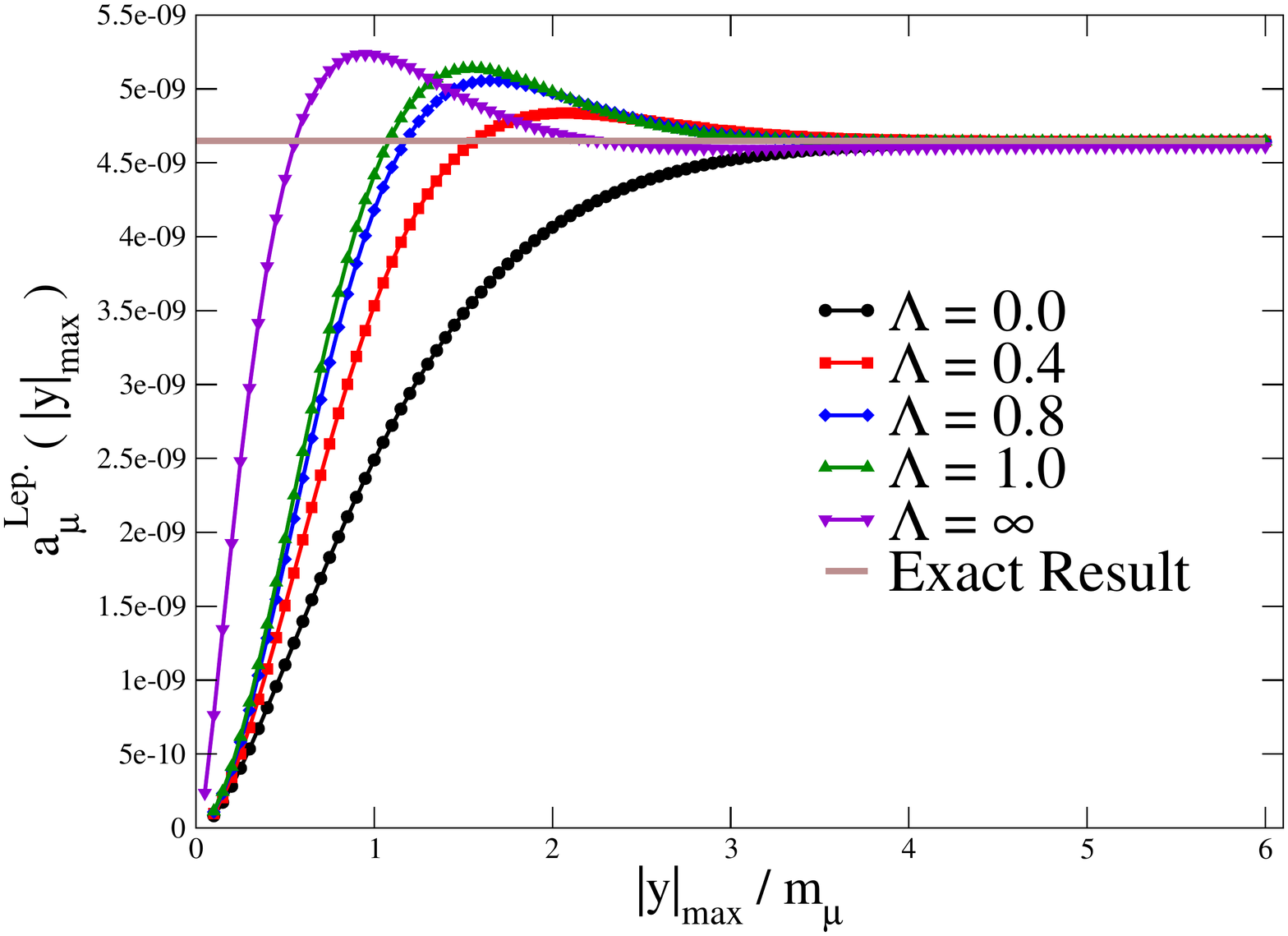}
(b) Method 2 integrals
\end{minipage}
\caption{Same as Fig.~\ref{fig:leploopM1} but for Method 2.}\label{fig:leploopM2}
\end{figure}

For both figures (\ref{fig:leploopM1} and \ref{fig:leploopM2}) we set the lepton mass equal to that of the muon.

Fig.~\ref{fig:leploopM1} shows the results for Method 1. There is little difference between the choices of $\Lambda$ for this calculation and they all saturate the integral reasonably quickly. The result when $\Lambda=\infty$ is too peaked at short distances to make this a useful kernel for the lattice computation. All of the peak positions for the integrands are roughly at the same (small) value of $|y|$.

In Fig.~\ref{fig:leploopM2} we show the results for Method 2. We see that the usual kernel ($\Lambda=\infty$) is very peaked at short distances, this will make it unsuitable for the lattice calculation. We also note that $\Lambda=0$ gives a broad integrand that very slowly approaches the exact result, this also makes it somewhat unsuitable for a lattice calculation unless a large physical volume and large values of $|y|$ are available.

The choice of $\Lambda=0.4$ has a slightly more pronounced peak at small $|y|$ than $\Lambda=0$ and it goes to zero at large distances rapidly. There is a small negative contribution to the integral at large $|y|$, but in the full QCD case can be corrected for by modelling the tail, as $\Lambda$ increases this negative correction grows. If our lepton-loop result translates to full QCD this choice of $\Lambda$ should allow for most of our integral to be encompassed by the lattice volume.

If we compare the two methods using the lepton-loop we see that the parameter $\Lambda$ makes a significant difference to the integrand for Method 2. Any of the choices $\Lambda=0.0,0.4,0.8,1.0$ seem usable for Method 1. It is interesting to note that Method 2 actively shifts the peak of the integrand to typically intermediate distances. So, with a careful choice of the parameter $\Lambda$ we can tune the kernel to peak in a region where the calculation is not so sensitive to either of the lattice cut-offs of lattice spacing and volume.
\section{Kernel independence of \texorpdfstring{$a_\mu^{\rm conn}$}{a-mu connected} and \texorpdfstring{$a_\mu^{\rm disc}$}{a-mu disconnected}}\label{app:WI}

The validity of the subtractions applied to the QED kernel originates from the Ward identity (current conservation) satisfied by the electromagnetic current. 
Thus, one will expect a kernel-independent result at the level of the total result of $\ahlbl$. 
In this paper, we have treated the contribution from the connected diagrams ($a_\mu^{\rm conn}$) and the contribution from the disconnected diagrams ($a_\mu^{\rm disc)}$) separately.
One natural question would be whether these two quantities themselves are also kernel-independent. 
The answer to this question is yes, due to the Ward identity satisfied by the vector current.
We will give a proof of this statement in the SU$(3)_f$ case.\\
In Euclidean space-time, if an infinitesimal transformation $\delta_\epsilon$ of the fields does not change the path integral measure, we have 
\begin{equation}{\label{eq:wi-euclidean}}
\langle \delta_\epsilon S \mathcal{O}\rangle  = \langle \delta_\epsilon \mathcal{O}\rangle,
\end{equation}
where $S$ is the QCD action and $\mathcal{O}$ is any operator.
Without changing the action, we can consider a partially-quenched theory by introducing a quenched quark $r$ and its ghost partner $\tilde{r}$ (see Appendix~\ref{app:pi0matching}).
If we consider an infinitesimal transformation generated by the local SU$(2)_f$ symmetry of the quark pair $(u,d)$ and their conjugates :    
\begin{equation}
\begin{split}
& \delta_\epsilon u(x_0) = i\epsilon(x_0) d(x_0) , \quad
\delta_\epsilon d(x_0) = i\epsilon(x_0) u(x_0),
\\
& \delta_\epsilon \bar{u}(x_0) = -i\epsilon(x_0) \bar{d}(x_0) , \quad
\delta_\epsilon \bar{d}(x_0) = -i\epsilon(x_0) \bar{u}(x_0),
\end{split}
\end{equation}
we have the following expression for the variation of the action related to the corresponding Noether current
\begin{equation}
\delta_\epsilon S = i \epsilon (x_0) \partial_{\lambda} (\bar{u}\gamma_\lambda d + \bar{d}\gamma_\lambda u)(x_0).
\end{equation}
We now consider the trilocal operator $\mathcal{O}_{\mu\nu\sigma}(x,y,z) := \bar{u}\gamma_\mu d (x) \bar{s}\gamma_\nu r(y) \bar{r}\gamma_\sigma s(z)$, which transforms as
\begin{equation}
\delta_\epsilon \mathcal{O}_{\mu\nu\sigma}(x) = i\epsilon( \bar{u}\gamma_\mu u - \bar{d} \gamma_\mu d)(x) \bar{s}\gamma_\nu r(y) \bar{r}\gamma_\lambda s(z).
\end{equation}
Plugging in everything in Eq.~(\ref{eq:wi-euclidean}), we have
\begin{equation}
\partial_{\lambda}\langle \bar{u}\gamma_\mu d (x) \bar{d}\gamma_\lambda u (x_0)  \bar{s}\gamma_\nu r(y) \bar{r}\gamma_\sigma s(z)\rangle = \partial_\lambda \langle \Pi_{\mu\lambda}(x,x_0)\Pi_{\nu\sigma}(y,z) \rangle_U
= 0,
\end{equation}
where $\Pi_{\mu\nu}(x,x_0)\Pi_{\nu\sigma}(y,z)$ gives one of the contractions needed for the disconnected computation.
One can then easily show that all the disconnected diagrams have similar current-conservation property. 
Consequently, the subtractions that we apply to the QED-kernel do not change the result on the contribution from the disconnected diagrams.
It follows directly that the connected contribution is also kernel-independent, because $a_\mu^{\rm conn} = \ahlbl - a_\mu^{\rm disc}$ in our SU$(3)_f$ case.
One can further show that $a_\mu^{\rm conn}$ is also itself kernel-independent in general flavor cases.

\bibliographystyle{apsrev4-1}
\nocite{bibtitles}
\bibliography{refs}

\begin{thebibliography}{48}%
\makeatletter
\providecommand \@ifxundefined [1]{%
 \@ifx{#1\undefined}
}%
\providecommand \@ifnum [1]{%
 \ifnum #1\expandafter \@firstoftwo
 \else \expandafter \@secondoftwo
 \fi
}%
\providecommand \@ifx [1]{%
 \ifx #1\expandafter \@firstoftwo
 \else \expandafter \@secondoftwo
 \fi
}%
\providecommand \natexlab [1]{#1}%
\providecommand \enquote  [1]{``#1''}%
\providecommand \bibnamefont  [1]{#1}%
\providecommand \bibfnamefont [1]{#1}%
\providecommand \citenamefont [1]{#1}%
\providecommand \href@noop [0]{\@secondoftwo}%
\providecommand \href [0]{\begingroup \@sanitize@url \@href}%
\providecommand \@href[1]{\@@startlink{#1}\@@href}%
\providecommand \@@href[1]{\endgroup#1\@@endlink}%
\providecommand \@sanitize@url [0]{\catcode `\\12\catcode `\$12\catcode
  `\&12\catcode `\#12\catcode `\^12\catcode `\_12\catcode `\%12\relax}%
\providecommand \@@startlink[1]{}%
\providecommand \@@endlink[0]{}%
\providecommand \url  [0]{\begingroup\@sanitize@url \@url }%
\providecommand \@url [1]{\endgroup\@href {#1}{\urlprefix }}%
\providecommand \urlprefix  [0]{URL }%
\providecommand \Eprint [0]{\href }%
\providecommand \doibase [0]{http://dx.doi.org/}%
\providecommand \selectlanguage [0]{\@gobble}%
\providecommand \bibinfo  [0]{\@secondoftwo}%
\providecommand \bibfield  [0]{\@secondoftwo}%
\providecommand \translation [1]{[#1]}%
\providecommand \BibitemOpen [0]{}%
\providecommand \bibitemStop [0]{}%
\providecommand \bibitemNoStop [0]{.\EOS\space}%
\providecommand \EOS [0]{\spacefactor3000\relax}%
\providecommand \BibitemShut  [1]{\csname bibitem#1\endcsname}%
\let\auto@bib@innerbib\@empty
\bibitem [{\citenamefont {Bennett}\ \emph {et~al.}(2006)\citenamefont {Bennett}
  \emph {et~al.}}]{Bennett:2006fi}%
  \BibitemOpen
  \bibfield  {author} {\bibinfo {author} {\bibfnamefont {G.~W.}\ \bibnamefont
  {Bennett}} \emph {et~al.} (\bibinfo {collaboration} {Muon $(g-2)$}),\
  }\bibfield  {title} {\enquote {\bibinfo {title} {Final report of the {E821}
  muon anomalous magnetic moment measurement at {BNL}},}\ }\href {\doibase
  10.1103/PhysRevD.73.072003} {\bibfield  {journal} {\bibinfo  {journal} {Phys.
  Rev. D}\ }\textbf {\bibinfo {volume} {73}},\ \bibinfo {pages} {072003}
  (\bibinfo {year} {2006})},\ \Eprint {http://arxiv.org/abs/hep-ex/0602035}
  {arXiv:hep-ex/0602035} \BibitemShut {NoStop}%
\bibitem [{\citenamefont {Jegerlehner}\ and\ \citenamefont
  {Nyffeler}(2009)}]{Jegerlehner:2009ry}%
  \BibitemOpen
  \bibfield  {author} {\bibinfo {author} {\bibfnamefont {F.}~\bibnamefont
  {Jegerlehner}}\ and\ \bibinfo {author} {\bibfnamefont {A.}~\bibnamefont
  {Nyffeler}},\ }\bibfield  {title} {\enquote {\bibinfo {title} {The muon
  $g-2$},}\ }\href {\doibase 10.1016/j.physrep.2009.04.003} {\bibfield
  {journal} {\bibinfo  {journal} {Phys. Rept.}\ }\textbf {\bibinfo {volume}
  {477}},\ \bibinfo {pages} {1} (\bibinfo {year} {2009})},\ \Eprint
  {http://arxiv.org/abs/0902.3360} {arXiv:0902.3360 [hep-ph]} \BibitemShut
  {NoStop}%
\bibitem [{\citenamefont {Blum}\ \emph {et~al.}(2013)\citenamefont {Blum},
  \citenamefont {Denig}, \citenamefont {Logashenko}, \citenamefont {de~Rafael},
  \citenamefont {Lee~Roberts} \emph {et~al.}}]{Blum:2013xva}%
  \BibitemOpen
  \bibfield  {author} {\bibinfo {author} {\bibfnamefont {T.}~\bibnamefont
  {Blum}}, \bibinfo {author} {\bibfnamefont {A.}~\bibnamefont {Denig}},
  \bibinfo {author} {\bibfnamefont {I.}~\bibnamefont {Logashenko}}, \bibinfo
  {author} {\bibfnamefont {E.}~\bibnamefont {de~Rafael}}, \bibinfo {author}
  {\bibfnamefont {B.}~\bibnamefont {Lee~Roberts}},  \emph {et~al.},\ }\bibfield
   {title} {\enquote {\bibinfo {title} {The muon $(g-2)$ theory value: Present
  and future},}\ }\href@noop {} {\  (\bibinfo {year} {2013})},\ \Eprint
  {http://arxiv.org/abs/1311.2198} {arXiv:1311.2198 [hep-ph]} \BibitemShut
  {NoStop}%
\bibitem [{\citenamefont {Jegerlehner}(2017)}]{Jegerlehner:2017gek}%
  \BibitemOpen
  \bibfield  {author} {\bibinfo {author} {\bibfnamefont {F.}~\bibnamefont
  {Jegerlehner}},\ }\href {\doibase 10.1007/978-3-319-63577-4} {\emph {\bibinfo
  {title} {{The Anomalous Magnetic Moment of the Muon}}}}\ (\bibinfo
  {publisher} {Springer Tracts Mod. Phys., vol. 274},\ \bibinfo {year}
  {2017})\BibitemShut {NoStop}%
\bibitem [{\citenamefont {Aoyama}\ \emph {et~al.}(2020)\citenamefont {Aoyama}
  \emph {et~al.}}]{Aoyama:2020ynm}%
  \BibitemOpen
  \bibfield  {author} {\bibinfo {author} {\bibfnamefont {T.}~\bibnamefont
  {Aoyama}} \emph {et~al.},\ }\bibfield  {title} {\enquote {\bibinfo {title}
  {{The anomalous magnetic moment of the muon in the Standard Model}},}\
  }\href@noop {} {\  (\bibinfo {year} {2020})},\ \Eprint
  {http://arxiv.org/abs/2006.04822} {arXiv:2006.04822 [hep-ph]} \BibitemShut
  {NoStop}%
\bibitem [{\citenamefont {Grange}\ \emph {et~al.}(2015)\citenamefont {Grange}
  \emph {et~al.}}]{Grange:2015fou}%
  \BibitemOpen
  \bibfield  {author} {\bibinfo {author} {\bibfnamefont {J.}~\bibnamefont
  {Grange}} \emph {et~al.} (\bibinfo {collaboration} {E989}),\ }\bibfield
  {title} {\enquote {\bibinfo {title} {Muon $(g-2)$ technical design report},}\
  }\href@noop {} {\  (\bibinfo {year} {2015})},\ \Eprint
  {http://arxiv.org/abs/1501.06858} {arXiv:1501.06858 [physics.ins-det]}
  \BibitemShut {NoStop}%
\bibitem [{\citenamefont {Mibe}(2011)}]{Mibe:2011zz}%
  \BibitemOpen
  \bibfield  {author} {\bibinfo {author} {\bibfnamefont {T.}~\bibnamefont
  {Mibe}} (\bibinfo {collaboration} {J-PARC $g-2$}),\ }\bibfield  {title}
  {\enquote {\bibinfo {title} {{Measurement of muon $g-2$ and EDM with an
  ultra-cold muon beam at J-PARC}},}\ }\bibfield  {booktitle} {\emph {\bibinfo
  {booktitle} {{Tau lepton physics. Proceedings, 11th International Workshop,
  TAU 2010, Manchester, UK, September 13-17, 2010}}},\ }\href {\doibase
  10.1016/j.nuclphysbps.2011.06.039} {\bibfield  {journal} {\bibinfo  {journal}
  {Nucl. Phys. B (Proc. Suppl.)}\ }\textbf {\bibinfo {volume} {218}},\ \bibinfo
  {pages} {242} (\bibinfo {year} {2011})}\BibitemShut {NoStop}%
\bibitem [{\citenamefont {Meyer}\ and\ \citenamefont
  {Wittig}(2019)}]{Meyer:2018til}%
  \BibitemOpen
  \bibfield  {author} {\bibinfo {author} {\bibfnamefont {H.~B.}\ \bibnamefont
  {Meyer}}\ and\ \bibinfo {author} {\bibfnamefont {H.}~\bibnamefont {Wittig}},\
  }\bibfield  {title} {\enquote {\bibinfo {title} {{Lattice QCD and the
  anomalous magnetic moment of the muon}},}\ }\href {\doibase
  10.1016/j.ppnp.2018.09.001} {\bibfield  {journal} {\bibinfo  {journal} {Prog.
  Part. Nucl. Phys.}\ }\textbf {\bibinfo {volume} {104}},\ \bibinfo {pages}
  {46} (\bibinfo {year} {2019})},\ \Eprint {http://arxiv.org/abs/1807.09370}
  {arXiv:1807.09370 [hep-lat]} \BibitemShut {NoStop}%
\bibitem [{\citenamefont {Blum}\ \emph {et~al.}(2017)\citenamefont {Blum},
  \citenamefont {Christ}, \citenamefont {Hayakawa}, \citenamefont {Izubuchi},
  \citenamefont {Jin}, \citenamefont {Jung},\ and\ \citenamefont
  {Lehner}}]{Blum:2016lnc}%
  \BibitemOpen
  \bibfield  {author} {\bibinfo {author} {\bibfnamefont {T.}~\bibnamefont
  {Blum}}, \bibinfo {author} {\bibfnamefont {N.}~\bibnamefont {Christ}},
  \bibinfo {author} {\bibfnamefont {M.}~\bibnamefont {Hayakawa}}, \bibinfo
  {author} {\bibfnamefont {T.}~\bibnamefont {Izubuchi}}, \bibinfo {author}
  {\bibfnamefont {L.}~\bibnamefont {Jin}}, \bibinfo {author} {\bibfnamefont
  {C.}~\bibnamefont {Jung}}, \ and\ \bibinfo {author} {\bibfnamefont
  {C.}~\bibnamefont {Lehner}},\ }\bibfield  {title} {\enquote {\bibinfo {title}
  {Connected and leading disconnected hadronic light-by-light contribution to
  the muon anomalous magnetic moment with a physical pion mass},}\ }\href
  {\doibase 10.1103/PhysRevLett.118.022005} {\bibfield  {journal} {\bibinfo
  {journal} {Phys. Rev. Lett.}\ }\textbf {\bibinfo {volume} {118}},\ \bibinfo
  {pages} {022005} (\bibinfo {year} {2017})},\ \Eprint
  {http://arxiv.org/abs/1610.04603} {arXiv:1610.04603 [hep-lat]} \BibitemShut
  {NoStop}%
\bibitem [{\citenamefont {Asmussen}\ \emph
  {et~al.}(2019{\natexlab{a}})\citenamefont {Asmussen}, \citenamefont
  {Gérardin}, \citenamefont {Nyffeler},\ and\ \citenamefont
  {Meyer}}]{Asmussen:2018oip}%
  \BibitemOpen
  \bibfield  {author} {\bibinfo {author} {\bibfnamefont {N.}~\bibnamefont
  {Asmussen}}, \bibinfo {author} {\bibfnamefont {A.}~\bibnamefont {Gérardin}},
  \bibinfo {author} {\bibfnamefont {A.}~\bibnamefont {Nyffeler}}, \ and\
  \bibinfo {author} {\bibfnamefont {H.~B.}\ \bibnamefont {Meyer}},\ }\bibfield
  {title} {\enquote {\bibinfo {title} {{Hadronic light-by-light scattering in
  the anomalous magnetic moment of the muon}},}\ }\href {\doibase
  10.21468/SciPostPhysProc.1.031} {\bibfield  {journal} {\bibinfo  {journal}
  {SciPost Phys. Proc.}\ }\textbf {\bibinfo {volume} {1}},\ \bibinfo {pages}
  {031} (\bibinfo {year} {2019}{\natexlab{a}})},\ \Eprint
  {http://arxiv.org/abs/1811.08320} {arXiv:1811.08320 [hep-lat]} \BibitemShut
  {NoStop}%
\bibitem [{\citenamefont {Colangelo}\ \emph {et~al.}(2018)\citenamefont
  {Colangelo}, \citenamefont {Hoferichter}, \citenamefont {Procura},\ and\
  \citenamefont {Stoffer}}]{Colangelo:2017urn}%
  \BibitemOpen
  \bibfield  {author} {\bibinfo {author} {\bibfnamefont {G.}~\bibnamefont
  {Colangelo}}, \bibinfo {author} {\bibfnamefont {M.}~\bibnamefont
  {Hoferichter}}, \bibinfo {author} {\bibfnamefont {M.}~\bibnamefont
  {Procura}}, \ and\ \bibinfo {author} {\bibfnamefont {P.}~\bibnamefont
  {Stoffer}},\ }\bibfield  {title} {\enquote {\bibinfo {title} {{Hadronic
  light-by-light contribution to $(g-2)_\mu$: a dispersive approach}},}\
  }\bibfield  {booktitle} {\emph {\bibinfo {booktitle} {{Proceedings, 35th
  International Symposium on Lattice Field Theory (Lattice 2017): Granada,
  Spain, June 18-24, 2017}}},\ }\href {\doibase 10.1051/epjconf/201817501025}
  {\bibfield  {journal} {\bibinfo  {journal} {EPJ Web Conf.}\ }\textbf
  {\bibinfo {volume} {175}},\ \bibinfo {pages} {01025} (\bibinfo {year}
  {2018})},\ \Eprint {http://arxiv.org/abs/1711.00281} {arXiv:1711.00281
  [hep-ph]} \BibitemShut {NoStop}%
\bibitem [{\citenamefont {Gérardin}\ \emph
  {et~al.}(2019{\natexlab{a}})\citenamefont {Gérardin}, \citenamefont
  {Meyer},\ and\ \citenamefont {Nyffeler}}]{Gerardin:2019vio}%
  \BibitemOpen
  \bibfield  {author} {\bibinfo {author} {\bibfnamefont {A.}~\bibnamefont
  {Gérardin}}, \bibinfo {author} {\bibfnamefont {H.~B.}\ \bibnamefont
  {Meyer}}, \ and\ \bibinfo {author} {\bibfnamefont {A.}~\bibnamefont
  {Nyffeler}},\ }\bibfield  {title} {\enquote {\bibinfo {title} {{Lattice
  calculation of the pion transition form factor with $N_f=2+1$ Wilson
  quarks}},}\ }\href {\doibase 10.1103/PhysRevD.100.034520} {\bibfield
  {journal} {\bibinfo  {journal} {Phys. Rev. D}\ }\textbf {\bibinfo {volume}
  {100}},\ \bibinfo {pages} {034520} (\bibinfo {year} {2019}{\natexlab{a}})},\
  \Eprint {http://arxiv.org/abs/1903.09471} {arXiv:1903.09471 [hep-lat]}
  \BibitemShut {NoStop}%
\bibitem [{\citenamefont {Gronberg}\ \emph {et~al.}(1998)\citenamefont
  {Gronberg} \emph {et~al.}}]{Gronberg:1997fj}%
  \BibitemOpen
  \bibfield  {author} {\bibinfo {author} {\bibfnamefont {J.}~\bibnamefont
  {Gronberg}} \emph {et~al.} (\bibinfo {collaboration} {CLEO}),\ }\bibfield
  {title} {\enquote {\bibinfo {title} {{Measurements of the meson-photon
  transition form factors of light pseudoscalar mesons at large momentum
  transfer}},}\ }\href {\doibase 10.1103/PhysRevD.57.33} {\bibfield  {journal}
  {\bibinfo  {journal} {Phys. Rev. D}\ }\textbf {\bibinfo {volume} {57}},\
  \bibinfo {pages} {33} (\bibinfo {year} {1998})},\ \Eprint
  {http://arxiv.org/abs/hep-ex/9707031} {arXiv:hep-ex/9707031} \BibitemShut
  {NoStop}%
\bibitem [{\citenamefont {del Amo~Sanchez}\ \emph {et~al.}(2011)\citenamefont
  {del Amo~Sanchez} \emph {et~al.}}]{BABAR:2011ad}%
  \BibitemOpen
  \bibfield  {author} {\bibinfo {author} {\bibfnamefont {P.}~\bibnamefont {del
  Amo~Sanchez}} \emph {et~al.} (\bibinfo {collaboration} {BaBar}),\ }\bibfield
  {title} {\enquote {\bibinfo {title} {{Measurement of the $\gamma \gamma^* \to
  \eta$ and $\gamma \gamma^* \to \eta'$ transition form factors}},}\ }\href
  {\doibase 10.1103/PhysRevD.84.052001} {\bibfield  {journal} {\bibinfo
  {journal} {Phys. Rev. D}\ }\textbf {\bibinfo {volume} {84}},\ \bibinfo
  {pages} {052001} (\bibinfo {year} {2011})},\ \Eprint
  {http://arxiv.org/abs/1101.1142} {arXiv:1101.1142 [hep-ex]} \BibitemShut
  {NoStop}%
\bibitem [{\citenamefont {Lees}\ \emph {et~al.}(2018)\citenamefont {Lees} \emph
  {et~al.}}]{BaBar:2018zpn}%
  \BibitemOpen
  \bibfield  {author} {\bibinfo {author} {\bibfnamefont {J.~P.}\ \bibnamefont
  {Lees}} \emph {et~al.} (\bibinfo {collaboration} {BaBar}),\ }\bibfield
  {title} {\enquote {\bibinfo {title} {{Measurement of the
  $\gamma^{\star}\gamma^{\star} \to \eta'$ transition form factor}},}\ }\href
  {\doibase 10.1103/PhysRevD.98.112002} {\bibfield  {journal} {\bibinfo
  {journal} {Phys. Rev. D}\ }\textbf {\bibinfo {volume} {98}},\ \bibinfo
  {pages} {112002} (\bibinfo {year} {2018})},\ \Eprint
  {http://arxiv.org/abs/1808.08038} {arXiv:1808.08038 [hep-ex]} \BibitemShut
  {NoStop}%
\bibitem [{\citenamefont {Blum}\ \emph {et~al.}(2020)\citenamefont {Blum},
  \citenamefont {Christ}, \citenamefont {Hayakawa}, \citenamefont {Izubuchi},
  \citenamefont {Jin}, \citenamefont {Jung},\ and\ \citenamefont
  {Lehner}}]{Blum:2019ugy}%
  \BibitemOpen
  \bibfield  {author} {\bibinfo {author} {\bibfnamefont {T.}~\bibnamefont
  {Blum}}, \bibinfo {author} {\bibfnamefont {N.}~\bibnamefont {Christ}},
  \bibinfo {author} {\bibfnamefont {M.}~\bibnamefont {Hayakawa}}, \bibinfo
  {author} {\bibfnamefont {T.}~\bibnamefont {Izubuchi}}, \bibinfo {author}
  {\bibfnamefont {L.}~\bibnamefont {Jin}}, \bibinfo {author} {\bibfnamefont
  {C.}~\bibnamefont {Jung}}, \ and\ \bibinfo {author} {\bibfnamefont
  {C.}~\bibnamefont {Lehner}},\ }\bibfield  {title} {\enquote {\bibinfo {title}
  {Hadronic light-by-light scattering contribution to the muon anomalous
  magnetic moment from lattice {QCD}},}\ }\href {\doibase
  10.1103/PhysRevLett.124.132002} {\bibfield  {journal} {\bibinfo  {journal}
  {Phys. Rev. Lett.}\ }\textbf {\bibinfo {volume} {124}},\ \bibinfo {pages}
  {132002} (\bibinfo {year} {2020})},\ \Eprint
  {http://arxiv.org/abs/1911.08123} {arXiv:1911.08123 [hep-lat]} \BibitemShut
  {NoStop}%
\bibitem [{\citenamefont {Asmussen}\ \emph {et~al.}(2020)\citenamefont
  {Asmussen}, \citenamefont {Gérardin}, \citenamefont {Green}, \citenamefont
  {Meyer},\ and\ \citenamefont {Nyffeler}}]{mainzHLbL1}%
  \BibitemOpen
  \bibfield  {author} {\bibinfo {author} {\bibfnamefont {N.}~\bibnamefont
  {Asmussen}}, \bibinfo {author} {\bibfnamefont {A.}~\bibnamefont {Gérardin}},
  \bibinfo {author} {\bibfnamefont {J.~R.}\ \bibnamefont {Green}}, \bibinfo
  {author} {\bibfnamefont {H.~B.}\ \bibnamefont {Meyer}}, \ and\ \bibinfo
  {author} {\bibfnamefont {A.}~\bibnamefont {Nyffeler}},\ }\bibfield  {title}
  {\enquote {\bibinfo {title} {{Hadronic light-by-light scattering contribution
  to the muon $g-2$ from lattice QCD: semi-analytical calculation of the QED
  kernel}},}\ }\href@noop {} {\  (\bibinfo {year} {2020})},\ \bibinfo {note}
  {in preparation}\BibitemShut {NoStop}%
\bibitem [{\citenamefont {Asmussen}\ \emph
  {et~al.}(2019{\natexlab{b}})\citenamefont {Asmussen}, \citenamefont {Chao},
  \citenamefont {Gérardin}, \citenamefont {Green}, \citenamefont {Hudspith},
  \citenamefont {Meyer},\ and\ \citenamefont {Nyffeler}}]{Asmussen:2019act}%
  \BibitemOpen
  \bibfield  {author} {\bibinfo {author} {\bibfnamefont {N.}~\bibnamefont
  {Asmussen}}, \bibinfo {author} {\bibfnamefont {E.-H.}\ \bibnamefont {Chao}},
  \bibinfo {author} {\bibfnamefont {A.}~\bibnamefont {Gérardin}}, \bibinfo
  {author} {\bibfnamefont {J.~R.}\ \bibnamefont {Green}}, \bibinfo {author}
  {\bibfnamefont {R.~J.}\ \bibnamefont {Hudspith}}, \bibinfo {author}
  {\bibfnamefont {H.~B.}\ \bibnamefont {Meyer}}, \ and\ \bibinfo {author}
  {\bibfnamefont {A.}~\bibnamefont {Nyffeler}},\ }\bibfield  {title} {\enquote
  {\bibinfo {title} {{Developments in the position-space approach to the HLbL
  contribution to the muon $g-2$ on the lattice}},}\ }\bibfield  {booktitle}
  {\emph {\bibinfo {booktitle} {{Proceedings, 37th International Symposium on
  Lattice Field Theory (Lattice 2019), Wuhan, China, 16--22 June 2019}}},\
  }\href@noop {} {\  (\bibinfo {year} {2019}{\natexlab{b}})},\ \Eprint
  {http://arxiv.org/abs/1911.05573} {arXiv:1911.05573 [hep-lat]} \BibitemShut
  {NoStop}%
\bibitem [{\citenamefont {Gérardin}\ \emph
  {et~al.}(2019{\natexlab{b}})\citenamefont {Gérardin}, \citenamefont
  {Harris},\ and\ \citenamefont {Meyer}}]{Gerardin:2018kpy}%
  \BibitemOpen
  \bibfield  {author} {\bibinfo {author} {\bibfnamefont {A.}~\bibnamefont
  {Gérardin}}, \bibinfo {author} {\bibfnamefont {T.}~\bibnamefont {Harris}}, \
  and\ \bibinfo {author} {\bibfnamefont {H.~B.}\ \bibnamefont {Meyer}},\
  }\bibfield  {title} {\enquote {\bibinfo {title} {{Nonperturbative
  renormalization and $O(a)$-improvement of the nonsinglet vector current with
  $N_f=2+1$ Wilson fermions and tree-level Symanzik improved gauge action}},}\
  }\href {\doibase 10.1103/PhysRevD.99.014519} {\bibfield  {journal} {\bibinfo
  {journal} {Phys. Rev. D}\ }\textbf {\bibinfo {volume} {99}},\ \bibinfo
  {pages} {014519} (\bibinfo {year} {2019}{\natexlab{b}})},\ \Eprint
  {http://arxiv.org/abs/1811.08209} {arXiv:1811.08209 [hep-lat]} \BibitemShut
  {NoStop}%
\bibitem [{\citenamefont {Bruno}\ \emph {et~al.}(2015)\citenamefont {Bruno}
  \emph {et~al.}}]{Bruno:2014jqa}%
  \BibitemOpen
  \bibfield  {author} {\bibinfo {author} {\bibfnamefont {M.}~\bibnamefont
  {Bruno}} \emph {et~al.},\ }\bibfield  {title} {\enquote {\bibinfo {title}
  {{Simulation of QCD with $N_\mathrm{f} = 2 + 1$ flavors of non-perturbatively
  improved Wilson fermions}},}\ }\href {\doibase 10.1007/JHEP02(2015)043}
  {\bibfield  {journal} {\bibinfo  {journal} {JHEP}\ }\textbf {\bibinfo
  {volume} {02}},\ \bibinfo {pages} {043} (\bibinfo {year} {2015})},\ \Eprint
  {http://arxiv.org/abs/1411.3982} {arXiv:1411.3982 [hep-lat]} \BibitemShut
  {NoStop}%
\bibitem [{\citenamefont {Bruno}\ \emph {et~al.}(2017)\citenamefont {Bruno},
  \citenamefont {Korzec},\ and\ \citenamefont {Schaefer}}]{Bruno:2016plf}%
  \BibitemOpen
  \bibfield  {author} {\bibinfo {author} {\bibfnamefont {M.}~\bibnamefont
  {Bruno}}, \bibinfo {author} {\bibfnamefont {T.}~\bibnamefont {Korzec}}, \
  and\ \bibinfo {author} {\bibfnamefont {S.}~\bibnamefont {Schaefer}},\
  }\bibfield  {title} {\enquote {\bibinfo {title} {{Setting the scale for the
  CLS $2 + 1$ flavor ensembles}},}\ }\href {\doibase
  10.1103/PhysRevD.95.074504} {\bibfield  {journal} {\bibinfo  {journal} {Phys.
  Rev. D}\ }\textbf {\bibinfo {volume} {95}},\ \bibinfo {pages} {074504}
  (\bibinfo {year} {2017})},\ \Eprint {http://arxiv.org/abs/1608.08900}
  {arXiv:1608.08900 [hep-lat]} \BibitemShut {NoStop}%
\bibitem [{\citenamefont {Peardon}\ \emph {et~al.}(2009)\citenamefont
  {Peardon}, \citenamefont {Bulava}, \citenamefont {Foley}, \citenamefont
  {Morningstar}, \citenamefont {Dudek}, \citenamefont {Edwards}, \citenamefont
  {Joó}, \citenamefont {Lin}, \citenamefont {Richards},\ and\ \citenamefont
  {Juge}}]{Peardon:2009gh}%
  \BibitemOpen
  \bibfield  {author} {\bibinfo {author} {\bibfnamefont {M.}~\bibnamefont
  {Peardon}}, \bibinfo {author} {\bibfnamefont {J.}~\bibnamefont {Bulava}},
  \bibinfo {author} {\bibfnamefont {J.}~\bibnamefont {Foley}}, \bibinfo
  {author} {\bibfnamefont {C.}~\bibnamefont {Morningstar}}, \bibinfo {author}
  {\bibfnamefont {J.}~\bibnamefont {Dudek}}, \bibinfo {author} {\bibfnamefont
  {R.~G.}\ \bibnamefont {Edwards}}, \bibinfo {author} {\bibfnamefont
  {B.}~\bibnamefont {Joó}}, \bibinfo {author} {\bibfnamefont {H.-W.}\
  \bibnamefont {Lin}}, \bibinfo {author} {\bibfnamefont {D.~G.}\ \bibnamefont
  {Richards}}, \ and\ \bibinfo {author} {\bibfnamefont {K.~J.}\ \bibnamefont
  {Juge}} (\bibinfo {collaboration} {Hadron Spectrum}),\ }\bibfield  {title}
  {\enquote {\bibinfo {title} {{Novel quark-field creation operator
  construction for hadronic physics in lattice QCD}},}\ }\href {\doibase
  10.1103/PhysRevD.80.054506} {\bibfield  {journal} {\bibinfo  {journal} {Phys.
  Rev. D}\ }\textbf {\bibinfo {volume} {80}},\ \bibinfo {pages} {054506}
  (\bibinfo {year} {2009})},\ \Eprint {http://arxiv.org/abs/0905.2160}
  {arXiv:0905.2160 [hep-lat]} \BibitemShut {NoStop}%
\bibitem [{\citenamefont {Morningstar}\ \emph {et~al.}(2011)\citenamefont
  {Morningstar}, \citenamefont {Bulava}, \citenamefont {Foley}, \citenamefont
  {Juge}, \citenamefont {Lenkner}, \citenamefont {Peardon},\ and\ \citenamefont
  {Wong}}]{Morningstar:2011ka}%
  \BibitemOpen
  \bibfield  {author} {\bibinfo {author} {\bibfnamefont {C.}~\bibnamefont
  {Morningstar}}, \bibinfo {author} {\bibfnamefont {J.}~\bibnamefont {Bulava}},
  \bibinfo {author} {\bibfnamefont {J.}~\bibnamefont {Foley}}, \bibinfo
  {author} {\bibfnamefont {K.~J.}\ \bibnamefont {Juge}}, \bibinfo {author}
  {\bibfnamefont {D.}~\bibnamefont {Lenkner}}, \bibinfo {author} {\bibfnamefont
  {M.}~\bibnamefont {Peardon}}, \ and\ \bibinfo {author} {\bibfnamefont
  {C.~H.}\ \bibnamefont {Wong}},\ }\bibfield  {title} {\enquote {\bibinfo
  {title} {{Improved stochastic estimation of quark propagation with Laplacian
  Heaviside smearing in lattice QCD}},}\ }\href {\doibase
  10.1103/PhysRevD.83.114505} {\bibfield  {journal} {\bibinfo  {journal} {Phys.
  Rev. D}\ }\textbf {\bibinfo {volume} {83}},\ \bibinfo {pages} {114505}
  (\bibinfo {year} {2011})},\ \Eprint {http://arxiv.org/abs/1104.3870}
  {arXiv:1104.3870 [hep-lat]} \BibitemShut {NoStop}%
\bibitem [{\citenamefont {Briceno}\ \emph {et~al.}(2017)\citenamefont
  {Briceno}, \citenamefont {Dudek}, \citenamefont {Edwards},\ and\
  \citenamefont {Wilson}}]{Briceno:2016mjc}%
  \BibitemOpen
  \bibfield  {author} {\bibinfo {author} {\bibfnamefont {R.~A.}\ \bibnamefont
  {Briceno}}, \bibinfo {author} {\bibfnamefont {J.~J.}\ \bibnamefont {Dudek}},
  \bibinfo {author} {\bibfnamefont {R.~G.}\ \bibnamefont {Edwards}}, \ and\
  \bibinfo {author} {\bibfnamefont {D.~J.}\ \bibnamefont {Wilson}},\ }\bibfield
   {title} {\enquote {\bibinfo {title} {{Isoscalar $\pi\pi$ scattering and the
  $\sigma$ meson resonance from QCD}},}\ }\href {\doibase
  10.1103/PhysRevLett.118.022002} {\bibfield  {journal} {\bibinfo  {journal}
  {Phys. Rev. Lett.}\ }\textbf {\bibinfo {volume} {118}},\ \bibinfo {pages}
  {022002} (\bibinfo {year} {2017})},\ \Eprint
  {http://arxiv.org/abs/1607.05900} {arXiv:1607.05900 [hep-ph]} \BibitemShut
  {NoStop}%
\bibitem [{\citenamefont {Nyffeler}(2016)}]{Nyffeler:2016gnb}%
  \BibitemOpen
  \bibfield  {author} {\bibinfo {author} {\bibfnamefont {A.}~\bibnamefont
  {Nyffeler}},\ }\bibfield  {title} {\enquote {\bibinfo {title} {{Precision of
  a data-driven estimate of hadronic light-by-light scattering in the muon
  $g-2$: Pseudoscalar-pole contribution}},}\ }\href {\doibase
  10.1103/PhysRevD.94.053006} {\bibfield  {journal} {\bibinfo  {journal} {Phys.
  Rev.}\ }\textbf {\bibinfo {volume} {D94}},\ \bibinfo {pages} {053006}
  (\bibinfo {year} {2016})},\ \Eprint {http://arxiv.org/abs/1602.03398}
  {arXiv:1602.03398 [hep-ph]} \BibitemShut {NoStop}%
\bibitem [{\citenamefont {Masjuan}\ and\ \citenamefont
  {Sanchez-Puertas}(2017)}]{Masjuan:2017tvw}%
  \BibitemOpen
  \bibfield  {author} {\bibinfo {author} {\bibfnamefont {P.}~\bibnamefont
  {Masjuan}}\ and\ \bibinfo {author} {\bibfnamefont {P.}~\bibnamefont
  {Sanchez-Puertas}},\ }\bibfield  {title} {\enquote {\bibinfo {title}
  {{Pseudoscalar-pole contribution to the $(g_{\mu}-2)$: a rational
  approach}},}\ }\href {\doibase 10.1103/PhysRevD.95.054026} {\bibfield
  {journal} {\bibinfo  {journal} {Phys. Rev. D}\ }\textbf {\bibinfo {volume}
  {95}},\ \bibinfo {pages} {054026} (\bibinfo {year} {2017})},\ \Eprint
  {http://arxiv.org/abs/1701.05829} {arXiv:1701.05829 [hep-ph]} \BibitemShut
  {NoStop}%
\bibitem [{\citenamefont {Asmussen}\ \emph {et~al.}(2018)\citenamefont
  {Asmussen}, \citenamefont {Gérardin}, \citenamefont {Meyer},\ and\
  \citenamefont {Nyffeler}}]{Asmussen:2017bup}%
  \BibitemOpen
  \bibfield  {author} {\bibinfo {author} {\bibfnamefont {N.}~\bibnamefont
  {Asmussen}}, \bibinfo {author} {\bibfnamefont {A.}~\bibnamefont {Gérardin}},
  \bibinfo {author} {\bibfnamefont {H.~B.}\ \bibnamefont {Meyer}}, \ and\
  \bibinfo {author} {\bibfnamefont {A.}~\bibnamefont {Nyffeler}},\ }\bibfield
  {title} {\enquote {\bibinfo {title} {{Exploratory studies for the
  position-space approach to hadronic light-by-light scattering in the muon
  $g-2$}},}\ }\bibfield  {booktitle} {\emph {\bibinfo {booktitle}
  {{Proceedings, 35th International Symposium on Lattice Field Theory (Lattice
  2017): Granada, Spain, June 18-24, 2017}}},\ }\href {\doibase
  10.1051/epjconf/201817506023} {\bibfield  {journal} {\bibinfo  {journal} {EPJ
  Web Conf.}\ }\textbf {\bibinfo {volume} {175}},\ \bibinfo {pages} {06023}
  (\bibinfo {year} {2018})},\ \Eprint {http://arxiv.org/abs/1711.02466}
  {arXiv:1711.02466 [hep-lat]} \BibitemShut {NoStop}%
\bibitem [{\citenamefont {Gérardin}\ \emph {et~al.}(2018)\citenamefont
  {Gérardin}, \citenamefont {Green}, \citenamefont {Gryniuk}, \citenamefont
  {von Hippel}, \citenamefont {Meyer}, \citenamefont {Pascalutsa},\ and\
  \citenamefont {Wittig}}]{Gerardin:2017ryf}%
  \BibitemOpen
  \bibfield  {author} {\bibinfo {author} {\bibfnamefont {A.}~\bibnamefont
  {Gérardin}}, \bibinfo {author} {\bibfnamefont {J.}~\bibnamefont {Green}},
  \bibinfo {author} {\bibfnamefont {O.}~\bibnamefont {Gryniuk}}, \bibinfo
  {author} {\bibfnamefont {G.}~\bibnamefont {von Hippel}}, \bibinfo {author}
  {\bibfnamefont {H.~B.}\ \bibnamefont {Meyer}}, \bibinfo {author}
  {\bibfnamefont {V.}~\bibnamefont {Pascalutsa}}, \ and\ \bibinfo {author}
  {\bibfnamefont {H.}~\bibnamefont {Wittig}},\ }\bibfield  {title} {\enquote
  {\bibinfo {title} {{Hadronic light-by-light scattering amplitudes from
  lattice QCD versus dispersive sum rules}},}\ }\href {\doibase
  10.1103/PhysRevD.98.074501} {\bibfield  {journal} {\bibinfo  {journal} {Phys.
  Rev. D}\ }\textbf {\bibinfo {volume} {98}},\ \bibinfo {pages} {074501}
  (\bibinfo {year} {2018})},\ \Eprint {http://arxiv.org/abs/1712.00421}
  {arXiv:1712.00421 [hep-lat]} \BibitemShut {NoStop}%
\bibitem [{\citenamefont {Bijnens}(2016)}]{Bijnens:2015jqa}%
  \BibitemOpen
  \bibfield  {author} {\bibinfo {author} {\bibfnamefont {J.}~\bibnamefont
  {Bijnens}},\ }\bibfield  {title} {\enquote {\bibinfo {title} {{Hadronic
  light-by-light contribution to $a_\mu$: extended Nambu-Jona-Lasinio, chiral
  quark models and chiral Lagrangians}},}\ }\href {\doibase
  10.1051/epjconf/201611801002} {\bibfield  {journal} {\bibinfo  {journal} {EPJ
  Web Conf.}\ }\textbf {\bibinfo {volume} {118}},\ \bibinfo {pages} {01002}
  (\bibinfo {year} {2016})},\ \Eprint {http://arxiv.org/abs/1510.05796}
  {arXiv:1510.05796 [hep-ph]} \BibitemShut {NoStop}%
\bibitem [{\citenamefont {Bijnens}\ and\ \citenamefont
  {Relefors}(2016)}]{Bijnens:2016hgx}%
  \BibitemOpen
  \bibfield  {author} {\bibinfo {author} {\bibfnamefont {J.}~\bibnamefont
  {Bijnens}}\ and\ \bibinfo {author} {\bibfnamefont {J.}~\bibnamefont
  {Relefors}},\ }\bibfield  {title} {\enquote {\bibinfo {title} {{Pion
  light-by-light contributions to the muon $g-2$}},}\ }\href {\doibase
  10.1007/JHEP09(2016)113} {\bibfield  {journal} {\bibinfo  {journal} {JHEP}\
  }\textbf {\bibinfo {volume} {09}},\ \bibinfo {pages} {113} (\bibinfo {year}
  {2016})},\ \Eprint {http://arxiv.org/abs/1608.01454} {arXiv:1608.01454
  [hep-ph]} \BibitemShut {NoStop}%
\bibitem [{\citenamefont {Bijnens}\ \emph {et~al.}(1996)\citenamefont
  {Bijnens}, \citenamefont {Pallante},\ and\ \citenamefont
  {Prades}}]{Bijnens:1995xf}%
  \BibitemOpen
  \bibfield  {author} {\bibinfo {author} {\bibfnamefont {J.}~\bibnamefont
  {Bijnens}}, \bibinfo {author} {\bibfnamefont {E.}~\bibnamefont {Pallante}}, \
  and\ \bibinfo {author} {\bibfnamefont {J.}~\bibnamefont {Prades}},\
  }\bibfield  {title} {\enquote {\bibinfo {title} {{Analysis of the hadronic
  light-by-light contributions to the muon $g-2$}},}\ }\href {\doibase
  10.1016/0550-3213(96)00288-X} {\bibfield  {journal} {\bibinfo  {journal}
  {Nucl. Phys. B}\ }\textbf {\bibinfo {volume} {474}},\ \bibinfo {pages} {379}
  (\bibinfo {year} {1996})},\ \Eprint {http://arxiv.org/abs/hep-ph/9511388}
  {arXiv:hep-ph/9511388} \BibitemShut {NoStop}%
\bibitem [{\citenamefont {Knecht}\ \emph {et~al.}(2018)\citenamefont {Knecht},
  \citenamefont {Narison}, \citenamefont {Rabemananjara},\ and\ \citenamefont
  {Rabetiarivony}}]{Knecht:2018sci}%
  \BibitemOpen
  \bibfield  {author} {\bibinfo {author} {\bibfnamefont {M.}~\bibnamefont
  {Knecht}}, \bibinfo {author} {\bibfnamefont {S.}~\bibnamefont {Narison}},
  \bibinfo {author} {\bibfnamefont {A.}~\bibnamefont {Rabemananjara}}, \ and\
  \bibinfo {author} {\bibfnamefont {D.}~\bibnamefont {Rabetiarivony}},\
  }\bibfield  {title} {\enquote {\bibinfo {title} {{Scalar meson contributions
  to $a_\mu$ from hadronic light-by-light scattering}},}\ }\href {\doibase
  10.1016/j.physletb.2018.10.048} {\bibfield  {journal} {\bibinfo  {journal}
  {Phys. Lett. B}\ }\textbf {\bibinfo {volume} {787}},\ \bibinfo {pages} {111}
  (\bibinfo {year} {2018})},\ \Eprint {http://arxiv.org/abs/1808.03848}
  {arXiv:1808.03848 [hep-ph]} \BibitemShut {NoStop}%
\bibitem [{\citenamefont {Prades}\ \emph {et~al.}(2009)\citenamefont {Prades},
  \citenamefont {de~Rafael},\ and\ \citenamefont {Vainshtein}}]{Prades:2009tw}%
  \BibitemOpen
  \bibfield  {author} {\bibinfo {author} {\bibfnamefont {J.}~\bibnamefont
  {Prades}}, \bibinfo {author} {\bibfnamefont {E.}~\bibnamefont {de~Rafael}}, \
  and\ \bibinfo {author} {\bibfnamefont {A.}~\bibnamefont {Vainshtein}},\
  }\bibfield  {title} {\enquote {\bibinfo {title} {The hadronic light-by-light
  scattering contribution to the muon and electron anomalous magnetic
  moments},}\ }\href {\doibase 10.1142/9789814271844_0009} {\bibfield
  {journal} {\bibinfo  {journal} {Adv. Ser. Direct. High Energy Phys.}\
  }\textbf {\bibinfo {volume} {20}},\ \bibinfo {pages} {303} (\bibinfo {year}
  {2009})},\ \Eprint {http://arxiv.org/abs/0901.0306} {arXiv:0901.0306
  [hep-ph]} \BibitemShut {NoStop}%
\bibitem [{\citenamefont {Gérardin}\ \emph
  {et~al.}(2019{\natexlab{c}})\citenamefont {Gérardin}, \citenamefont {Cè},
  \citenamefont {von Hippel}, \citenamefont {Hörz}, \citenamefont {Meyer},
  \citenamefont {Mohler}, \citenamefont {Ottnad}, \citenamefont {Wilhelm},\
  and\ \citenamefont {Wittig}}]{Gerardin:2019rua}%
  \BibitemOpen
  \bibfield  {author} {\bibinfo {author} {\bibfnamefont {A.}~\bibnamefont
  {Gérardin}}, \bibinfo {author} {\bibfnamefont {M.}~\bibnamefont {Cè}},
  \bibinfo {author} {\bibfnamefont {G.}~\bibnamefont {von Hippel}}, \bibinfo
  {author} {\bibfnamefont {B.}~\bibnamefont {Hörz}}, \bibinfo {author}
  {\bibfnamefont {H.~B.}\ \bibnamefont {Meyer}}, \bibinfo {author}
  {\bibfnamefont {D.}~\bibnamefont {Mohler}}, \bibinfo {author} {\bibfnamefont
  {K.}~\bibnamefont {Ottnad}}, \bibinfo {author} {\bibfnamefont
  {J.}~\bibnamefont {Wilhelm}}, \ and\ \bibinfo {author} {\bibfnamefont
  {H.}~\bibnamefont {Wittig}},\ }\bibfield  {title} {\enquote {\bibinfo {title}
  {{The leading hadronic contribution to $(g-2)_\mu$ from lattice QCD with
  $N_{\rm f}=2+1$ flavours of O($a$) improved Wilson quarks}},}\ }\href
  {\doibase 10.1103/PhysRevD.100.014510} {\bibfield  {journal} {\bibinfo
  {journal} {Phys. Rev. D}\ }\textbf {\bibinfo {volume} {100}},\ \bibinfo
  {pages} {014510} (\bibinfo {year} {2019}{\natexlab{c}})},\ \Eprint
  {http://arxiv.org/abs/1904.03120} {arXiv:1904.03120 [hep-lat]} \BibitemShut
  {NoStop}%
\bibitem [{\citenamefont {Hoferichter}\ \emph {et~al.}(2018)\citenamefont
  {Hoferichter}, \citenamefont {Hoid}, \citenamefont {Kubis}, \citenamefont
  {Leupold},\ and\ \citenamefont {Schneider}}]{Hoferichter:2018dmo}%
  \BibitemOpen
  \bibfield  {author} {\bibinfo {author} {\bibfnamefont {M.}~\bibnamefont
  {Hoferichter}}, \bibinfo {author} {\bibfnamefont {B.-L.}\ \bibnamefont
  {Hoid}}, \bibinfo {author} {\bibfnamefont {B.}~\bibnamefont {Kubis}},
  \bibinfo {author} {\bibfnamefont {S.}~\bibnamefont {Leupold}}, \ and\
  \bibinfo {author} {\bibfnamefont {S.~P.}\ \bibnamefont {Schneider}},\
  }\bibfield  {title} {\enquote {\bibinfo {title} {{Pion-pole contribution to
  hadronic light-by-light scattering in the anomalous magnetic moment of the
  muon}},}\ }\href {\doibase 10.1103/PhysRevLett.121.112002} {\bibfield
  {journal} {\bibinfo  {journal} {Phys. Rev. Lett.}\ }\textbf {\bibinfo
  {volume} {121}},\ \bibinfo {pages} {112002} (\bibinfo {year} {2018})},\
  \Eprint {http://arxiv.org/abs/1805.01471} {arXiv:1805.01471 [hep-ph]}
  \BibitemShut {NoStop}%
\bibitem [{\citenamefont {Colangelo}\ \emph {et~al.}(2017)\citenamefont
  {Colangelo}, \citenamefont {Hoferichter}, \citenamefont {Procura},\ and\
  \citenamefont {Stoffer}}]{Colangelo:2017fiz}%
  \BibitemOpen
  \bibfield  {author} {\bibinfo {author} {\bibfnamefont {G.}~\bibnamefont
  {Colangelo}}, \bibinfo {author} {\bibfnamefont {M.}~\bibnamefont
  {Hoferichter}}, \bibinfo {author} {\bibfnamefont {M.}~\bibnamefont
  {Procura}}, \ and\ \bibinfo {author} {\bibfnamefont {P.}~\bibnamefont
  {Stoffer}},\ }\bibfield  {title} {\enquote {\bibinfo {title} {{Dispersion
  relation for hadronic light-by-light scattering: two-pion contributions}},}\
  }\href {\doibase 10.1007/JHEP04(2017)161} {\bibfield  {journal} {\bibinfo
  {journal} {JHEP}\ }\textbf {\bibinfo {volume} {04}},\ \bibinfo {pages} {161}
  (\bibinfo {year} {2017})},\ \Eprint {http://arxiv.org/abs/1702.07347}
  {arXiv:1702.07347 [hep-ph]} \BibitemShut {NoStop}%
\bibitem [{\citenamefont {L{\"u}scher}\ and\ \citenamefont
  {Schaefer}(2013)}]{Luscher:2012av}%
  \BibitemOpen
  \bibfield  {author} {\bibinfo {author} {\bibfnamefont {M.}~\bibnamefont
  {L{\"u}scher}}\ and\ \bibinfo {author} {\bibfnamefont {S.}~\bibnamefont
  {Schaefer}},\ }\bibfield  {title} {\enquote {\bibinfo {title} {{Lattice QCD
  with open boundary conditions and twisted-mass reweighting}},}\ }\href
  {\doibase 10.1016/j.cpc.2012.10.003} {\bibfield  {journal} {\bibinfo
  {journal} {Comput. Phys. Commun.}\ }\textbf {\bibinfo {volume} {184}},\
  \bibinfo {pages} {519} (\bibinfo {year} {2013})},\ \Eprint
  {http://arxiv.org/abs/1206.2809} {arXiv:1206.2809 [hep-lat]} \BibitemShut
  {NoStop}%
\bibitem [{\citenamefont {Edwards}\ and\ \citenamefont
  {Joó}(2005)}]{Edwards:2004sx}%
  \BibitemOpen
  \bibfield  {author} {\bibinfo {author} {\bibfnamefont {R.~G.}\ \bibnamefont
  {Edwards}}\ and\ \bibinfo {author} {\bibfnamefont {B.}~\bibnamefont {Joó}}
  (\bibinfo {collaboration} {SciDAC, LHPC, UKQCD}),\ }\bibfield  {title}
  {\enquote {\bibinfo {title} {{The Chroma software system for lattice QCD}},}\
  }\bibfield  {booktitle} {\emph {\bibinfo {booktitle} {{Lattice field theory.
  Proceedings, 22nd International Symposium, Lattice 2004, Batavia, USA, June
  21-26, 2004}}},\ }\href {\doibase 10.1016/j.nuclphysbps.2004.11.254}
  {\bibfield  {journal} {\bibinfo  {journal} {Nucl. Phys. B (Proc. Suppl.)}\
  }\textbf {\bibinfo {volume} {140}},\ \bibinfo {pages} {832} (\bibinfo {year}
  {2005})},\ \Eprint {http://arxiv.org/abs/hep-lat/0409003}
  {arXiv:hep-lat/0409003} \BibitemShut {NoStop}%
\bibitem [{\citenamefont {Stathopoulos}\ and\ \citenamefont
  {McCombs}(2010)}]{PRIMME}%
  \BibitemOpen
  \bibfield  {author} {\bibinfo {author} {\bibfnamefont {A.}~\bibnamefont
  {Stathopoulos}}\ and\ \bibinfo {author} {\bibfnamefont {J.~R.}\ \bibnamefont
  {McCombs}},\ }\bibfield  {title} {\enquote {\bibinfo {title} {{PRIMME}:
  {PR}econditioned {I}terative {M}ulti{M}ethod {E}igensolver---methods and
  software description},}\ }\href {\doibase 10.1145/1731022.1731031} {\bibfield
   {journal} {\bibinfo  {journal} {ACM Trans. Math. Softw.}\ }\textbf {\bibinfo
  {volume} {37}},\ \bibinfo {pages} {21:1} (\bibinfo {year}
  {2010})}\BibitemShut {NoStop}%
\bibitem [{\citenamefont {{Jülich Supercomputing Centre}}(2015)}]{juqueen}%
  \BibitemOpen
  \bibfield  {author} {\bibinfo {author} {\bibnamefont {{Jülich Supercomputing
  Centre}}},\ }\bibfield  {title} {\enquote {\bibinfo {title} {{JUQUEEN: IBM
  Blue Gene/Q} supercomputer system at the {Jülich Supercomputing Centre}},}\
  }\href {\doibase 10.17815/jlsrf-1-18} {\bibfield  {journal} {\bibinfo
  {journal} {J. Large-Scale Res. Facil.}\ }\textbf {\bibinfo {volume} {1}},\
  \bibinfo {pages} {A1} (\bibinfo {year} {2015})}\BibitemShut {NoStop}%
\bibitem [{\citenamefont {Bernard}\ and\ \citenamefont
  {Golterman}(1992)}]{Bernard:1992mk}%
  \BibitemOpen
  \bibfield  {author} {\bibinfo {author} {\bibfnamefont {C.~W.}\ \bibnamefont
  {Bernard}}\ and\ \bibinfo {author} {\bibfnamefont {M.~F.}\ \bibnamefont
  {Golterman}},\ }\bibfield  {title} {\enquote {\bibinfo {title} {{Chiral
  perturbation theory for the quenched approximation of QCD}},}\ }\href
  {\doibase 10.1103/PhysRevD.46.853} {\bibfield  {journal} {\bibinfo  {journal}
  {Phys. Rev. D}\ }\textbf {\bibinfo {volume} {46}},\ \bibinfo {pages} {853}
  (\bibinfo {year} {1992})},\ \Eprint {http://arxiv.org/abs/hep-lat/9204007}
  {arXiv:hep-lat/9204007} \BibitemShut {NoStop}%
\bibitem [{\citenamefont {Sharpe}\ and\ \citenamefont
  {Shoresh}(2000)}]{Sharpe:2000bc}%
  \BibitemOpen
  \bibfield  {author} {\bibinfo {author} {\bibfnamefont {S.~R.}\ \bibnamefont
  {Sharpe}}\ and\ \bibinfo {author} {\bibfnamefont {N.}~\bibnamefont
  {Shoresh}},\ }\bibfield  {title} {\enquote {\bibinfo {title} {{Physical
  results from unphysical simulations}},}\ }\href {\doibase
  10.1103/PhysRevD.62.094503} {\bibfield  {journal} {\bibinfo  {journal} {Phys.
  Rev. D}\ }\textbf {\bibinfo {volume} {62}},\ \bibinfo {pages} {094503}
  (\bibinfo {year} {2000})},\ \Eprint {http://arxiv.org/abs/hep-lat/0006017}
  {arXiv:hep-lat/0006017} \BibitemShut {NoStop}%
\bibitem [{\citenamefont {Giusti}\ and\ \citenamefont
  {Lüscher}(2009)}]{Giusti:2008vb}%
  \BibitemOpen
  \bibfield  {author} {\bibinfo {author} {\bibfnamefont {L.}~\bibnamefont
  {Giusti}}\ and\ \bibinfo {author} {\bibfnamefont {M.}~\bibnamefont
  {Lüscher}},\ }\bibfield  {title} {\enquote {\bibinfo {title} {{Chiral
  symmetry breaking and the Banks-Casher relation in lattice QCD with Wilson
  quarks}},}\ }\href {\doibase 10.1088/1126-6708/2009/03/013} {\bibfield
  {journal} {\bibinfo  {journal} {JHEP}\ }\textbf {\bibinfo {volume} {03}},\
  \bibinfo {pages} {013} (\bibinfo {year} {2009})},\ \Eprint
  {http://arxiv.org/abs/0812.3638} {arXiv:0812.3638 [hep-lat]} \BibitemShut
  {NoStop}%
\bibitem [{\citenamefont {Della~Morte}\ and\ \citenamefont
  {Jüttner}(2010)}]{DellaMorte:2010aq}%
  \BibitemOpen
  \bibfield  {author} {\bibinfo {author} {\bibfnamefont {M.}~\bibnamefont
  {Della~Morte}}\ and\ \bibinfo {author} {\bibfnamefont {A.}~\bibnamefont
  {Jüttner}},\ }\bibfield  {title} {\enquote {\bibinfo {title} {{Quark
  disconnected diagrams in chiral perturbation theory}},}\ }\href {\doibase
  10.1007/JHEP11(2010)154} {\bibfield  {journal} {\bibinfo  {journal} {JHEP}\
  }\textbf {\bibinfo {volume} {11}},\ \bibinfo {pages} {154} (\bibinfo {year}
  {2010})},\ \Eprint {http://arxiv.org/abs/1009.3783} {arXiv:1009.3783
  [hep-lat]} \BibitemShut {NoStop}%
\bibitem [{\citenamefont {Wess}\ and\ \citenamefont
  {Zumino}(1971)}]{Wess:1971yu}%
  \BibitemOpen
  \bibfield  {author} {\bibinfo {author} {\bibfnamefont {J.}~\bibnamefont
  {Wess}}\ and\ \bibinfo {author} {\bibfnamefont {B.}~\bibnamefont {Zumino}},\
  }\bibfield  {title} {\enquote {\bibinfo {title} {{Consequences of anomalous
  Ward identities}},}\ }\href {\doibase 10.1016/0370-2693(71)90582-X}
  {\bibfield  {journal} {\bibinfo  {journal} {Phys. Lett. B}\ }\textbf
  {\bibinfo {volume} {37}},\ \bibinfo {pages} {95} (\bibinfo {year}
  {1971})}\BibitemShut {NoStop}%
\bibitem [{\citenamefont {Witten}(1983)}]{Witten:1983tw}%
  \BibitemOpen
  \bibfield  {author} {\bibinfo {author} {\bibfnamefont {E.}~\bibnamefont
  {Witten}},\ }\bibfield  {title} {\enquote {\bibinfo {title} {Global aspects
  of current algebra},}\ }\href {\doibase 10.1016/0550-3213(83)90063-9}
  {\bibfield  {journal} {\bibinfo  {journal} {Nucl. Phys. B}\ }\textbf
  {\bibinfo {volume} {223}},\ \bibinfo {pages} {422} (\bibinfo {year}
  {1983})}\BibitemShut {NoStop}%
\bibitem [{\citenamefont {Detmold}\ \emph {et~al.}(2006)\citenamefont
  {Detmold}, \citenamefont {Tiburzi},\ and\ \citenamefont
  {Walker-Loud}}]{Detmold:2006vu}%
  \BibitemOpen
  \bibfield  {author} {\bibinfo {author} {\bibfnamefont {W.}~\bibnamefont
  {Detmold}}, \bibinfo {author} {\bibfnamefont {B.}~\bibnamefont {Tiburzi}}, \
  and\ \bibinfo {author} {\bibfnamefont {A.}~\bibnamefont {Walker-Loud}},\
  }\bibfield  {title} {\enquote {\bibinfo {title} {{Electromagnetic and spin
  polarizabilities in lattice QCD}},}\ }\href {\doibase
  10.1103/PhysRevD.73.114505} {\bibfield  {journal} {\bibinfo  {journal} {Phys.
  Rev. D}\ }\textbf {\bibinfo {volume} {73}},\ \bibinfo {pages} {114505}
  (\bibinfo {year} {2006})},\ \Eprint {http://arxiv.org/abs/hep-lat/0603026}
  {arXiv:hep-lat/0603026} \BibitemShut {NoStop}%
\bibitem [{\citenamefont {Laporta}\ and\ \citenamefont
  {Remiddi}(1993)}]{Laporta:1992pa}%
  \BibitemOpen
  \bibfield  {author} {\bibinfo {author} {\bibfnamefont {S.}~\bibnamefont
  {Laporta}}\ and\ \bibinfo {author} {\bibfnamefont {E.}~\bibnamefont
  {Remiddi}},\ }\bibfield  {title} {\enquote {\bibinfo {title} {The analytical
  value of the electron light-light graphs contribution to the muon $(g-2)$ in
  {QED}},}\ }\href {\doibase 10.1016/0370-2693(93)91176-N} {\bibfield
  {journal} {\bibinfo  {journal} {Phys. Lett. B}\ }\textbf {\bibinfo {volume}
  {301}},\ \bibinfo {pages} {440} (\bibinfo {year} {1993})}\BibitemShut
  {NoStop}%
\end{thebibliography}%

\end{document}